\documentclass{article}
\usepackage{amsmath}
\usepackage{amsfonts}
\usepackage{amssymb}
\usepackage{amscd}
\usepackage[small]{titlesec}
\usepackage{a4wide}
\usepackage{float}
\usepackage{array, multirow}

\usepackage{graphicx}
\usepackage{caption}
\usepackage{bm}
\usepackage{latexsym}
\usepackage{color}
\usepackage{multirow}
\usepackage{dsfont}
\usepackage{indentfirst}
\usepackage[latin1]{inputenc}
\usepackage[justification=centering]{caption}

\def\qed{\hfill $\square$}

\newtheorem{definition}{Definition}[section]
\newtheorem{proposition}[definition]{Proposition}
\newtheorem{theorem}[definition]{Theorem}

\newtheorem{remark}[definition]{Remark}
\newtheorem{lema}[definition]{Lemma}
\newtheorem{corollary}[definition]{Corollary}

\numberwithin{equation}{section}

\newcolumntype{P}[1]{>{\centering\arraybackslash}p{#1}}
\newcolumntype{M}[1]{>{\centering\arraybackslash}m{#1}}

\DeclareMathAlphabet{\mathpzc}{OT1}{pzc}{m}{it}

\begin{document}
\begin{center}
{\Large{ \textbf{On the Wasserstein distance \\ and the Dobrushin uniqueness theorem.}}}

\medskip

\today
\end{center}

\begin{center}
\small{T.C. Dorlas\footnote{Dublin Institute for Advanced Studies, School of Theoretical Physics, 10 Burlington road, Dublin 04, Ireland}, B. Savoie$^{*}$}

\end{center}
\vspace{0.5cm}

\begin{abstract}
In this paper, we revisit the Dobrushin uniqueness theorem for Gibbs measures of lattice systems of interacting particles at thermal equilibrium. In a nutshell, Dobrushin's uniqueness theorem provides a practical way to derive sufficient conditions on the inverse temperature and/or model parameters assuring uniqueness of Gibbs measures by reducing the uniqueness problem to a suitable estimate of the Wasserstein distance between pairs of 1-point Gibbs measures with different boundary conditions. After proving a general result of completeness for the Wasserstein distance, we reformulate the Dobrushin uniqueness theorem in a convenient form for lattice systems of interacting particles described by Hamiltonians that are not necessarily translation-invariant with possibly infinite-range pair-potentials and with general complete metric spaces as single-spin spaces. Our reformulation includes existence, and covers both classical lattice sytems and the Euclidean
version of quantum lattice systems. Further, a generalization to the Dobrushin-Shlosman theorem for translation-invariant lattice systems is given. Subsequently, we give a series of applications of these uniqueness criteria to some high-temperature
 classical lattice systems including the Heisenberg, Potts and Ising models. An application to  classical lattice systems for which the local Gibbs measures are convex perturbations of Gaussian measures is also given. 
\end{abstract}
\vspace{1.0cm}


\noindent \textbf{MSC-2010}: 82B05, 82B10, 82B20, 82B26 (Primary); 28C20, 46G10, 60B05, 60B10 (Secondary).

\medskip

\noindent \textbf{Keywords}: (limit) Gibbs distributions, (Euclidean) Gibbs measures, DLR approach, Dobrushin's uniqueness theorem, Dobrushin-Shlosman theorem, Gaussian measures.

\medskip

\tableofcontents

\medskip

\section{Introduction.}
\label{intro}

A core problem of classical and quantum equilibrium statistical mechanics is the description of equilibrium thermodynamic properties of large systems of interacting particles, see, e.g., \cite{Wig}. The literature on mathematical characterizations of the equilibrium thermodynamics of interacting particle systems is extensive; for an account of rigorous methods, see, e.g., \cite{Mu,Ru2,Isr,BR2,Sim2,Ru3}.\\
\indent A standard way to describe the equilibrium thermodynamics of a large system of interacting particles consists in constructing the equilibrium Gibbs states of the associated infinite system at a given inverse temperature $\beta>0$ and given values of the system parameters. For systems undergoing structural phase transitions, it is expected that the set of equilibrium Gibbs states consists of more than one element corresponding to the different phases. Below, we give a brief account of some rigorous constructions of equilibrium Gibbs states in classical and quantum lattice systems, with an emphasis on lattice systems with unbounded single-spin spaces (also called state spaces).\\
\indent Equilibrium Gibbs states of classical lattice systems of interacting particles may be represented by \textit{Gibbs measures}, also called \textit{limit Gibbs distributions} or \textit{DLR} (Dobrushin-Lanford-Ruelle) \textit{measures}. A Gibbs measure is the distribution of a random field on the infinite lattice admitting a prescribed family of conditional distributions, see, e.g.,
\cite{D0, M1, M2, D1, D2, D3, D4, D5, LR} for pioneering works. Local Gibbs measures may describe local equilibrium Gibbs states of systems of interacting particles in finite domains (the particle interactions are not limited to the interior of these domains) at thermal equilibrium with their exteriors where the configurations of particles are held fixed. The latter play the role of boundary conditions and, as such, determine conditions for the distributions. Gibbs measures are then defined by means of the family of local Gibbs measures as solutions of the equilibrium DLR equation. This approach, called \textit{DLR approach}, is standard, see, e.g., \cite{Ge} and references therein. Note that this approach allows one to define Gibbs states of infinite lattice systems without resorting to any limiting procedures.\\
\indent Local equilibrium Gibbs states of finite-volume quantum systems are traditionally defined as positive normalised linear functionals on the C*-algebra of bounded operators on a Hilbert space, satisfying the KMS (Kubo-Martin-Schwinger) boundary condition relative to the time evolution consisting of a one-parameter group of *-automorphisms, see, e.g., \cite{HHW, Hug, BR2,Ru3}. Constructing from the local Gibbs states equilibrium Gibbs states of infinite-volume systems as KMS states requires the limiting time evolution to exist (in some sense) on the C*-algebra of quasi-local observables. For  applications to bounded quantum spin systems, see, e.g., \cite[Sec. 6.2]{BR2}. In the case of quantum lattice systems described by unbounded operators, the time evolution as a group of *-automorphisms of the quasi-local algebra may not exist however. See, e.g., \cite{BR2} where the case of non-interacting boson systems is discussed, and see, e.g., \cite{Pa,Pa1}  for an application of the Green's function method to this case. For a characterization of equilibrium states of infinite-volume lattice systems with the Gibbs equilibrium condition \cite[Def. 6.2.16]{BR2}, see, e.g.,  \cite{Arak1,Hug}. For bounded quantum spin systems, the Gibbs equilibrium condition is equivalent to the KMS condition, see, e.g., \cite[Sec. 6.2.2]{BR2}, and reduces to the DLR equation if the interaction is classical.\\
An alternative to the above algebraic approaches to construct equilibrium Gibbs states of quantum lattice systems of interacting particles is the so-called \textit{Euclidean approach}, see, e.g., \cite{HK,AHK,Alber1}. The Euclidean approach relies on a one-to-one correspondence between local Gibbs states, as functionals on  C*-algebras of observables, and local Gibbs measures, as Feynman-Kac measures on $\beta$-periodic path spaces. To connect both, the Matsubara functions (temperature Green functions) play a key role as they uniquely determine the local Gibbs states. For further details, see, e.g., \cite[Sec. 6]{Alber2} and \cite[Sec. 2.5]{KP}. As such, equilibrium Gibbs states of quantum lattice systems may be represented, analogously to classical lattice systems, by Gibbs measures defined through the DLR approach but with local Gibbs measures living on infinite-dimensional path spaces. They are often called \textit{Euclidean Gibbs measures}, see, e.g., \cite{Alber1}.\\ 
\indent In the study of Gibbs measures in lattice systems with unbounded single-spin spaces, two main mathematical problems arise: existence and uniqueness. For the case of compact single-spin spaces the existence problem is simpler, see e.g., \cite{D1,Ge,Sim2} and \cite[Chap. 6]{FV}. At this stage, it should be noted that, as shown in \cite{GRS}, when the set of Gibbs measures consists of several elements, some may have no physical relevance. As suggested by Euclidean quantum field theories, see, e.g., \cite{Sim1,GRS}, the Gibbs measures of interest are those for which the sequence of their moments satisfies some \textit{a priori} growth limitations at infinity. They are the so-called \textit{tempered Gibbs measures}. Uniqueness of Gibbs measures characterizes the absence of (first-order) phase transitions.  \\
\indent \textit{The existence problem.} This may be solved by constructing Gibbs measures as thermodynamic limits. This may be achieved by proving that the family of local Gibbs measures, indexed by an increasing sequence of bounded regions filling the whole lattice, has at least one limit point in the weak topology, and that this limit point is a Gibbs measure. Sufficient conditions were derived by Dobrushin \cite{D1,D2,D4} and \cite[Thm. 1]{D5}. See also \cite[Thm. 1.3]{Si}. Dobrushin's existence criteria have been applied to some classical lattice models of Euclidean lattice field theories with single-spin space $\mathbb{R}$ in \cite{COPP,BK}. Therein, some restrictions on the boundary conditions are necessary. One of the key methods is Ruelle's technique of superstability estimates in \cite{Ru0,Ru1,LP} which requires the interactions to be translation invariant, superstable and with pair-potentials growing at most quadratically. Applied to some models of Euclidean lattice field theories with single-spin space $\mathbb{R}$, it is proved in \cite{COPP} that the family of local Gibbs measures associated with a wide class of boundary conditions has at least one accumulation point in the set of \textit{'superstable' Gibbs measures}, a subset of the set of tempered Gibbs measures, see \cite[Thm. 1.2]{COPP}. This result was extended to some quantum anharmonic lattice systems with superstable interactions in \cite[Thm 2.6]{PaYo1} by extending Ruelle's technique to quantum statistical mechanics, see \cite{Pa2,Pa}.
See also \cite[Thm. 3.1]{KP} covering a wide class of quantum anharmonic lattice systems. In general, for quantum lattice systems where the single-spin spaces are infinite-dimensional, verifying Dobrushin's existence criteria turns out to be challenging, see, e.g., \cite[Sec. I.5]{Si}. A brief review of methods can be found in \cite[Sec. 2]{Alber2}.\\
\indent \textit{The uniqueness problem.} Sufficient conditions were derived by Dobrushin in \cite{D1,D3} in the case of compact single-spin spaces. Dobrushin's uniqueness theorem typically holds for a general class of high-temperature lattice systems with short/long-range, many-body interaction potentials and compact metrizable single-spin spaces. 
The domain of validity for interaction potentials and temperatures is often referred to as the Dobrushin uniqueness region. Note that Simon showed in \cite{Sim3} that there are interactions not in the Dobrushin uniqueness region but arbitrary close to (in some sense) giving rise to multiple Gibbs measures. A general statement of Dobrushin's uniqueness theorem including arbitrary single-spin spaces (typically, Polish spaces) can be found in \cite[Thm. 4]{D5}. In a nutshell, the Dobrushin uniqueness theorem reduces the uniqueness problem to a suitable estimate of the Wasserstein distance between pairs of 1-lattice point  Gibbs measures subject to different boundary conditions. In particular, uniqueness holds provided that the Dobrushin matrix coefficients satisfy the so-called condition of weak dependence  \cite[Eq. (5.2)]{D5}. In practice, the Dobrushin criterion allows one to derive sufficient conditions on the inverse temperature and/or model parameters for uniqueness to hold, from which a lower-bound for the critical inverse temperature can be inferred if phase transitions occur. In the case of  translation-invariant lattice models with short-range interactions, Dobrushin and Shlosman  generalized the 1-point Dobrushin condition by introducing a set of finite-volume conditions on the interaction, referred to as Dobrushin-Shlosman conditions, ensuring uniqueness to hold, see \cite[Thm 2.1]{DS1} and  \cite{DS2,DS3}. Note that the complexity of the Dobrushin-Shlosman conditions increases with the volume, and as a result, applying  the Dobrushin-Schlosman criterion in practice requires  numerical estimations. For completeness, we mention another uniqueness criterion in \cite[Thm. 1]{D6}, referred to as Dobrushin-Pecherski criterion, in the situation where the 1-point Gibbs measures essentially depend on large values of the boundary conditions (e.g., case of attractive potentials with polynomial asymptotics).\\   We mention that  Dobrushin's uniqueness criterion has been applied to some classical lattice models of Euclidean lattice field theories with single-spin space $\mathbb{R}$ in \cite[Sec. 2]{COPP}. The key ingredient is an expression of the Wasserstein distance for probability measures on $\mathbb{R}$ in terms of their distribution functions, see \cite[Theorem]{Val}. Dobrushin's uniqueness criterion has also been applied  to some quantum anharmonic lattice models of quantum crystals in \cite{AKRT1, AKRT2}. See also \cite[Thm 3.4]{KP}. Therein, the main ingredient is the representation of the Wasserstein distance by the Kantorovich-Rubinstein duality theorem, see, e.g., \cite{Ed}.  
For completeness, we mention some other techniques used to prove uniqueness of Gibbs measures. Cluster expansion techniques have been  used in  \cite{PaYo2} to prove uniqueness in the high-temperature regime for a class of quantum anharmonic lattice systems with superstable interactions. Note that in 1-dimensional lattices, uniqueness holds at all temperatures. See also, e.g., \cite{AKMR, Rbk, VEVH}. In \cite{AKR}, a quantum crystal with double-well potential is shown to have a phase transition using a Peierls-type argument.\\
\indent In high-temperature translation-invariant lattice systems with interactions in the Dobrushin uniqueness region, Dobrushin's uniqueness theorem assures the  non-existence of first-order phase transitions. Gross showed in fact the absence of second-order phase transitions. More precisely, Gross proved in \cite{Gro1} that the pressure is twice continuously Gateaux differentiable in the Dobrushin uniqueness region (differentiability is in parameters occurring linearly in the interactions). Prakash extended the Gateaux differentiability to any order in a subset of the Dobrushin uniqueness region, see \cite{Pra}. Turning to complex analyticity, Gallavotti and Miracle-Sole showed in the case of single-site spin space consisting of  two points, that the pressure is analytic at high temperature for a broader class of potentials than that to which the Dobrushin uniqueness theorem applies, see \cite{GMS}. The techniques used seem to be limited to spin spaces having a finite number of points. For some results on high-temperature analyticity with continuous spin spaces and a general class of many-body long-range interactions, see, e.g.,  \cite{Isr0}. Herein, Dobrushin' uniqueness techniques are used. See also \cite{DS2,DS3}. For some general results on analyticity via high-temperature expansions, see, e.g., \cite{DS,IS} and references therein.\\ 

\indent In this paper, we revisit Dobrushin's uniqueness theorem of Gibbs measures for lattice systems at thermal equilibrium that are not necessarily translation-invariant, consisting of interacting particles (with possibly long-range interactions)  and with a general complete metric space as single-spin space. Both the existence and uniqueness problems are proved in this general setting, along with the Dobrushin-Shlosman generalization of Dobrushin's uniqueness theorem for translation-invariant lattice systems. Our formulation of the uniqueness theorems covers both classical lattice systems and the Euclidean version of quantum lattice systems.\\
\indent To prove the existence of limit Gibbs distributions, the key ingredient is the completeness of the Wasserstein metric. We show that that completeness is guaranteed when the underlying single-spin space is a general  complete metric space, see Theorem \ref{complete}, extending the results in \cite{Bol} where separability is assumed. Our proof relies on the construction of a well-chosen Prokhorov compact to show the uniform tightness of the probability measures, after showing the equivalence between convergence in Wasserstein metric and weak convergence with modified Prokhorov condition in Proposition \ref{prop1}.  Our method would allow us to extend this equivalence result to uniform spaces, replacing the Wasserstein metric by a Wasserstein uniformity. Turning to uniqueness results, our proofs are essentially based on Dobrushin and Dobrushin-Shlosman techniques respectively. \\
\indent To illustrate the effectiveness of the Dobrushin and Dobrushin-Shlosman criteria, two series of applications to classical lattice systems are given. The emphasis  is on techniques to estimate the Wasserstein distance, rather than optimal lower bounds for critical inverse temperatures.\\
\indent Our first series of applications includes the classical Heisenberg model, $q$-state Potts model and Ising model. With a view to applying the Dobrushin and Dobrushin-Shlosman criteria, our focus is on different techniques to estimate the Wasserstein distance between two probability measures. We start by proving a general result of uniqueness at high temperatures with direct application to the classical Heisenberg model, see Proposition \ref{highT}. Our estimation of the Wasserstein distance is based on an approximation procedure for probability measures. Subsequently, we derive an explicit lower bound for the critical inverse temperature for the $q$-state Potts models in dimension $d=2,3$ (asymptotic for large $q$), followed by the $d$-dimensional Ising model. The results are based on an explicit formula of the Wasserstein distance for probability measures on $q$-point spaces, see Lemma \ref{Pottlem}. Finally, the effectiveness of the  Dobrushin-Shlosman criteria is illustrated with the square lattice Ising model in dimension $d=2,3$. The estimation of the Wasserstein distance is based on linear programming techniques, see, e.g., \cite{Gas} and Section \ref{linprogr} for a brief overview. This algorithm is used to numerically compute a lower bound for the critical inverse temperature.\\
\indent Our second series of applications focuses on convex perturbations of the Gaussian free field model. After applying the Dobrushin criterion to the Gaussian free field model with $d$-dimensional spins, the uniqueness result is extended when including a perturbation by a convex self-interaction. Proposition \ref{convex} covers the case of 1-dimensonial spins while Proposition \ref{convex2} covers the case of $d$-dimensional spins for a particular class of convex potentials. We show that uniqueness holds at all temperatures by constructing suitable couplings with the 1-point Gibbs measures as marginals. \\ 

\indent Our paper is organized as follows. In Section \ref{Wapro}, we recall some properties of the Wasserstein distance and give the completeness result in Theorem \ref{complete}. After recalling some definitions of local and limit Gibbs distributions, Theorem \ref{Dobrushin} in Section \ref{Dobs} is our reformulation of Dobrushin's uniqueness theorem for general lattice systems of interacting particles when the single-spin space is a general metric space. This covers both classical lattice systems and the Euclidean version of quantum lattice systems. Two corollaries related to the decay of correlations in the case of nearest-neighbour interactions are also given. The proof of Theorem \ref{Dobrushin} and proofs of the corollaries are given in Section \ref{DobSecc}. Theorem \ref{DSthm} in Section \ref{DobShlos} is our reformulation of Dobrushin-Shlosman's uniqueness theorem for translation-invariant lattice systems. In Section \ref{App}, we focus on a first series of applications of the uniqueness criteria to classical lattice systems at high-temperatures. Dobrushin's criterion is applied to the classical Heisenberg model in Section \ref{app11}, the $q$-state Potts model in Section \ref{secPott} and the $d$-dimensional Ising model in Section \ref{Isingdim}. Before applying the Dobrushin-Shlosman criterion to the 2 and 3-dimensional square lattice Ising model in Section \ref{DSapp}, we recall some results from  linear programming techniques in Section \ref{linprogr} including the algorithm used for the numerical estimations. In Section \ref{classiC}, we focus on a second series of applications to classical lattice systems with convex perturbations. Dobrushin's criterion is applied to the Gaussian free field model with $d$-dimensional spins in Section \ref{Gaussfr}. The perturbation by a convex self-interaction potential is then treated in Section \ref{convxx}, see Propositions \ref{convex} and \ref{convex2}. Section \ref{compl} contains the proof of Theorem \ref{complete}. The key result for the proof of Theorem \ref{complete} $(ii)$ is Proposition \ref{prop1}. Section \ref{appX} is our Appendix. The disintegration theorem for measures is recalled in Section \ref{disintg}. The dual problem corresponding to the linear programming problem is discussed in Section \ref{duality}. By way of application, a proof of the Kantorovich-Rubinstein duality theorem is given in Section \ref{Kanto}. The quantum harmonic crystal lattice model is revisited in Section \ref{appdx2}, see Proposition \ref{quantc}.

\section{Wasserstein distance and completeness.}
\label{Wapro}

Let $(\mathcal{X},\rho)$ be a metric space. Denote by $\mathcal{B}(\mathcal{X})$ the Borel $\sigma$-algebra of subsets of $\mathcal{X}$. Let $\mathcal{M}^{+}(\mathcal{X})$ denote the set of all finite non-negative Radon measures on $\mathcal{X}$ and let
\begin{equation*}
\mathcal{P}(\mathcal{X}) := \{\mu \in \mathcal{M}^{+}(\mathcal{X})\,:\, \mu(\mathcal{X})=1\},
\end{equation*}
be the set of all Radon probability measures on $\mathcal{X}$. For any $\mu,\nu \in \mathcal{P}(\mathcal{X})$, define the set of Radon couplings of $\mu$ and $\nu$ as
\begin{equation*}
\Xi_{\mathcal{X}}(\mu,\nu) := \left\{\sigma \in \mathcal{P}(\mathcal{X}\times\mathcal{X})\,:\, \sigma(A\times \mathcal{X})=\mu(A),\, \sigma(\mathcal{X}\times A)=\nu(A)\,\,\textrm{for all $A \in \mathcal{B}(\mathcal{X})$}\right\}.
\end{equation*}
Let $\rho_{W}: \mathcal{P}(\mathcal{X})\times \mathcal{P}(\mathcal{X}) \rightarrow \mathbb{R}_{+} \cup \{\infty\}$ be the Wasserstein distance defined as (see, e.g., \cite{Wass})
\begin{equation}
\label{rhoW}
\rho_{W}(\mu,\nu) := \inf_{\sigma \in \Xi_{\mathcal{X}}(\mu,\nu)} \int_{\mathcal{X}\times\mathcal{X}} \rho(x,y) \sigma(dx,dy).
\end{equation}
Introduce the set
\begin{equation}
\label{P1def}
\mathcal{P}_{1}(\mathcal{X}) :=\left\{ \mu \in \mathcal{P}(\mathcal{X})\, :\, \int_{\mathcal{X}} \rho(x,x_{0}) \mu(dx) < \infty,\,\, \textrm{for some (and hence all) $x_{0} \in \mathcal{X}$}\right\},
\end{equation}
which is independent of $x_{0}$. The main result of this section is the following
\begin{theorem}
\label{complete}
$(i)$. The Wasserstein distance $\rho_{W}$ is a metric on $\mathcal{P}_{1}(\mathcal{X})$.\\
$(ii)$. If $(\mathcal{X},\rho)$ is a complete metric space, then so is $(\mathcal{P}_{1}(\mathcal{X}), \rho_{W})$.
\end{theorem}
The proof of Theorem \ref{complete} is postponed to Sec. \ref{compl} for reader's convenience. We point out that the proof of Theorem \ref{complete} $(ii)$ relies on Prokhorov's theorem, see, e.g.,  \cite[Sec. IX.5.5]{Bou}. For a proof with the separability assumption, see, e.g., \cite{Bol}. We note that Theorem \ref{complete} is used in the proof of existence in Dobrushin's theorem in Sec. \ref{exist}. Hereafter, $\mathcal{X}$ as in Theorem \ref{complete} will play the role of state space (i.e., single-spin space).


\section{Dobrushin's uniqueness theorem revisited.}
\label{Dobs}

\subsection{Gibbs distributions.}
\label{secnot}

In this section, we recall the definition of local Gibbs distributions and limit Gibbs distributions. Our definitions below are taken from \cite{Si}. See also, e.g., \cite{M1,D1,D2,D3,D4,LR, D5, LP, Ge}.\\
\indent \textit{Notations}. For dimension $d \in \mathbb{N}$, let $\mathcal{S} := \{\Gamma \subset \mathbb{Z}^{d}\, :\, 0 < \vert \Gamma \vert < \infty\}$ be the (countably infinite) set of all non-empty finite subsets of $\mathbb{Z}^{d}$. Here and hereafter, $\vert \Lambda \vert$ denotes the cardinality of $\Lambda \in \mathcal{S}$, $\Lambda^{c} := \mathbb{Z}^{d} \setminus \Lambda$ its complement and $\partial \Lambda := \{j' \in \Lambda^{c}\,:\, \exists j \in \Lambda,\, \vert j - j'\vert =1\}$ its boundary. In the following, the state space $\mathcal{X}$ is assumed to be a metric space (note that all the definitions below hold for $\mathcal{X}$ a topological space). The space of finite configurations in $\Lambda \in \mathcal{S}$ and the space of all possible configurations are respectively $\mathcal{X}^{\Lambda}$ and $\mathcal{X}^{\mathbb{Z}^{d}}$ endowed with the product topology and equipped with the Borel $\sigma$-algebra $\mathcal{B}(\mathcal{X}^{\Lambda})$ and $\mathcal{B}(\mathcal{X}^{\mathbb{Z}^{d}})$ respectively. Note that the latter coincides with the $\sigma$-algebra generated by the cylinder sets. Configurations in $\mathcal{X}^{\Lambda}$, $\Lambda \in \mathcal{S}$ will be often denoted by $\underline{\xi}_{\Lambda}$ and a configuration $\underline{\xi}_{\Lambda}$ will be often decomposed as the concatenation $\underline{\xi}_{\Lambda} = \underline{\xi}_{\Delta} \underline{\xi}_{\Lambda \setminus \Delta}$ for a given $\Delta \subset \Lambda$. Let $\mathcal{P}(\mathcal{X}^{\Lambda})$, $\Lambda \in \mathcal{S}$ or $\Lambda = \mathbb{Z}^{d}$ denote the set of all Radon probability measures on $\mathcal{X}^{\Lambda}$. Let $\mathcal{P}_{1}(\mathcal{X}^{\Lambda})$, $\Lambda \in \mathcal{S}$ denote the subset similar to \eqref{P1def} but with a metric $\rho^{(\Lambda)}$ on $\mathcal{X}^{\Lambda}$. Let $\{\Pi_{\Lambda}\}_{\Lambda \in \mathcal{S}}$ denote the family of projection maps $\Pi_{\Lambda}: \mathcal{X}^{\mathbb{Z}^{d}} \rightarrow \mathcal{X}^{\Lambda}$.\\
\indent Given $\Lambda \in \mathcal{S}$, the energy of a configuration $\underline{\xi}_{\Lambda} \in \mathcal{X}^{\Lambda}$ is defined by
\begin{equation}
\label{HLam}
H(\underline{\xi}_{\Lambda}) := \sum_{X \in \mathcal{S}: X\subset \Lambda} \mathcal{V}_{X}(\underline{\xi}_{X}),
\end{equation}
where for every $X \in \mathcal{S}$, the function $\mathcal{V}_{X}: \mathcal{X}^{X} \rightarrow \mathbb{R}$ stands for the joint interaction energy of the $\underline{\xi}$'s inside $X$. The family $\{\mathcal{V}_{X}\}_{X \in \mathcal{S}}$ is commonly called a \textit{potential}, and typically, the $\mathcal{V}_{X}$'s are assumed to be $\mathcal{B}(\mathcal{X}^{X})$-measurable and such that the series in \eqref{HLam}, and in \eqref{interact} below, exist. \\
\indent The interaction energy between a configuration $\underline{\xi}_{\Lambda} \in \mathcal{X}^{\Lambda}$ and $\underline{\eta}_{\Lambda^{c}} \in \mathcal{X}^{\Lambda^{c}}$ is defined as
\begin{equation}
\label{interact}
H(\underline{\xi}_{\Lambda},\underline{\eta}_{\Lambda^{c}}) := \sum_{\substack{X\in \mathcal{S}:X\cap \Lambda \neq \emptyset, \\ X \cap \Lambda^{c} \neq \emptyset}} \mathcal{V}_{X}(\underline{\xi}_{X\cap\Lambda} \underline{\eta}_{X \cap \Lambda^{c}}).
\end{equation}
In the following, given $\Lambda \in \mathcal{S}$, configurations in $\mathcal{X}^{\Lambda^{c}}$ will play the role of boundary conditions.\\
\indent The total energy of a configuration $\underline{\xi}_{\Lambda}$ under the boundary condition $\underline{\eta}_{\Lambda^{c}}$ is then defined as
\begin{equation}
\label{totHam}
H_{\Lambda}(\underline{\xi}_{\Lambda} \vert \underline{\eta}_{\Lambda^{c}}) := H(\underline{\xi}_{\Lambda}) + H(\underline{\xi}_{\Lambda},\underline{\eta}_{\Lambda^{c}}).
\end{equation}

\begin{definition}
The local Gibbs distribution at inverse temperature $\beta>0$ for the domain $\Lambda \in \mathcal{S}$ under the boundary condition $\underline{\eta}_{\Lambda^{c}} \in \mathcal{X}^{\Lambda^{c}}$ is a probability measure on $(\mathcal{X}^{\Lambda},\mathcal{B}(\mathcal{X}^{\Lambda}))$ defined as
\begin{equation}
\label{condexp}
\mu_{\Lambda}^{\beta}(A \vert \underline{\eta}_{\Lambda^{c}}) := \frac{1}{Z_{\Lambda}^{\beta}(\underline{\eta}_{\Lambda^{c}})} \int_{A} \exp\left(- \beta H_{\Lambda}(\underline{\xi}_{\Lambda} \vert \underline{\eta}_{\Lambda^{c}})\right) \prod_{j \in \Lambda} \mu_{0}(d\xi_{j}),\quad A \in \mathcal{B}(\mathcal{X}^{\Lambda}),
\end{equation}
where $H_{\Lambda}(\cdot\,\vert \underline{\eta}_{\Lambda^{c}})$ is defined in \eqref{totHam}, $Z_{\Lambda}^{\beta}(\underline{\eta}_{\Lambda^{c}})$ is a normalisation constant (called partition function)
\begin{equation}
\label{partfunc}
Z_{\Lambda}^{\beta}(\underline{\eta}_{\Lambda^{c}}) :=  \int_{\mathcal{X}^{\Lambda}} \exp\left(- \beta H_{\Lambda}(\underline{\xi}_{\Lambda} \vert \underline{\eta}_{\Lambda^{c}})\right) \prod_{j \in \Lambda} \mu_{0}(d\xi_{j}),
\end{equation}
and the (single-spin) measure $\mu_{0}$ is a given a priori measure on $(\mathcal{X},\mathcal{B}(\mathcal{X}))$, not necessarily bounded.
\end{definition}

We now turn to the definition of limit Gibbs distributions. We refer the reader to the disintegration theorem in Sec. \ref{disintg} for the existence of conditional probability measures.

\begin{definition}
\label{defG}
A limit Gibbs distribution at inverse temperature $\beta>0$ corresponding to the formal Hamiltonian $H:  \mathcal{X}^{\mathbb{Z}^{d}} \rightarrow \mathbb{R}$ given by a potential as follows
\begin{equation*}
H(\underline{\xi}) := \sum_{X \subset \mathbb{Z}^{d}} \mathcal{V}_{X}(\underline{\xi}_{X}),
\end{equation*}
is a probability measure $\mu^{\beta}$ on $(\mathcal{X}^{\mathbb{Z}^{d}},\mathcal{B}(\mathcal{X}^{\mathbb{Z}^{d}}))$ such that, for any domain $\Lambda \in \mathcal{S}$,\\
$(i)$. For $\mu^{\beta}$- a.e. $\underline{\xi}=(\underline{\xi}_{\Lambda},\underline{\xi}_{\Lambda^{c}})$,  $H_{\Lambda}(\underline{\xi}_{\Lambda}\vert \underline{\xi}_{\Lambda^{c}})$ in \eqref{totHam} and $Z_{\Lambda}^{\beta}(\underline{\xi}_{\Lambda^{c}})$ in \eqref{partfunc} are finite;\\
$(ii)$. The conditional probability measure induced by $\mu^{\beta}$ on $(\mathcal{X}^{\Lambda},\mathcal{B}(\mathcal{X}^{\Lambda}))$ under the boundary condition $\underline{\eta}_{\Lambda^{c}} \in \mathcal{X}^{\Lambda^{c}}$ coincides $\mu_{\Lambda^{c}}^{\beta}$- a.e. with the local Gibbs distribution in \eqref{condexp}.
\end{definition}

\begin{remark}
\label{frek1}
The conditional probability measure in $(ii)$ above is relative to the projection map $\Pi_{\Lambda^{c}}$, see Sec. \ref{disintg}. Denoting it by $\mu^{\beta}(\cdot\,\vert \underline{\eta}_{\Lambda^{c}})$ for a fixed configuration $\underline{\eta}_{\Lambda^{c}} \in \mathcal{X}^{\Lambda^{c}}$, $(ii)$ reads
\begin{equation*}
\mu^{\beta}(\cdot\,\vert \underline{\eta}_{\Lambda^{c}}) \circ \Pi_{\Lambda}^{-1} = \mu_{\Lambda}^{\beta}(\cdot\,\vert \underline{\eta}_{\Lambda^{c}}).
\end{equation*}
Note that the measure $\mu^{\beta}(\cdot\,\vert \underline{\eta}_{\Lambda^{c}})$ is concentrated on the set $\{\underline{\xi} \in \mathcal{X}^{\mathbb{Z}^{d}}\,:\, \Pi_{\Lambda^{c}} \underline{\xi} = \underline{\eta}_{\Lambda^{c}}\}$.
\end{remark}

\begin{remark}
\label{equiDLR}
Connection with Gibbsian specification and equilibrium DLR equation. From the local Gibbs distribution in \eqref{condexp}, we can associate on $(\mathcal{X}^{\mathbb{Z}^{d}}, \mathcal{B}(\mathcal{X}^{\mathbb{Z}^{d}}))$ the probability measure
\begin{equation*}
\pi_{\Lambda}^{\beta}(A\vert \underline{\eta}) := \int_{\mathcal{X}^{\Lambda}} \mathbb{I}_{A}(\underline{\xi}_{\Lambda} \Pi_{\Lambda^{c}} \underline{\eta}) \mu_{\Lambda}^{\beta}(d\underline{\xi}_{\Lambda}\vert \Pi_{\Lambda^{c}} \underline{\eta}),\quad A \in \mathcal{B}(\mathcal{X}^{\mathbb{Z}^{d}}),\,\, \underline{\eta} \in \mathcal{X}^{\mathbb{Z}^{d}}.
\end{equation*}
$\{\pi_{\Lambda}^{\beta}\}_{\Lambda \in \mathcal{S}}$ forms a family of proper probability kernels from $\mathcal{B}(\mathcal{X}^{\Lambda^{c}})$ to $\mathcal{B}(\mathcal{X}^{\mathbb{Z}^{d}})$, see, e.g., \cite{Ge}. By virtue of the additive structure of the Hamiltonian, the family satisfies the consistency relation
\begin{equation*}
\pi_{\Lambda'}^{\beta} \pi_{\Lambda}^{\beta}(A\vert \underline{\eta}) :=
\int_{\mathcal{X}^{\mathbb{Z}^{d}}} \pi_{\Lambda}^{\beta}(A \vert \underline{\xi})  \pi_{\Lambda'}^{\beta}(d \underline{\xi}\vert \underline{\eta})  = \pi_{\Lambda'}^{\beta}(A \vert \underline{\eta}),\quad \Lambda \subset \Lambda',\,\,\,\Lambda,\Lambda' \in \mathcal{S}.
\end{equation*}
Due to this feature, $\{\pi_{\Lambda}^{\beta}\}_{\Lambda \in \mathcal{S}}$ is called a Gibbsian specification in Georgii's terminology.  In the DLR formalism, a limit Gibbs distribution at $\beta>0$ is defined as a measure $\mu^{\beta} \in  \mathcal{P}(\mathcal{X}^{\mathbb{Z}^{d}})$ satisfying
\begin{equation}
\label{DLReq}
\mu^{\beta} \pi_{\Lambda}^{\beta}(A) := \int_{\mathcal{X}^{\mathbb{Z}^{d}}} \pi_{\Lambda}^{\beta}(A \vert \underline{\xi}) \mu^{\beta}(d\underline{\xi}) = \mu^{\beta}(A),
\end{equation}
for all $\Lambda \in \mathcal{S}$ and all $A \in \mathcal{B}(\mathcal{X}^{\mathbb{Z}^{d}})$. \eqref{DLReq} is  the equilibrium DLR equation. $\mu^{\beta}$ is said to be specified by $\{\pi_{\Lambda}^{\beta}\}_{\Lambda \in \mathcal{S}}$. This definition is equivalent to Definition \ref{defG} $(ii)$, see, e.g., \cite[Rem. 1.24]{Ge}.
\end{remark}

\subsection{The Dobrushin uniqueness and existence theorem.}
\label{Secthm}

Theorem \ref{Dobrushin} below is our reformulation of Dobrushin's uniqueness theorem \cite[Thm. 4]{D5} for general lattice systems of interacting particles at thermal equilibrium, and covers both classical lattice systems and the Euclidean version of quantum lattice systems. Our reformulation includes a statement about existence which can also be proved using the Dobrushin condition in \eqref{Dcondd}.

\begin{theorem}
\label{Dobrushin}
Let $(\mathcal{X},\rho)$ be a complete metric space and $\mu_{0}$ a given a priori measure on $(\mathcal{X},\mathcal{B}(\mathcal{X}))$. Let $H: \mathcal{X}^{\mathbb{Z}^{d}} \rightarrow \mathbb{R}$ be a formal Hamiltonian of the form
\begin{equation}
\label{Hamm}
H(\underline{\xi}) = \sum_{X \subset \mathbb{Z}^{d}} \mathcal{V}_{X} (\underline{\xi}_{X}).
\end{equation}
Let $\beta >0$ be fixed. Assume the following, \\
$(C1)$. For all $X \subset \mathbb{Z}^{d}$, the functions $\mathcal{V}_{X}: \mathcal{X}^{X} \rightarrow \mathbb{R}$ are continuous;\\
$(C2)$. Given $j \in \mathbb{Z}^{d}$ and $\underline{\eta} \in \mathcal{X}^{\mathbb{Z}^{d} \setminus \{j\}}$, the local partition functions defined as
\begin{equation}
\label{locpartf}
Z_{j}^{\beta}(\underline{\eta}) := \int_{\mathcal{X}} \exp\left(-\beta \sum_{X \subset \mathbb{Z}^{d} : j \in X} \mathcal{V}_{X}\left(\xi_{j}\,\underline{\eta}_{X \setminus \{j\}}\right) \right) \mu_{0}(d \xi_{j}),
\end{equation}
are finite and bounded uniformly in $j \in \mathbb{Z}^{d}$;\\
$(C3)$. There exists $\xi^{*} \in \mathcal{X}$ such that
\begin{equation}
\label{unifi}
c_{0} := \sup_{j \in \mathbb{Z}^{d}} \int_{\mathcal{X}} \rho(\xi,\xi^{*}) \mu_{j}^{\beta}(d\xi \vert  \underline{\xi}_{\mathbb{Z}^{d} \setminus \{j\}}^{*})  < \infty,
\end{equation}
where $\xi_{k}^{*} = \xi^{*}$ for all $k \in \mathbb{Z}^{d}$.\\
Given $j \in \mathbb{Z}^{d}$ and $\underline{\eta} \in \mathcal{X}^{\mathbb{Z}^{d} \setminus \{j\}}$, the 1-point Gibbs distribution reads
\begin{equation}
\label{condma}
\mu_{j}^{\beta}(A\vert \underline{\eta}) := \frac{1}{Z_{j}^{\beta}(\underline{\eta})} \int_{A} \exp \left(- \beta \sum_{X \subset \mathbb{Z}^{d}: j \in X} \mathcal{V}_{X}\left(\xi_{j}\,\underline{\eta}_{X \setminus \{j\}}\right)\right) \mu_{0}(d\xi_{j}),\quad A \in \mathcal{B}(\mathcal{X}).
\end{equation}
Then there exists a unique limit Gibbs distribution $\mu^{\beta} \in \mathcal{P}(\mathcal{X}^{\mathbb{Z}^{d}})$, associated with the Hamiltonian \eqref{Hamm}, with marginal distributions satisfying
\begin{equation}
\label{schcond}
\sup_{j \in \mathbb{Z}^{d}} \int_{\mathcal{X}} \rho(\xi,\xi^{*}) \mu_{j}^{\beta}(d\xi) < \infty,
\end{equation}
provided that, for all $j \in \mathbb{Z}^{d}$ and for all $(\underline{\eta},\underline{\eta}') \in \mathcal{X}^{\mathbb{Z}^{d} \setminus \{j\}} \times \mathcal{X}^{\mathbb{Z}^{d} \setminus \{j\}}$,
\begin{equation}
\label{Dcondd}
\rho_{W}\left(\mu_{j}^{\beta}\left(\cdot\,\vert \underline{\eta}_{\mathbb{Z}^{d}\setminus \{j\}}\right),\mu_{j}^{\beta}\left(\cdot\,\vert \underline{\eta}_{\mathbb{Z}^{d}\setminus \{j\}}'\right)\right) \leq \sum_{\substack{l \in \mathbb{Z}^{d} \\ l \neq j}} r(l,j) \rho(\eta_{l},\eta_{l}'),
\end{equation}
holds for some constants $r(l,j) \geq 0$ satisfying, for all $j \in \mathbb{Z}^{d}$,
\begin{equation}
\label{cond}
\sum_{\substack{l \in \mathbb{Z}^{d} \\ l \neq j}} r(l,j) \leq \lambda < 1.
\end{equation}
\end{theorem}

\begin{remark}
For  lattice systems described by formal Hamiltonians of the form \eqref{Hamm}, uniqueness of the limit Gibbs distribution assures the non-existence of first-order phase transitions. \\ If uniqueness holds for all $\beta>0$, the system is said to be stable.
\end{remark}

\begin{remark}
\label{extqutm}
In Theorem \ref{Dobrushin}, $(\mathcal{X},\rho)$ is a general complete metric space. The latter could be the infinite-dimensional space of all continuous and periodic functions on $[0,\beta]$ equipped with the supremum norm, so that Theorem \ref{Dobrushin} also applies to quantum lattice systems. See Sec. \ref{appdx2}.
\end{remark}

\begin{remark}
\label{DLReqaa}
Under the conditions of Theorem \ref{Dobrushin}, the equilibrium DLR equation (see Remark \ref{equiDLR}) is satisfied in the following sense. For any $\beta>0$, for all $\Lambda_{0}, \Lambda \in \mathcal{S}$ such that $\Lambda_{0}\subset \Lambda$, for all $A \in \mathcal{B}(\mathcal{X}^{\Lambda_{0}})$ and all $B \in \mathcal{B}(\mathcal{X}^{\Lambda \setminus \Lambda_{0}})$ bounded,
\begin{equation}
\label{DLRequ}
\int_{B} \mu_{\Lambda_{0}}^{\beta}(A \vert \underline{\xi}_{\Lambda\setminus \Lambda_{0}})\,\overline{\mu}_{\Lambda \setminus \Lambda_{0}}^{\beta}(d\underline{\xi}_{\Lambda \setminus \Lambda_{0}}) = \overline{\mu}_{\Lambda}^{\beta}(A \times B),
\end{equation}
where $\overline{\mu}^{\beta}$ denotes the limit Gibbs distribution from Theorem \ref{Dobrushin} (to distinguish in the notation the marginals of the limit Gibbs distribution from the local Gibbs distributions).
\end{remark}

\begin{remark}
\label{nearestn}
In the case of nearest-neighbour interactions only, the conditions \eqref{Dcondd}-\eqref{cond} can be simply replaced with
\begin{equation}
\label{nniupbd}
\rho_{W}\left(\mu_{j}^{\beta}(\cdot\,\vert \underline{\eta}), \mu_{j}^{\beta}(\cdot\,\vert \underline{\eta}')\right) \leq \frac{\lambda}{2d} \sum_{l \in N_{1}(j)} \rho(\eta_{l},\eta_{l}'),
\end{equation}
where $0<\lambda < 1$ and $N_{1}(j) := \{j' \in \mathbb{Z}^{d}: \vert j'-j\vert = 1\}$ is the set of nearest-neighbours of $j \in \mathbb{Z}^{d}$.
\end{remark}

Below, we give a remark on the assumptions of Theorem \ref{Dobrushin}. By a bounded boundary condition, we mean any configuration $\underline{\vartheta} \in \mathcal{X}^{\mathbb{Z}^{d}}$ such that
$\sup_{j \in \mathbb{Z}^{d}} \rho(\xi^{*},\vartheta_{j}) < \infty$. Given $\beta>0$ and $\Delta, \Gamma \in \mathcal{S}$ such that $\Delta \subset \Gamma$ and $\underline{\vartheta} \in \mathcal{X}^{\mathbb{Z}^{d}}$, introduce on $(\mathcal{X}^{\Delta},\mathcal{B}(\mathcal{X}^{\Delta}))$ the probability measure
\begin{equation}
\label{tildemud}
\mu_{\Delta}^{\Gamma}(\cdot\,\vert \underline{\vartheta}_{\Gamma^{c}}) := \mu_{\Gamma}^{\beta}(\cdot\,\vert \underline{\vartheta}_{\Gamma^{c}}) \circ (\Pi_{\Delta}^{\Gamma})^{-1},
\end{equation}
where $\mu_{\Gamma}^{\beta}(\cdot\,\vert \underline{\vartheta}_{\Gamma^{c}})$ is the local Gibbs distribution for the domain $\Gamma$ and $\Pi^{\Gamma}_{\Delta}: \mathcal{X}^{\Gamma} \rightarrow \mathcal{X}^{\Delta}$ the projection map. Given $\Lambda \in \mathcal{S}$, define the metric $\rho^{(\Lambda)}$ on $\mathcal{X}^{\Lambda}$ as
\begin{equation}
\label{metriclambda}
\rho^{(\Lambda)}(\underline{\xi}_{\Lambda},\underline{\xi}_{\Lambda}') := \sum_{j \in \Lambda} \rho(\xi_{j},\xi_{j}'),
\end{equation}
and let $\rho_{W}^{(\Lambda)}$ denote the corresponding Wasserstein distance on $\mathcal{P}(\mathcal{X}^{\Lambda})$ defined similarly to \eqref{rhoW} but with the metric $\rho^{(\Lambda)}$.

\begin{remark}
\label{recondt}
Assumptions (C1)-(C3) together guarantee that, for any domain $\Lambda \in \mathcal{S}$, there exists a constant $c_{\infty} >0$ such that,
\begin{equation*}
\sup_{j \in \Lambda} \int_{\mathcal{X}} \rho(\xi, \xi^{*}) \mu_{j}^{\Lambda}(d\xi\vert \underline{\xi}_{\Lambda^{c}}^*) \leq c_{\infty}.
\end{equation*}
See Sec. \ref{exist} for further details. In particular, the above guarantees that $\mu_{\Lambda}^{\beta}(\cdot\,\vert  \underline{\xi}_{\Lambda^{c}}^*) \in
\mathcal{P}_{1}(\mathcal{X}^{\Lambda})$.
\end{remark}

\subsection{Some corollaries.}
\label{corollaries}

The two corollaries of Theorem \ref{Dobrushin} below follow from the proof given in Sec. \ref{exist}. The proofs of the corollaries are deferred to Sec. \ref{coroproo}.

\begin{corollary}
\label{corol1}
Consider the special case of nearest-neighbour interactions, i.e., where $\mathcal{V}_{\mathcal{X}} = 0$ unless $\vert X \vert = 1$, or $\vert X \vert = 2$ and $X = \{k,l\}$ with $\vert k - l \vert =1$.
Let $\beta>0$ be fixed and assume that the conditions of Theorem \ref{Dobrushin} hold.  Then, given a finite domain $\Delta \in \mathcal{S}$, there exists a constant $C>0$ such that, for any $\Lambda,\Lambda' \in \mathcal{S}$ with $\Delta \subset \Lambda \subset \Lambda'$ and for any bounded boundary condition $\underline{\vartheta} \in \mathcal{X}^{\mathbb{Z}^{d}}$,
\begin{equation*}
\rho_{W}^{(\Delta)}\left(\mu_{\Delta}^{\Lambda}(\cdot\, \vert \underline{\vartheta}_{\Lambda^{c}}),\mu_{\Delta}^{\Lambda'}(\cdot\, \vert \underline{\vartheta}_{\Lambda'^{c}})\right) \leq C \lambda^{\mathrm{dist}(\Delta, \Lambda' \setminus \Lambda)},
\end{equation*}
where $\mathrm{dist}(\Delta, \Lambda' \setminus \Lambda) := \inf_{(j,j') \in \Delta \times \Lambda'\setminus \Lambda} \vert j - j'\vert>0$, and the two probability measures, $\mu_{\Delta}^{\Lambda}(\cdot\, \vert \underline{\vartheta}_{\Lambda^{c}})$ and $\mu_{\Delta}^{\Lambda'}(\cdot\, \vert \underline{\vartheta}_{\Lambda'^{c}})$ on $(\mathcal{X}^{\Delta},\mathcal{B}(\mathcal{X}^{\Delta}))$ are defined similarly to \eqref{tildemud}. In particular, given a Lipschitz-continuous function $f : \mathcal{X}^{\Delta} \rightarrow \mathbb{R}$, i.e. such that,
\begin{equation*}
\left \vert f(\underline{\xi}_{\Delta}) - f(\underline{\xi}_{\Delta}') \right \vert \leq c_{f} \sum_{j \in \Delta} \rho(\xi_{j},\xi_{j}'),
\end{equation*}
for a constant $c_{f}>0$, we have,
\begin{equation*}
\left\vert \int_{\mathcal{X}^{\Lambda}} f(\underline{\xi}_{\Delta}) \mu_{\Lambda}^{\beta}(d\underline{\xi}_{\Lambda}\vert \underline{\vartheta}_{\Lambda^{c}}) - \int_{\mathcal{X}^{\Lambda'}} f(\underline{\xi}_{\Delta}') \mu_{\Lambda'}^{\beta}(d\underline{\xi}_{\Lambda'}'\vert \underline{\vartheta}_{\Lambda'^{c}}) \right \vert \leq c_{f} C \lambda^{\mathrm{dist}(\Delta, \Lambda' \setminus \Lambda)}.
\end{equation*}
\end{corollary}

A similar statement can be made concerning the decay of correlations. For further related results, see, e.g., \cite{Gro,Kl}, \cite[Sec. 7]{Ku},  and references therein.

\begin{corollary}
\label{corol2}
Consider the special case of nearest-neighbour interactions, i.e., where $\mathcal{V}_{\mathcal{X}} = 0$ unless $\vert X \vert = 1$, or $\vert X \vert = 2$ and $X = \{k,l\}$ with $\vert k - l \vert =1$.
Let $\beta>0$ be fixed and assume that the conditions of Theorem \ref{Dobrushin} hold.
Let $\Delta, \Delta' \in \mathcal{S}$ such that $\Delta \cap \Delta' = \emptyset$. Let $f: \mathcal{X}^{\Delta} \rightarrow \mathbb{R}$ and $g: \mathcal{X}^{\Delta'} \rightarrow \mathbb{R}$ be bounded continuous functions. Then, there exists a constant $C>0$ such that, if $\mu^{\beta}$ is the limit-Gibbs measure at inverse temperature $\beta$, then
\begin{multline*}
\left \vert \int_{\mathcal{X}^{\Delta \cup \Delta'}} f(\underline{\xi}_{\Delta}) g(\underline{\xi}_{\Delta'}) \mu_{\Delta \cup \Delta'}^{\beta}(d\underline{\xi}_{\Delta \cup \Delta'}) - \int_{\mathcal{X}^{\Delta}} f(\underline{\xi}_{\Delta}) \mu_{\Delta}^{\beta}(d\underline{\xi}_{\Delta}) \int_{\mathcal{X}^{\Delta'}} g(\underline{\xi}_{\Delta'}) \mu_{\Delta'}^{\beta}(d\underline{\xi}_{\Delta'}) \right\vert \\
\leq C  \lambda^{\mathrm{dist}(\Delta,\Delta')}.
\end{multline*}
\end{corollary}

\section{Proof of Theorem \ref{Dobrushin}.}
\label{DobSecc}

\indent In this section, we give a proof for existence and a proof for uniqueness separately. While the proof of existence relies on Theorem \ref{complete}, the proof of uniqueness is based on Proposition \ref{propDobp}. Before turning to the actual proofs in Sec. \ref{exist} and \ref{unique}, we collect in Sec. \ref{prelrest} some preliminary results.\\

\indent For reader's convenience, we recall below some notations from Sec. \ref{secnot}.\\
Given $\Delta,\Lambda\in \mathcal{S}$ such that $\Delta \subset \Lambda$, let $\Pi_{\Lambda} : \mathcal{X}^{\mathbb{Z}^{d}} \rightarrow \mathcal{X}^{\Lambda}$ and $\Pi_{\Delta}^{\Lambda}: \mathcal{X}^{\Lambda} \rightarrow \mathcal{X}^{\Delta}$ be the projection maps. Given a probability measure $\nu \in \mathcal{P}(\mathcal{X}^{\mathbb{Z}^{d}})$, let $\{\nu_{\Lambda}\}_{\Lambda \in \mathcal{S}}$ be the family of marginal distributions of $\nu$ defined as $\nu_{\Lambda} := \nu  \circ \Pi_{\Lambda}^{-1} \in \mathcal{P}(\mathcal{X}^{\Lambda})$. This means that $\nu_{\Lambda}(A) = \nu(A\times \mathcal{X}^{\Lambda^{c}})$ for any $A \in \mathcal{B}(\mathcal{X}^{\Lambda})$. Let $\{\nu(\cdot\,\vert \underline{\eta}_{\Lambda^{c}}),\,\underline{\eta}_{\Lambda^{c}} \in \mathcal{X}^{\Lambda^{c}}\}$ be the  conditional probability measures in $\mathcal{P}(\mathcal{X}^{\mathbb{Z}^{d}})$ relative to $\Pi_{\Lambda^{c}}$. Note that $\nu(\{\underline{\xi} \in \mathcal{X}^{\mathbb{Z}^{d}}: \Pi_{\Lambda^{c}} \underline{\xi} = \underline{\eta}_{\Lambda^{c}}\}\vert \underline{\eta}_{\Lambda^{c}})=1$.

\subsection{Some preliminary results.}
\label{prelrest}

\begin{lema}
\label{Dobpr}
Let $\mu, \nu \in \mathcal{P}(\mathcal{X}^{\mathbb{Z}^{d}})$ be probability measures. Given $\xi^{*} \in \mathcal{X}$, assume that there exists a constant $C>0$ such that
\begin{equation}
\label{bd2C}
\max\left\{\sup_{j \in \mathbb{Z}^{d}} \int_{\mathcal{X}^{\mathbb{Z}^{d}}} \rho(\xi_{j},\xi^{*}) \mu(d\underline{\xi}), \sup_{j \in \mathbb{Z}^{d}}  \int_{\mathcal{X}^{\mathbb{Z}^{d}}} \rho(\xi_{j},\xi^{*}) \nu(d\underline{\xi})\right\} \leq C.
\end{equation}
Further, assume that for all $j \in \mathbb{Z}^{d}$ and all $(\underline{\xi},\underline{\xi}') \in \mathcal{X}^{\mathbb{Z}^{d}\setminus \{j\}} \times \mathcal{X}^{\mathbb{Z}^{d}\setminus \{j\}}$,
\begin{equation}
\label{iily}
\rho_{W}\left(\mu_{j}\left(\cdot\,\vert \underline{\xi}_{\mathbb{Z}^{d} \setminus \{j\}}\right),\nu_{j}\left(\cdot\,\vert \underline{\xi}_{\mathbb{Z}^{d} \setminus \{j\}}'\right)\right) \leq \sum_{\substack{l \in \mathbb{Z}^{d} \\ l\neq j}} r(l,j) \rho(\xi_{l},\xi_{l}'),
\end{equation}
holds for some constants $r(l,j) \geq 0$ satisfying, for all $j \in \mathbb{Z}^{d}$,
\begin{equation}
\label{consumf}
\sum_{\substack{l \in \mathbb{Z}^{d} \\ l\neq j}} r(l,j) \leq \lambda < 1.
\end{equation} 
Then, for any $\delta > 0$, there exists a coupling $\sigma \in  \Xi_{\mathcal{X}^{\mathbb{Z}^{d}}}(\mu,\nu)$ such that, for every $j \in \mathbb{Z}^{d}$,
\begin{equation} 
\label{ineqgam}
\gamma(j) \leq \sum_{\substack{l \in \mathbb{Z}^{d} \\ l\neq j}} r(l,j) \gamma(l) + \delta,
\end{equation}
where $\gamma$ is defined by 
\begin{equation} 
\label{gammadef}
\gamma(j) := \int_{\mathcal{X}^{\mathbb{Z}^{d}} \times \mathcal{X}^{\mathbb{Z}^{d}}} \rho(\xi_{j}, \xi_{j}')\,\sigma(d\underline{\xi}, d\underline{\xi}'). 
\end{equation} 
Further, given an arbitrary finite domain $\Lambda \in \mathcal{S}$ and  $\epsilon>0$, there exists another coupling $\tilde{\sigma} \in \Xi_{\mathcal{X}^{\mathbb{Z}^{d}}}(\mu,\nu)$ such that, for every $j \in \mathbb{Z}^{d}$,
\begin{equation}
\label{lem2es1}
\int_{\mathcal{X}^{\mathbb{Z}^{d}}\times \mathcal{X}^{\mathbb{Z}^{d}}} \rho(\xi_{j},\xi_{j}') \tilde{\sigma}(d\underline{\xi},d\underline{\xi}') \leq \chi(j) + \epsilon,
\end{equation}
where $\chi(j) = 2C$ for $j \notin \Lambda$, and for $j \in \Lambda$,
\begin{equation*}
\chi(j) = \sum_{l \in \Lambda^{c}} \gamma_{l} \sum_{k \in \Lambda} \sum_{n=0}^{\infty} r(l,k)(R_{\Lambda}^{n})_{k,j},
\end{equation*}
where the matrix $R_{\Lambda}$ is defined as
\begin{equation}
\label{mattRlbd}
(R_{\Lambda})_{k,j} := \left\{ \begin{array}{ll}	r(k,j),
	\quad &\textrm{if $j,k \in \Lambda$ and $j\neq k$},\\
0,\quad &\textrm{if $j,k \in \Lambda$ and $j=k$}
\end{array}\right.
\end{equation}
\end{lema}

\noindent \textbf{Proof of Lemma \ref{Dobpr}.} Let $j \in \mathbb{Z}^{d}$  and $\delta >0$ be fixed. By definition of $\rho_{W}$ in \eqref{rhoW}, there exists a coupling $\sigma \in \Xi_{\mathcal{X}^{\mathbb{Z}^{d}}}(\mu,\nu)$  such that 
\begin{equation}
\label{defsiga}
\int_{\mathcal{X}^{\mathbb{Z}^{d}} \times \mathcal{X}^{\mathbb{Z}^{d}}} \rho(\xi_{j}, \xi_{j}')\,\sigma(d\underline{\xi}, d\underline{\xi}') \leq \rho_{W}\left(\mu_{j}, \nu_{j}\right) + \frac{\delta}{2}. 
\end{equation}
Note that, from \eqref{bd2C}, it follows that, uniformly in $k \in \mathbb{Z}^d$, 
\begin{equation}
\label{gammalC}
\gamma(k)\leq \int_{\mathcal{X}^{\mathbb{Z}^{d}} \times \mathcal{X}^{\mathbb{Z}^{d}}} \left(\rho(\xi_{k},\xi^{*}) + \rho(\xi^{*},\xi_{k}')\right) \sigma(d\underline{\xi},d\underline{\xi}')\leq 2C.
\end{equation}
We now prove that \eqref{ineqgam} holds for $\sigma$ as in \eqref{defsiga}. By Dobrushin's condition in \eqref{iily} there exist, for all ($\underline{\eta}, \underline{\eta}') \in \mathcal{X}^{\mathbb{Z}^{d} \setminus\{j\}} \times \mathcal{X}^{\mathbb{Z}^{d} \setminus\{j\}}$, couplings  
$\sigma_{j;\underline{\eta}, \underline{\eta}'} \in \Xi_{\mathcal{X}}(\mu_{j}(\cdot\, \vert \underline{\eta}), \nu_{j}(\cdot\, \vert \underline{\eta}'))$ such that
\begin{equation}
\label{need2}	
\int_{{\mathcal X} \times {\mathcal X}} \rho(\xi,\xi')\,\sigma_{j;\underline{\eta}, \underline{\eta}'}(d\xi,d\xi')
\leq  \sum_{\substack{k \in \mathbb{Z}^{d} \\ k \neq j}} r(k,j) \rho(\eta_{k}, \eta_{k}') + \frac{\delta}{2}.
\end{equation} 
For $A,B \in \mathcal{B}(\mathcal{X}^{\mathbb{Z}^{d}})$, define $\tilde{\sigma}^{(j)} \in \mathcal{P}(\mathcal{X}^{\mathbb{Z}^{d}} \times \mathcal{X}^{\mathbb{Z}^{d}})$ as follows
\begin{equation*}
\tilde{\sigma}^{(j)}\left(A \times B\right) := \int_{\mathcal{X}^{\mathbb{Z}^{d}} \times \mathcal{X}^{\mathbb{Z}^{d}}} \mathbb{I}_{A \times B}\left((\xi,\underline{\eta}),(\xi',\underline{\eta}')\right)  \sigma_{j;\underline{\eta},\underline{\eta}'}(d\xi,d\xi') \sigma_{\mathcal{X}^{\mathbb{Z}^{d}\setminus \{j\}}} (d\underline{\eta},d\underline{\eta}').
\end{equation*}
Then taking $B= \mathcal{X}^{\mathbb{Z}^{d}}$,
\begin{equation*}
\begin{split}
\tilde{\sigma}^{(j)}\left(A \times \mathcal{X}^{\mathbb{Z}^{d}}\right) &= \int_{\mathcal{X}^{\mathbb{Z}^{d}}} \mathbb{I}_{A}\left((\xi,\underline{\eta})\right) \int_{\mathcal{X}^{\mathbb{Z}^{d}\setminus\{j\}}} \sigma_{\mathcal{X}^{\mathbb{Z}^{d}\setminus \{j\}}} (d\underline{\eta},d\underline{\eta}') \int_{\mathcal{X}} \sigma_{j;\underline{\eta},\underline{\eta}'}(d\xi,d\xi') \\
&= \int_{\mathcal{X}^{\mathbb{Z}^{d}}} \mathbb{I}_{A}\left((\xi,\underline{\eta})\right)  \mu_{\mathcal{X}^{\mathbb{Z}^{d}\setminus\{j\}}}(d\underline{\eta})
\mu_{j}(d\xi\vert \underline{\eta}) = \mu(A).
\end{split}
\end{equation*}
Similarly, $\tilde{\sigma}^{(j)}(\mathcal{X}^{\mathbb{Z}^{d}} \times B) = \nu(B)$. As a result, $\tilde{\sigma}^{(j)} \in \Xi_{\mathcal{X}^{\mathbb{Z}^{d}}}(\mu,\nu)$. Now define
\begin{equation*}
\gamma^{(j)}(l) := \int_{\mathcal{X}^{\mathbb{Z}^{d}}\times \mathcal{X}^{\mathbb{Z}^{d}}} \rho(\xi_{l},\xi_{l}')\,\tilde{\sigma}^{(j)}(d\underline{\xi},d\underline{\xi}'),\quad l,j \in \mathbb{Z}^{d}.
\end{equation*}
Clearly, for $l \neq j$,
\begin{equation*}
\begin{split}
\gamma^{(j)}(l) &= \int_{\mathcal{X}^{\mathbb{Z}^{d}}\times \mathcal{X}^{\mathbb{Z}^{d}}} \rho(\eta_{l}, \eta_{l}')\, \sigma_{j;\underline{\eta},\underline{\eta}'}(d\xi,d\xi') \sigma_{\mathcal{X}^{\mathbb{Z}^{d}\setminus \{j\}}} (d\underline{\eta},d\underline{\eta}') \\&= \int_{\mathcal{X}^{\mathbb{Z}^{d}\setminus \{j\}}\times \mathcal{X}^{\mathbb{Z}^{d}\setminus \{j\}}} \rho(\eta_{l}, \eta_{l}')\,\sigma_{\mathcal{X}^{\mathbb{Z}^{d}\setminus \{j\}}}(d\underline{\eta},d\underline{\eta}') = \gamma(l).
\end{split}
\end{equation*}
Note that for $l=j$, we have,
\begin{equation*}
\gamma^{(j)}(j) \geq \rho_{W}\left(\mu_{j},\nu_{j}\right).
\end{equation*} 
It then follows from \eqref{defsiga} that 
\begin{equation}
\label{need2b} 
\gamma(j) - \gamma^{(j)}(j) \leq  \rho_{W}\left(\mu_{j},\nu_{j}\right) + \frac{\delta}{2} - \rho_{W}\left(\mu_{j},\nu_{j}\right) = \frac{\delta}{2}.
\end{equation}
By using \eqref{need2}, we have,
\begin{equation*} 
\gamma^{(j)}(j) 
\leq  \sum_{\substack{k \in \mathbb{Z}^{d} \\ k \neq j}} r(k,j)  \int_{\mathcal{X}^{\mathbb{Z}^{d}\setminus \{j\}} \times \mathcal{X}^{\mathbb{Z}^{d}\setminus \{j\}}} \rho(\eta_{k}, \eta_{k}') \sigma_{\mathcal{X}^{\mathbb{Z}^{d}\setminus \{j\}}}(d\underline{\eta}, d\underline{\eta}') + \frac{\delta}{2} 
= \sum_{\substack{k \in \mathbb{Z}^{d} \\ k \neq j}} r(k,j) \gamma(k) 
		+ \frac{\delta}{2}.
\end{equation*}
To complete the proof of \eqref{ineqgam}, it suffices to use \eqref{need2b} yielding 
\begin{equation*} 
\gamma(j) \leq \gamma^{(j)}(j) + \frac{\delta}{2} \leq
\sum_{\substack{k \in \mathbb{Z}^{d} \\ k \neq j}} r(k,j) \gamma(k) + \delta.
\end{equation*} 
We now pick $\Lambda \in \mathcal{S}$. Applying the above result to each $j \in \Lambda$, we find that \eqref{lem2es1} holds with
\begin{equation*}
\chi(j) = \sum_{\substack{l \in \mathbb{Z}^{d} \\l\neq j}} r(l,j)\gamma(l) ,\quad j \in \Lambda.
\end{equation*}
We split the sum over $l\in \mathbb{Z}^{d}\setminus \{j\}$ into $l \in (\Lambda\setminus \{j\}) \cup \Lambda^{c}$. Then, for each $l \in \Lambda\setminus \{j\}$, we can replace $\gamma(l)$ by $\chi(l)$ to get an improved bound. In view of \eqref{mattRlbd}, this yields
\begin{equation*}
\begin{split}
\chi(j) &= \sum_{l \in \Lambda^{c}} r(l,j)\gamma(l) + \sum_{\substack{l \in \Lambda \\ l\neq j}} \sum_{\substack{k \in \mathbb{Z}^{d}\\ k \neq l}}  r(k,l) \gamma(k) (R_{\Lambda})_{l,j} \\
&= \sum_{l \in \Lambda^{c}} \gamma(l) \sum_{k \in \Lambda} r(l,k) \left((\mathbb{I}_{\Lambda})_{k,j} + (R_{\Lambda})_{k,j}\right) + \sum_{l\in \Lambda} \gamma(l) (R_{\Lambda}^{2})_{l,j}.
\end{split}
\end{equation*}
Continuing this way, we obtain at the $m$-th stage
\begin{equation*}
\chi(j) = X_{m}(j) + \mathcal{R}_{m+1}(j),
\end{equation*}
with
\begin{gather*}
X_{m}(j) := \sum_{l \in \Lambda^{c}} \gamma(l) \sum_{k \in \Lambda} r(l,k) \sum_{n=0}^{m} (R_{\Lambda}^{n})_{k,j} \\
\mathcal{R}_{m+1}(j):= \sum_{l \in \Lambda} \gamma(l) (R_{\Lambda}^{m+1})_{l,j}.
\end{gather*}
Defining the matrix norm of $R_{\Lambda}$ by
\begin{equation}
	\label{nommatx}
	\Vert R_{\Lambda} \Vert := \sup_{k\in \Lambda} \sum_{j \in \Lambda} \vert (R_{\Lambda})_{j,k} \vert,
\end{equation} 
$\Vert R_{\Lambda}\Vert \leq \lambda < 1$ by assumption, and hence $\Vert R_{\Lambda}^{m} \Vert \rightarrow 0$ as $m \rightarrow \infty$. As a result, $\mathcal{R}_{m+1}(j) \rightarrow 0$ in the limit $m \rightarrow \infty$. \qed

\subsection{Proof of existence.}
\label{exist}

Let $\beta>0$ be fixed. For convenience, we hereafter drop the $\beta$-dependence in our notations. Given $\Gamma \in \mathcal{S}$ and $\underline{\vartheta} \in \mathcal{X}^{\mathbb{Z}^{d}}$ such that $\sup_{k \in \mathbb{Z}^{d}} \rho(\xi^{*}, \vartheta_{k}) < \infty$, let $\mu_{\Gamma}(\cdot\,\vert \underline{\vartheta}_{\Gamma^{c}})$ denote the local Gibbs distribution on $(\mathcal{X}^{\Gamma},\mathcal{B}(\mathcal{X}^{\Gamma}))$ with bounded boundary condition $\underline{\vartheta}_{\Gamma^{c}}$ held fixed outside of $\Gamma$. Given $\Delta \in \mathcal{S}$ such that $\Delta \subset \Gamma$, we introduce on $(\mathcal{X}^{\Delta},\mathcal{B}(\mathcal{X}^{\Delta}))$ the probability measure
\begin{equation}
\label{tildmud}
\mu_{\Delta}^{\Gamma}(\cdot\,\vert \underline{\vartheta}_{\Gamma^{c}}) := \mu_{\Gamma}(\cdot\,\vert \underline{\vartheta}_{\Gamma^{c}}) \circ (\Pi_{\Delta}^{\Gamma})^{-1},
\end{equation}
where $\Pi_{\Delta}^{\Gamma} : \mathcal{X}^{\Gamma} \rightarrow \mathcal{X}^{\Delta}$ is the projection map. \\
Let $\Lambda \in \mathcal{S}$ be fixed. The first part of the proof consists in showing that
\begin{equation}
\label{cinfty}
\sup_{j \in \Lambda} \int_{\mathcal{X}} \rho(\xi,\xi^{*}) \mu_{j}^{\Lambda}(d\xi\vert \underline{\xi}_{\Lambda^{c}}^{*}) \leq c_{\infty}:= \frac{c_{0}}{1 - \lambda},
\end{equation}
with $c_{0}>0$ defined in \eqref{unifi}. Let $j\in \Lambda$ be fixed. By assumption, given $\epsilon>0$ and $\underline{\eta} \in \mathcal{X}^{\Lambda}$, there exists a coupling $\sigma_{j;\underline{\eta},\underline{\xi}^{*}}\in \Xi_{\mathcal{X}}(\mu_{j}(\cdot\,\vert \underline{\eta}_{\Lambda \setminus \{j\}} \underline{\xi}_{\Lambda^{c}}^{*}), \mu_{j}(\cdot\,\vert \underline{\xi}_{\mathbb{Z}^{d} \setminus \{j\}}^{*}))$ such that
\begin{equation*}
\int_{\mathcal{X}\times \mathcal{X}} \rho(\xi,\xi') \sigma_{j;\underline{\eta},\underline{\xi}^{*}}(d\xi, d\xi') < \sum_{\substack{l \in \Lambda \\ l \neq j}} r(l,j) \rho(\eta_{l},\xi^{*}) + \epsilon.
\end{equation*}
Then, we have,
\begin{equation*}
\begin{split}
\int_{\mathcal{X}} \rho(\xi,\xi^{*}) \mu_{j}(d\xi\vert \underline{\eta}_{\Lambda \setminus \{j\}} \underline{\xi}_{\Lambda^{c}}^{*}) &=  \int_{\mathcal{X} \times \mathcal{X}} \rho(\xi,\xi^{*}) \sigma_{j;\underline{\eta},\underline{\xi}^{*}}(d\xi, d\xi') \\
&\leq  \int_{\mathcal{X}\times \mathcal{X}} \left(\rho(\xi,\xi') + \rho(\xi',\xi^{*})\right) \sigma_{j;\underline{\eta},\underline{\xi}^{*}}(d\xi, d\xi') \\
&< \sum_{\substack{l \in \Lambda \\ l \neq j}} r(l,j) \rho(\eta_{l},\xi^{*})  + \epsilon + c_{0}.
\end{split}
\end{equation*}
Letting $\epsilon \rightarrow 0$, we are left with
\begin{equation}
\label{interm1}
\int_{\mathcal{X}} \rho(\xi,\xi^{*}) \mu_{j}(d\xi\vert \underline{\eta}_{\Lambda \setminus \{j\}} \underline{\xi}_{\Lambda^{c}}^{*}) \leq \sum_{\substack{ l \in \Lambda \\ l \neq j}} r(l,j) \rho(\eta_{l},\xi^{*}) +  c_{0}.
\end{equation}
Next, let $(K_{n})_{n \in \mathbb{N}}\subset \mathcal{X}$ be an increasing sequence of compact sets such that $\mu_{\Lambda}(K_{n}^{\Lambda}\vert \underline{\xi}_{\Lambda^{c}}^*) \rightarrow 1$ when $n \rightarrow \infty$. Define the sequence of $K_{n}$-restrictions of \eqref{tildmud} as $\mu_{\Delta,n}^{\Gamma}(A\vert  \underline{\vartheta}_{\Gamma^{c}}) := \mu_{\Delta}^{\Gamma}(A \cap K^{\Delta}_{n} \vert \underline{\vartheta}_{\Gamma^{c}})$, $A \in \mathcal{B}(\mathcal{X}^{\Delta})$. Obviously, for all $n \in \mathbb{N}$, we have,
\begin{equation*}
\sup_{k \in \Lambda}\int_{\mathcal{X}} \rho(\xi,\xi^{*}) \mu_{k,n}^{\Lambda}(d\xi\vert \underline{\xi}_{\Lambda^{c}}^*) < \infty.
\end{equation*}
Then, we have,
\begin{align}
\int_{\mathcal{X}} \rho(\xi,\xi^{*}) \mu^{\Lambda}_{j,n}(d\xi\vert \underline{\xi}_{\Lambda^{c}}^*) &=   \int_{\mathcal{X}^{\Lambda \setminus \{j\}}} \int_{\mathcal{X}} \mathbb{I}_{K_{n}}(\xi)\rho(\xi,\xi^{*}) \mu_{j}(d\xi\vert \underline{\eta}_{\Lambda \setminus \{j\}} \underline{\xi}_{\Lambda^{c}}^{*})\, \mu_{\Lambda,n}(d\underline{\eta}_{\Lambda \setminus \{j\}} \vert \underline{\xi}_{\Lambda^{c}}^*) \nonumber \\
&\leq  \sum_{\substack{l \in \Lambda \\l \neq j}} r(l,j) \int_{\mathcal{X}^{\Lambda \setminus \{j\}}} \rho(\eta_{l},\xi^{*}) \mu_{\Lambda,n}(d\underline{\eta}_{\Lambda \setminus \{j\}}\vert \underline{\xi}_{\Lambda^{c}}^*)+ c_{0} \nonumber \\
\label{itterr}
&= \sum_{l \in \Lambda} (R_{\Lambda})_{l,j} \int_{\mathcal{X}}\rho(\eta_{l},\xi^{*}) \mu_{l,n}^{\Lambda}(d\eta_{l}\vert \underline{\xi}_{\Lambda^{c}}^*) + c_{0},
\end{align}
where we used \eqref{interm1}. Iterating \eqref{itterr}, we have,
\begin{equation}
\label{fromthr}
\int_{\mathcal{X}} \rho(\xi,\xi^{*}) \mu^{\Lambda}_{j,n}(d\xi\vert \underline{\xi}_{\Lambda^{c}}^*)  \leq \sum_{l \in \Lambda} (R_{\Lambda})_{l,j}^{p} \int_{\mathcal{X}}\rho(\eta_{l},\xi^{*}) \mu_{l,n}^{\Lambda}(d\eta_{l}\vert \underline{\xi}_{\Lambda^{c}}^*) + c_{p-1}
\end{equation}
where, for all integers $p \geq 2$,
\begin{equation*}
c_{p-1} := c_{0} \sum_{s=0}^{p-1} \sup_{j \in \Lambda} \sum_{l \in \Lambda} (R_{\Lambda})^{s}_{l,j}  = c_{0} \sum_{s=0}^{p-1} \Vert R_{\Lambda}^{s} \Vert \leq \frac{c_{0}}{1 - \lambda}.
\end{equation*}
Clearly, the first term in the r.h.s. of \eqref{fromthr} tends to 0 in the limit $p \rightarrow \infty$. To conclude the proof of \eqref{cinfty}, we use the monotone convergence theorem to take $n\rightarrow \infty$. \\

\indent To complete the proof of existence, consider a sequence of subsets $(\Lambda_{n})_{n \in \mathbb{N}}$ ordered by inclusion and exhausting $\mathbb{Z}^{d}$ and let us show that the sequence of probability measures $(\mu_{\Lambda_{n}}(\cdot\,\vert \underline{\xi}_{\Lambda_{n}^{c}}^*))_{n \in \mathbb{N}}$ converges weakly in $\mathcal{P}(\mathcal{X}^{\mathbb{Z}^{d}})$. Let $\Lambda_{0} \in \mathcal{S}$ be fixed. Let $n,n' \in \mathbb{N}$ with $n<n'$ sufficiently large such that $\Lambda_{0} \subset \Lambda_{n} \subset \Lambda_{n'}$. In view of \eqref{cinfty}, by applying Lemma \ref{Dobpr}, for all $\epsilon>0$, there exists a coupling $\tilde{\sigma}\in \Xi_{\mathcal{X}^{\mathbb{Z}^{d}}} (\mu_{\Lambda_{n}}(\cdot\,\vert \underline{\xi}_{\Lambda_{n}^{c}}^*), \mu_{\Lambda_{n'}}(\cdot\, \vert \underline{\xi}_{\Lambda_{n'}^{c}}^*))$ such that, for all $j \in \mathbb{Z}^{d}$,
\begin{equation*}
\int_{\mathcal{X}^{\mathbb{Z}^{d}} \times \mathcal{X}^{\mathbb{Z}^{d}}} \rho(\xi_{j},\xi_{j}') \tilde{\sigma}(d\underline{\xi}, d\underline{\xi}') < \chi(j) + \frac{\epsilon}{2},
\end{equation*}
where $\chi(j)= 2 c_{\infty}$ for $j \notin \Lambda_{n}$ and where $\chi(j)$ for $j \in \Lambda_{n}$ is given by
\begin{equation*}
\chi(j) = 2 c_{\infty} \sum_{l \in \Lambda_{n}^{c}} \sum_{k \in \Lambda_{n}} \sum_{p=0}^{\infty} r(l,j) (R_{\Lambda_{n}}^{p})_{k,j}.
\end{equation*}
We now claim that $\chi(j)$ is arbitrarily small for $j \in \Lambda_{0}$ provided that $\mathrm{dist}(\Lambda_{0},\Lambda_{n}^{c})$ is large enough (see the proof of Proposition \ref{propDobp} for further details). Restricting $\tilde{\sigma}$ to $\mathcal{X}^{\Lambda_{0}} \times \mathcal{X}^{\Lambda_{0}}$, it follows that, for any $\epsilon>0$,
\begin{equation*}
\rho_{W}^{(\Lambda_{0})} \left (\mu_{\Lambda_{0}}^{\Lambda_{n}}(\cdot\,\vert \underline{\xi}_{\Lambda_{n}^{c}}^*), \mu_{\Lambda_{0}}^{\Lambda_{n'}}(\cdot\,\vert \underline{\xi}_{\Lambda_{n'}^{c}}^*)\right) < \epsilon,
\end{equation*}
provided that $\mathrm{dist}(\Lambda_{0},\Lambda_{n}^{c})$ is large enough. This means that $(\mu_{\Lambda_{0}}^{\Lambda_{n}}(\cdot\,\vert \underline{\xi}_{\Lambda_{n}^{c}}^*))_{n \in \mathbb{N}}$ is a Cauchy sequence in $\mathcal{P}_{1}(\mathcal{X}^{\Lambda_{0}})$, and hence converges since $(\mathcal{P}_{1}(\mathcal{X}^{\Lambda_{0}}),\rho^{(\Lambda_{0})})$ is complete by Theorem \ref{complete}.  As this holds for all finite domains $\Lambda_{0}$, the sequence $(\mu_{\Lambda_{n}}(\cdot\,\vert \underline{\xi}_{\Lambda_{n}^{c}}^*))_{n \in \mathbb{N}}$ converges weakly in $\mathcal{P}(\mathcal{X}^{\mathbb{Z}^{d}})$.\\
\indent Let $\overline{\mu}^{(\underline{\xi}^{*})}$ denote the limit. We end the proof of existence by showing that, under our conditions, the equilibrium DLR equation follows from the above, see Remarks \ref{equiDLR} and \ref{DLReqaa}. Let $\Lambda_{0}, \Lambda \in \mathcal{S}$ such that $\Lambda_{0} \subset \Lambda$. Let $(\Lambda_{n})_{n}$ be the sequence of subsets as above. Let $A \in \mathcal{B}(\mathcal{X}^{\Lambda_{0}})$ and $B \in \mathcal{B}(\mathcal{X}^{\Lambda \setminus \Lambda_{0}})$ bounded. For all $n$ large enough so that $\Lambda \subset \Lambda_{n}$, the consistency relation (see Remark \ref{equiDLR}) yields
\begin{equation*}
\int_{B} \mu_{\Lambda_{0}}(A \vert \underline{\xi}_{\Lambda \setminus \Lambda_{0}}) \mu_{\Lambda \setminus \Lambda_{0}}^{\Lambda_{n}}(d \underline{\xi}_{\Lambda \setminus \Lambda_{0}} \vert \underline{\xi}_{\Lambda_{n}^{c}}) = \mu_{\Lambda}^{\Lambda_{n}}(A \times B \vert \underline{\xi}_{\Lambda_{n}^{c}}).
\end{equation*}
By using that $(\mu_{X}^{\Lambda_{n}}(\cdot\,\vert \underline{\xi}_{\Lambda_{n}^{c}}))_{n}$ converges weakly to $\overline{\mu}_{X}^{(\underline{\xi})}:= \overline{\mu}^{(\underline{\xi})} \circ \Pi_{X}^{-1}$, $X \subseteq \Lambda$, \eqref{DLRequ} follows. \qed

\subsection{Proof of uniqueness.}
\label{unique}

The proof of uniqueness directly follows from the following Proposition

\begin{proposition}
\label{propDobp}
Let $\mu, \nu \in \mathcal{P}(\mathcal{X}^{\mathbb{Z}^{d}})$ be probability measures with the same family of 1-point conditional distributions $\mu_{k}(\cdot\,\vert \underline{\xi})$, with $k\in \mathbb{Z}^{d}$ and $\underline{\xi} \in \mathcal{X}^{\mathbb{Z}^{d} \setminus \{k\}}$. Then, under the conditions of Lemma \ref{Dobpr}, $\mu = \nu$.
\end{proposition}

\noindent \textbf{Proof.} Let $\Lambda_{0} \in \mathcal{S}$ be fixed. For any $D>0$ define the set
\begin{equation*}
\Lambda_{D} := \{ k \in \mathbb{Z}^{d}: \exists j \in \Lambda_{0},\, \vert k - j \vert \leq D\}.
\end{equation*}
Fix $0 < \epsilon < 1$ and let $N \in \mathbb{N}$ be large enough so that
\begin{equation}
\label{usef1}
2C \frac{\lambda^{N+2}}{1 -\lambda} < \frac{\epsilon}{2},
\end{equation}
where $C>0$ is defined in \eqref{bd2C}, see \eqref{gammalC}. By assumption, there exists $D_{1}>0$ such that
\begin{equation}
\label{sumeps1}
\sup_{j \in \Lambda_{0}} \sum_{l \in \Lambda_{D_{1}}^c} r(l,j) < \frac{\epsilon}{2C}.
\end{equation}
Continuing by recursion, there exist $D_{1}<D_{2}< \dotsb <D_{N+1}$ such that
\begin{equation}
\label{sumeps2}
\sup_{j \in \Lambda_{D_{m}}} \sum_{l \in \Lambda_{D_{m+1}}^c} r(l,j) < \frac{\epsilon}{2C},\quad m=1,\ldots,N.
\end{equation}
Set $\Lambda = \Lambda_{D_{N+1}}$. Applying Lemma \ref{Dobpr}, there exists $\tilde{\sigma} \in \Xi_{\mathcal{X}^{\mathbb{Z}^{d}}}(\mu,\nu)$ such that,
\begin{equation*}
\int_{\mathcal{X}^{\mathbb{Z}^{d}} \times \mathcal{X}^{\mathbb{Z}^{d}}} \rho(\xi_{j},\xi_{j}') \tilde{\sigma}(d\underline{\xi},d\underline{\xi}') < \chi(j) + \frac{\epsilon}{2},\quad j \in \mathbb{Z}^{d},
\end{equation*}
where $\chi(j)=2C$ for $j \notin \Lambda$, and,
\begin{equation}
\label{chide}
\chi(j) = 2C \sum_{l \in \Lambda^{c}} \sum_{k \in \Lambda} r(l,k) \sum_{n=0}^{\infty} (R_{\Lambda}^{n})_{k,j},\quad j \in \Lambda.
\end{equation}
Define the matrix norm of $R_{\Lambda}$ by
\begin{equation*}
\Vert R_{\Lambda} \Vert := \sup_{k\in \Lambda} \sum_{j \in \Lambda} \vert (R_{\Lambda})_{j,k} \vert.
\end{equation*}
Note that $\Vert R_{\Lambda} \Vert \leq \lambda < 1$ by assumption. To estimate \eqref{chide} with $j \in \Lambda_{0}$, we split the sum over $n$ into the terms with $n\leq N$ and the rest, and note first that
\begin{equation*}
2C \sum_{l \in \Lambda^{c}} \sum_{k \in \Lambda} r(l,k) \sum_{n=N+1}^{\infty} (R_{\Lambda}^{n})_{k,j} \leq 2C \sup_{k \in \Lambda} \sum_{\substack{l \in \mathbb{Z}^{d} \\ l \neq k}} r(l,k) \sum_{n=N+1}^{\infty} \Vert R_{\Lambda} \Vert^{n} \leq 2C \frac{\lambda^{N+2}}{1 - \lambda}< \frac{\epsilon}{2},
\end{equation*}
uniformly in $j \in \Lambda_{0}$. Here, we used \eqref{usef1} in the right-hand side of the last inequality.\\ The remaining terms, we write as follows
\begin{equation*}
2C \sum_{l \in \Lambda^{c}} \left(r(l,j) + \sum_{n=1}^{N} \sum_{k_{1},\ldots, k_{n} \in \Lambda} r(l,k_{n}) \left(\prod_{i=1}^{n-1} r(k_{n-i+1},k_{n-i})\right)r(k_{1},j)\right),\quad j \in \Lambda_{0}.
\end{equation*}
Now,
\begin{equation*}
2C \sup_{j \in \Lambda_{0}} \sum_{l \in \Lambda^{c}} r(l,j) < \epsilon,
\end{equation*}
by the definition of $\Lambda$ and \eqref{sumeps1} since $D_{N+1} > D_{1}$. Similarly, for $n \geq1$, if $k_{n} \in \Lambda_{D_{n}}$, we obtain
\begin{equation*}
2C \sum_{ l\in \Lambda^{c}} \sum_{n=1}^{N} \sum_{k_{n} \in \Lambda_{D_n}} r(l,k_{n}) (R_{\Lambda}^{n})_{k_{n},j} \leq 2C \sup_{k\in \Lambda} \sum_{\substack{ l \in \mathbb{Z}^{d} \\ l \neq k}} r(l,k) \sum_{n=1}^{N} \Vert R_{\Lambda}\Vert^{n} \leq \frac{\lambda}{1 - \lambda} \epsilon,
\end{equation*}
uniformly in $j \in \Lambda_{0}$. The term where $k_{n} \in \Lambda \setminus \Lambda_{D_{n}}$ can be written as
\begin{multline*}
2C \sum_{l\in \Lambda^{c}} \sum_{n=1}^{N} \sum_{p=1}^{n-1} \sum_{k_{n} \in \Lambda \setminus \Lambda_{D_{n}}} \dotsb \sum_{k_{n-p+1} \in \Lambda \setminus \Lambda_{D_{n-p+1}}} \sum_{k_{n-p} \in \Lambda_{D_{n-p}}} r(l,k_{n}) \\
\times \left(\prod_{s=1}^{p} r(k_{n-s+1},k_{n-s})\right) (R_{\Lambda}^{n-p})_{k_{n-p},j}.
\end{multline*}
This is bounded by
\begin{multline*}
2C \sum_{n=1}^{N} \sum_{p=1}^{n-1}\left(\sup_{k \in \Lambda} \sum_{\substack{l \in \mathbb{Z}^{d} \\ l \neq k}} r(l,k)\right) \Vert R_{\Lambda}^{p-1} \Vert \sup_{k' \in \Lambda_{D_{n-p}}} \left(\sum_{k'' \in \Lambda_{D_{n-p+1}}^{c}} r(k'',k')\right) \Vert R_{\Lambda}^{n-p}\Vert
\\ \leq \sum_{n=1}^{N} \sum_{p=1}^{n-1} \lambda^{n} \epsilon < \frac{\lambda}{(1-\lambda)^{2}} \epsilon,
\end{multline*}
where we used \eqref{sumeps2} in the right-hand side of the first inequality. In total, we obtain the following upper bound
\begin{equation*}
\begin{split}
\rho^{(\Lambda_{0})} (\mu_{\Lambda_{0}},\nu_{\Lambda_{0}}) &\leq \sum_{j \in \Lambda_{0}} \int_{\mathcal{X}^{\mathbb{Z}^{d}} \times \mathcal{X}^{\mathbb{Z}^{d}}} \rho(\xi_{j},\xi_{j}') \tilde{\sigma}(d\underline{\xi}, d\underline{\xi}') \\
&< \left(1 + \frac{\lambda}{1 - \lambda} + \frac{\lambda}{(1 - \lambda)^{2}}\right) \vert \Lambda_{0} \vert \epsilon = \frac{1}{(1-\lambda)^{2}} \vert \Lambda_{0} \vert \epsilon.
\end{split}
\end{equation*}
Letting $\epsilon \rightarrow 0$, we conclude that  $\mu_{\Lambda_{0}} = \nu_{\Lambda_{0}}$. Since $\Lambda_{0} \subset \mathbb{Z}^{d}$ is an arbitrary finite subset and the measures are entirely determined by their marginals, $\mu = \nu$. \qed

\subsection{Proof of Corollaries \ref{corol1} and \ref{corol2}.}
\label{coroproo}

\noindent \textbf{Proof of Corollary \ref{corol1}.} We have, as in the proof of Proposition \ref{propDobp}, that there exists a coupling $\tilde{\sigma} \in \Xi_{\mathcal{X}^{\mathbb{Z}^{d}}}(\mu_{\Delta}^{\Lambda}(\cdot\,\vert \underline{\vartheta}_{\Lambda^{c}}), \mu_{\Delta}^{\Lambda'}(\cdot\,\vert \underline{\vartheta}_{\Lambda'^{c}}))$ such that, for all $j \in \mathbb{Z}^{d}$,
\begin{equation*}
\int_{\mathcal{X}^{\mathbb{Z}^{d}} \times \mathcal{X}^{\mathbb{Z}^{d}}} \rho(\xi_{j},\xi_{j}') \tilde{\sigma}(d \underline{\xi}, d\underline{\xi}') < \chi(j) + \epsilon,
\end{equation*}
where $\chi(j) = 2 c_{\infty}$ for $j \notin \Lambda$, and where $\chi(j)$ for $j \in \Lambda$ is given by
\begin{equation*}
\chi(j) = 2 c_{\infty} \sum_{l \in \Lambda^{c}} \sum_{k \in \Lambda} \sum_{p=0}^{\infty} r(l,k) (R_{\Lambda}^{p})_{k,j}.
\end{equation*}
In the particular case of nearest-neighbour interactions, $r(l,k)=0$ unless $\vert k-l\vert=1$, and hence, if $\mathrm{dist}(\Delta, \Lambda^{c}) = D>0$, $r(l,k) (R_{\Lambda}^{p})_{k,j} = 0$ unless $p \geq D-1$. Since $\Vert R_{\Lambda} \Vert = \lambda < 1$ by assumption, we conclude that
\begin{equation*}
\chi(j) \leq 2 c_{\infty} \sum_{p=D}^{\infty} \lambda^{p} = \frac{2 c_{\infty}}{1 - \lambda}\lambda^{D},
\end{equation*}
from which it follows that
\begin{equation*}
\rho_{W}^{(\Delta)} \left(\mu_{\Delta}^{\Lambda}(\cdot\, \vert \underline{\vartheta}_{\Lambda^{c}}), \mu_{\Delta}^{\Lambda'}(\cdot\, \vert \underline{\vartheta}_{\Lambda'^{c}})\right) \leq \frac{2 c_{\infty}}{1 - \lambda} \vert \Delta \vert \lambda^{D}.
\end{equation*}
The second part of the corollary is now straightforward
\begin{multline*}
\left\vert \int_{\mathcal{X}^{\Lambda}} f(\underline{\xi}_{\Delta}) \mu_{\Lambda}^{\beta}(d\underline{\xi}_{\Lambda}\vert \underline{\vartheta}_{\Lambda^{c}}) - \int_{\mathcal{X}^{\Lambda'}} f(\underline{\xi}_{\Delta}') \mu_{\Lambda'}^{\beta}(d\underline{\xi}_{\Lambda'}' \vert \underline{\vartheta}_{\Lambda'^{c}}) \right \vert \\
\begin{split}
& \leq \int_{\mathcal{X}^{\mathbb{Z}^{d}} \times \mathcal{X}^{\mathbb{Z}^{d}}} \left\vert f(\underline{\xi}_{\Delta})  -  f(\underline{\xi}_{\Delta}') \right \vert \tilde{\sigma}(d\underline{\xi}, d\underline{\xi}') \\
&\leq  c_{f}  \int_{\mathcal{X}^{\mathbb{Z}^{d}} \times \mathcal{X}^{\mathbb{Z}^{d}}}\sum_{j \in \Delta} \rho(\xi_{j}, \xi_{j}') \tilde{\sigma}(d\underline{\xi}, d\underline{\xi}') \leq \vert \Delta \vert (C \lambda^{D} + \epsilon),
\end{split}
\end{multline*}
where $c_{f}>0$ is the Lipschitz constant. Taking $\epsilon \rightarrow 0$ this proves the second statement. \qed \\

\noindent \textbf{Proof of Corollary \ref{corol2}.} Let $\Lambda \subset \Delta'^{c}$ be finite such that $\Delta \subset \Lambda$. We can write
\begin{equation*}
\int_{\mathcal{X}^{\Delta \cup \Delta'}} f(\underline{\xi}_{\Delta}) g(\underline{\xi}_{\Delta'}) \mu_{\Delta \cup \Delta'}(d\underline{\xi}_{\Delta \cup \Delta'}) = \int_{\mathcal{X}^{\Lambda^{c}}}  g(\underline{\xi}_{\Delta'}) \mu_{\Lambda^{c}}(d\underline{\xi}_{\Lambda^{c}}) \int_{\mathcal{X}^{\Lambda}} f(\underline{\xi}_{\Delta}) \mu_{\Delta}^{\Lambda}(d\underline{\xi}_{\Lambda} \vert \underline{\xi}_{\Lambda^{c}}).
\end{equation*}
Hence, it suffices to show that
\begin{equation*}
\rho_{W}^{(\Delta)}\left(\mu_{\Delta}, \mu_{\Delta}^{\Lambda}(\cdot\,\vert \underline{\xi}_{\Lambda^{c}})\right) < C \lambda^{\mathrm{dist}(\Delta,\Lambda^{c})}.
\end{equation*}
This is analogous to the previous corollary. \qed

\section{A generalization: The Dobrushin-Shlosman Theorem}
\label{DobShlos}

The Dobrushin-Shlosman Theorem \cite[Thm 2.1]{DS1} generalizes the 1-point Dobrushin condition in \eqref{Dcondd} to a set of finite-volume conditions, see \eqref{DScond0}-\eqref{DScond} below, referred to as the Dobrushin-Shlosman conditions (also known as $\mathrm{C}_{\mathrm{V}}$ conditions), ensuring uniqueness of limit Gibbs distributions. 

\begin{theorem}
\label{DSthm}
Let $\beta >0$. Suppose that the conditions $\mathrm{(C1)}$-$\mathrm{(C2)}$ in Theorem \ref{Dobrushin} are satisfied and assume that the potentials $\{\mathcal{V}_{X}\}_{X \in \mathcal{S}}$ in \eqref{Hamm} are translation-invariant. Further, assume that the following condition holds. There exists a finite domain $\Lambda_{0} \in \mathcal{S}$ such that, for all $k \in \Lambda_{0}^{c}$ and for all $(\underline{\eta},\underline{\eta}') \in \mathcal{X}^{\Lambda_{0}^{c}} \times \mathcal{X}^{\Lambda_{0}^{c}}$, there exists a non-negative function $r$ satisfying \\ 
\noindent $\mathrm{(i)}$. 
\begin{equation}
\label{exponR}
r(k) \leq R\,e^{-\alpha \vert k\vert}, \quad \textrm{for some $\alpha, R > 0$};
\end{equation}
\noindent $\mathrm{(ii)}$.
\begin{equation}
\label{DScond0}	 
\frac{1}{\vert \Lambda_{0}\vert} \sum_{k \in \Lambda_{0}^{c}} r(k) = \lambda < 1; 
\end{equation}
such that
\begin{equation} 
\label{DScond}
\rho_{W}^{(\Lambda_{0})}\left(\mu_{\Lambda_{0}}\left( \cdot\,\vert\,\underline{\eta}_{\Lambda_{0}^{c}}\right), \mu_{\Lambda_{0}}\left(\cdot\,\vert\, \underline{\eta}_{\Lambda_{0}^{c}}'\right)\right) \leq \sum_{k \in \Lambda_{0}^{c}} r(k) \rho(\eta_{k}, \eta_{k}'). 		 
\end{equation} 
Then, there exists at most one limit Gibbs distribution $\mu^{\beta} \in \mathcal{P}(\mathcal{X}^{\mathbb{Z}^{d}})$, associated with the Hamiltonian \eqref{Hamm}, with marginal distributions satisfying \eqref{schcond}.
\end{theorem} 

\begin{remark}
$r$ in \eqref{exponR}-\eqref{DScond} may depend on the boundary condition $\eta_{\Lambda_{0}^{c}}$.
\end{remark}

We start the proof with a generalization of Lemma \ref{Dobpr}

\begin{lema} 
Let $\mu, \nu \in \mathcal{P}(\mathcal{X}^{\mathbb{Z}^{d}})$ be probability measures satisfying \eqref{bd2C} for some constant $C>0$. Assume that $\mu, \nu$ satisfy the DLR condition in \eqref{DLRequ} and the Dobrushin-Shlosman condition \eqref{DScond}.\\
Fix a finite domain $\Lambda \in \mathcal{S}$. Then, for any $\delta > 0$, there exists a coupling $\sigma \in  \Xi_{\mathcal{X}^{\mathbb{Z}^{d}}}(\mu,\nu)$ such that, 
\begin{equation} 
\label{ineqgam2}
\sum_{j \in \Lambda_{0}} \gamma(j+l) \leq \sum_{k \in \Lambda_{0}^{c}} r(k) \gamma(k+l) + \delta,
\end{equation}
holds for all $l \in \Lambda$, and where $\gamma$ is defined by 
\begin{equation} 
\label{gammaj}
\gamma(j) := \int_{\mathcal{X}^{\mathbb{Z}^{d}} \times \mathcal{X}^{\mathbb{Z}^{d}}} \rho(\xi_{j}, \xi_{j}')\,\sigma(d\underline{\xi}, d\underline{\xi}'). 
\end{equation} 
\end{lema}

\noindent \textbf{Proof.} The proof is similar to the proof of Lemma \ref{Dobpr}.  Set $\Lambda' := \Lambda + \Lambda_{0}$ and let $\delta >0$ be fixed. By definition of $\rho_{W}^{(\Lambda')}$, see \eqref{metriclambda} along with  \eqref{rhoW}, there exists a coupling $\sigma \in \Xi_{\mathcal{X}^{\mathbb{Z}^{d}}}(\mu,\nu)$  such that 
\begin{equation}
\label{nott1}
\sum_{k \in \Lambda'} \int_{\mathcal{X}^{\mathbb{Z}^{d}} \times \mathcal{X}^{\mathbb{Z}^{d}}} \rho(\xi_{k}, \xi_{k}')\,\sigma(d\underline{\xi}, d\underline{\xi}') \leq \rho_{W}^{(\Lambda')}\left(\mu_{\Lambda'}, \nu_{\Lambda'}\right) + \frac{\delta}{2}. 
\end{equation}
Now, we fix $l \in \Lambda$. By the Dobrushin-Shlosman condition in \eqref{DScond}, there exist, for all ($\underline{\eta}, \underline{\eta}') \in \mathcal{X}^{l+ \Lambda_{0}^{c}} \times \mathcal{X}^{l+ \Lambda_{0}^{c}}$, couplings  
$\sigma_{l+\Lambda_{0};\underline{\eta}, \underline{\eta}'} \in \Xi_{\mathcal{X}^{l+\Lambda_{0}}}(\mu_{l+\Lambda_{0}}(\cdot\, \vert \underline{\eta}), \nu_{l+\Lambda_{0}}(\cdot\, \vert \underline{\eta}'))$ such that
\begin{equation}
\label{need22}	
\sum_{j \in \Lambda_{0}} \int_{\mathcal{X}^{l+\Lambda_{0}} \times \mathcal{X}^{l+\Lambda_{0}}} \rho(\xi_{l+j},\xi_{l+j}')\,\sigma_{l+\Lambda_{0};\underline{\eta}, \underline{\eta}'}(d\underline{\xi},d\underline{\xi}')
\leq  \sum_{k \in \Lambda_{0}^{c}} r(k) \rho(\eta_{l+k}, \eta_{l+k}') + \frac{\delta}{2}.
\end{equation} 
For $A,B \in \mathcal{B}(\mathcal{X}^{\mathbb{Z}^{d}})$, define the coupling $\tilde{\sigma}^{(l)} \in \mathcal{P}(\mathcal{X}^{\mathbb{Z}^{d}} \times \mathcal{X}^{\mathbb{Z}^{d}})$ as follows
\begin{equation*}
\tilde{\sigma}^{(l)}(A \times B) := \int_{\mathcal{X}^{\mathbb{Z}^{d}} \times \mathcal{X}^{\mathbb{Z}^{d}}} \mathbb{I}_{A \times B}\left(\left(\underline{\xi}_{l+\Lambda_{0}},\underline{\eta}_{l+ \Lambda_{0}^{c}}\right),\left(\underline{\xi}_{l+\Lambda_{0}}',\underline{\eta}_{l+\Lambda_{0}^{c}}'\right)\right)  \sigma_{l+\Lambda_{0};\underline{\eta},\underline{\eta}'}(d\underline{\xi},d\underline{\xi}') \sigma_{l+\Lambda_{0}^{c}} (d\underline{\eta},d\underline{\eta}').
\end{equation*}
By mimicking the arguments in the proof of Lemma \ref{Dobpr}, we can show that
$\tilde{\sigma}^{(l)}\in \Xi_{\mathcal{X}^{\mathbb{Z}^{d}}}(\mu,\nu)$. Next define
\begin{equation*}
\tilde{\gamma}^{(l)}(k) := \int_{\mathcal{X}^{\mathbb{Z}^{d}}\times \mathcal{X}^{\mathbb{Z}^{d}}} \rho(\xi_{k},\xi_{k}')\,\tilde{\sigma}^{(l)}(d\underline{\xi},d\underline{\xi}').
\end{equation*}
Clearly, for $k \in \Lambda' \setminus (l + \Lambda_{0})$,
\begin{equation*}
\begin{split}
\tilde{\gamma}^{(l)}(k) &= \int_{\mathcal{X}^{\mathbb{Z}^{d}}\times \mathcal{X}^{\mathbb{Z}^{d}}} \rho(\eta_{k}, \eta_{k}')\, \sigma_{l+\Lambda_{0};\underline{\eta},\underline{\eta}'}
(d\underline{\xi}_{l+\Lambda_{0}},
d\underline{\xi}_{l+\Lambda_{0}}') \sigma_{l+\Lambda_{0}^{c}} (d\underline{\eta}_{l+\Lambda_{0}^{c}},d\underline{\eta}_{l+ \Lambda_{0}^{c}}') \\
&= \int_{\mathcal{X}^{l+\Lambda_{0}^{c}}\times \mathcal{X}^{l + \Lambda_{0}^{c}}} \rho(\eta_{k}, \eta_{k}')\,\sigma_{l + \Lambda_{0}^{c}}(d\underline{\eta},d\underline{\eta}') = \gamma(k).
\end{split}
\end{equation*}
Note that, since $\tilde{\sigma}^{(l)}\in \Xi_{\mathcal{X}^{\mathbb{Z}^{d}}}(\mu,\nu)$, we have,
\begin{equation}
\label{need2b3}
\sum_{k \in \Lambda'}\tilde{\gamma}^{(l)}(k) \geq \rho_{W}^{(\Lambda')}\left(\mu_{\Lambda'},\nu_{\Lambda'}\right).
\end{equation} 
It then follows from \eqref{nott1} along with \eqref{need2b3} that 
\begin{equation}
\label{need2b2} 
\sum_{j \in \Lambda_{0}} \left(\gamma(j+l) - \tilde{\gamma}^{(l)}(j+l) \right) = \sum_{k \in \Lambda'} \gamma(k) - \sum_{k \in \Lambda'} \tilde{\gamma}^{(l)}(k) \leq \frac{\delta}{2}.
\end{equation}
Integrating \eqref{need22} with respect to $\sigma_{l + \Lambda_{0}^{c}}$, we have,
\begin{equation*} 
\begin{split}
\sum_{j \in \Lambda_{0}} \tilde{\gamma}^{(l)}(j+l) 
&\leq  \sum_{k \in \Lambda_{0}^{c}} r(k)  \int_{\mathcal{X}^{l + \Lambda_{0}^{c}} \times \mathcal{X}^{l + \Lambda_{0}^{c}}} \rho(\eta_{l+k}, \eta_{l+k}') \sigma_{l + \Lambda_{0}^{c}}(d\underline{\eta},d\underline{\eta}') + \frac{\delta}{2}  \\
&= \sum_{k \in \Lambda_{0}^{c}} r(k) \gamma(k+l) 
+ \frac{\delta}{2}.
\end{split}
\end{equation*}
To complete the proof of \eqref{ineqgam2}, it suffices to use \eqref{need2b2} yielding 
\begin{equation*} 
\sum_{j \in \Lambda_{0}} \gamma(j+l) \leq \sum_{j \in \Lambda_{0}}  \tilde{\gamma}^{(l)}(j+l) + \frac{\delta}{2} \leq
\sum_{k \in \Lambda_{0}^c} r(k) \gamma(k+l) + \delta. \tag*{\qed}
\end{equation*} 

We conclude this section with the proof of Theorem \ref{DSthm}\\

\noindent \textbf{Proof of Theorem \ref{DSthm}.} Under the conditions of the Theorem, let $\Lambda_{0} \in \mathcal{S}$ be such that the Dobrushin-Shlosman condition in \eqref{DScond}  holds. Let  $\Lambda_{1} \in \mathcal{S}$ be a given finite subset. Let $\Lambda \subset \mathbb{Z}^{d}$ be large such that $\Lambda_{1} \subset \Lambda$. Define, for $\epsilon > 0$,
\begin{equation*} 
c(k) := \mathrm{e}^{-\epsilon\,\mathrm{dist}(k,\Lambda_{1})} >0. 
\end{equation*}
Multiplying the inequality \eqref{ineqgam2} by $c(l)$ and summing over $l \in \Lambda$, we have,
\begin{equation} 
\label{gammsuminq}
\sum_{l \in \Lambda} c(l) \sum_{j \in \Lambda_{0}} \gamma(j+l) \leq 
\sum_{l \in \Lambda} c(l) \sum_{k \in \Lambda_{0}^{c}} r(k) \gamma(k+l) + \delta \sum_{l \in \Lambda} c(l). 
\end{equation}
We rewrite the left-hand side of \eqref{gammsuminq} as 
\begin{equation*} 
\sum_{l \in \Lambda} c(l) \sum_{j \in \Lambda_{0}} \gamma(j+l)  = 
\sum_{l \in \Lambda + \Lambda_{0}} \gamma(l) \sum_{j \in \Lambda_{0}\,:\, l-j \in \Lambda} c(l-j). 
\end{equation*}
Denote $D_{0} := \mathrm{diam}(\Lambda_{0})$. Since
\begin{equation*}
\mathrm{dist}(l-j,\Lambda_{1}) \leq  \mathrm{dist}(l,\Lambda_{1}) + \vert j\vert \leq \mathrm{dist}(l,\Lambda_{1}) + D_{0},
\end{equation*}
we have,
\begin{align} 
\sum_{l \in \Lambda} c(l) \sum_{j \in \Lambda_{0}} \gamma(j+l)  &\geq
\sum_{l \in \Lambda + \Lambda_{0}} \gamma(l) \,\mathrm{e}^{-\epsilon D_{0}} c(l) \big\vert \{j \in \Lambda_0:\,l-j \in \Lambda\}\big\vert \nonumber \\
\label{leftinqlwbd} 
&\geq  \vert \Lambda_{0} \vert\, \mathrm{e}^{-\epsilon D_{0}}\sum_{l \in \bigcap_{j \in \Lambda_{0}} (\Lambda + j)} \gamma(l)  c(l). 
\end{align}
In the right-hand side of \eqref{gammsuminq}, we have similarly,
\begin{equation*}
\sum_{l \in \Lambda} c(l) \sum_{k \in \Lambda_{0}^{c}} r(k) \gamma(k+l) = 
\sum_{l \in \mathbb{Z}^{d}} \gamma(l)
\sum_{k \in \Lambda_{0}^{c} \,:\, l-k \in \Lambda} r(k) c(l-k).
\end{equation*}
Assume now that $\epsilon < \frac{\alpha}{2}$, see \eqref{exponR}, and choose $\Lambda' \subset \mathbb{Z}^{d}$ such that  $\Lambda + \Lambda_{0} \subset \Lambda'$ and so large that
\begin{equation}
\label{alphlada} 
R \sum_{k \in \Lambda'^{c}} \mathrm{e}^{-\frac{\alpha}{2} \vert k \vert} < \frac{1}{2}(1-\lambda). 
\end{equation}
Denote $D' := \mathrm{diam}(\Lambda')$. Since for all $k \in \Lambda' \setminus \Lambda_{0}$,
\begin{equation*}
\mathrm{dist}(l-k,\Lambda_{1}) \geq \mathrm{dist}(l,\Lambda_{1}) - \vert k \vert \geq \mathrm{dist}(l,\Lambda_{1}) - D',
\end{equation*}
we have,
\begin{align} 
\sum_{l \in \Lambda} c(l) \sum_{k \in \Lambda_{0}^{c}} r(k) \gamma(k+l) &=
\sum_{l \in \mathbb{Z}^{d}} \gamma(l) \sum_{k \in \Lambda_{0}^{c}\,:\, l-k \in \Lambda} r(k) c(l-k) \nonumber \\ 
&\leq \sum_{l \in \mathbb{Z}^{d}} \gamma(l) c(l) 
\left( \mathrm{e}^{\epsilon D'} \sum_{k \in \Lambda' \setminus \Lambda_{0}} r(k)
+ R \sum_{k \in \Lambda'^{c}} \mathrm{e}^{-(\alpha-\epsilon) \vert k \vert} \right) \nonumber \\ 
\label{rightingupbd}
&\leq \sum_{l \in \mathbb{Z}^{d}} \gamma(l) c(l) 
\left( \vert \Lambda_{0}\vert\,\lambda \mathrm{e}^{\epsilon D'} + \frac{1}{2}(1-\lambda) \right).
\end{align} 
Here, we used \eqref{exponR} and \eqref{alphlada} in the right-hand side of the first and second inequality respectively.  From \eqref{gammsuminq} together with \eqref{leftinqlwbd} and \eqref{rightingupbd}, we eventually have
\begin{equation}
\label{finnek}
\vert \Lambda_{0} \vert\, \mathrm{e}^{-\epsilon D_{0}}\sum_{l \in \bigcap_{j \in \Lambda_{0}} (\Lambda + j)} \gamma(l)  c(l) \leq \sum_{l \in \mathbb{Z}^{d}} \gamma(l) c(l) 
\left( \vert \Lambda_{0}\vert\,\lambda \mathrm{e}^{\epsilon D'} + \frac{1}{2}(1-\lambda) \right) + \delta \sum_{l \in \Lambda} c(l).
\end{equation}
We split the sum over $l\in \mathbb{Z}^{d}$ in the right-hand side of \eqref{finnek} into $l \in \bigcap_{j \in \Lambda_{0}} (\Lambda + j)$ and $l \in \bigcup_{j \in \Lambda_{0}} (\Lambda + j)^{c} = \Lambda^{c} + \Lambda_{0}$, and move the former to the left-hand side. The resulting inequality is
\begin{multline} 
\label{finnek2}
\vert \Lambda_{0}\vert \left(\mathrm{e}^{-\epsilon D_{0}} - \lambda \mathrm{e}^{\epsilon D'} - \frac{1}{2}(1-\lambda)\right)
\sum_{l \in \cap_{j \in \Lambda_{0}} (\Lambda + j)} \gamma(l) c(l) \\
\leq  
\left(\lambda \mathrm{e}^{\epsilon D'} + \frac{1}{2}(1-\lambda)\right) \sum_{l \in \Lambda^{c} + \Lambda_{0}} \gamma(l) c(l) + \delta \sum_{l \in \Lambda} c(l). 
\end{multline}
Note that for $\Lambda$ large enough, $\Lambda_{1} \subset \bigcap_{j \in \Lambda_{0}} (\Lambda + j)$, so in the left-hand side of \eqref{finnek2}  we can restrict the sum over $l \in \bigcap_{j \in \Lambda_{0}} (\Lambda + j)$ to $l \in \Lambda_{1}$. We now let $\epsilon$ be so small that 
\begin{equation*}
\kappa := \mathrm{e}^{-\epsilon D_{0}} - \frac{1}{2} - \lambda \left( \mathrm{e}^{\epsilon D'} -\frac{1}{2}\right) > 0.
\end{equation*} 
As a result, \eqref{finnek2} becomes
\begin{equation} 
\label{finnek3}	
\sum_{l \in \Lambda_{1}} \gamma(l)\, c(l) \leq \frac{1}{\kappa \vert \Lambda_{0}\vert} \left( \sum_{l \in \Lambda^{c} + \Lambda_{0}} \gamma(l) c(l) + \delta \sum_{l \in \mathbb{Z}^{d}} c(l)\right).
\end{equation} 
Letting $\Lambda$ tend to $\mathbb{Z}^{d}$, the first sum in the right-hand side of \eqref{finnek3} tends to zero. Subsequently, letting $\delta \to 0$,
we find that the right-hand side of \eqref{finnek3} tends to zero. Therefore, in view of \eqref{gammaj},
\begin{equation*}
\sum_{l \in \Lambda_{1}} \gamma(l) =  \rho_{W}^{(\Lambda_{1})}(\mu_{\Lambda_{1}}, \nu_{\Lambda_{1}}) = 0,
\end{equation*}
which implies that $\mu_{\Lambda_{1}} = \nu_{\Lambda_{1}}$. Since $\Lambda_{1} \subset \mathbb{Z}^{d}$ is an arbitrary finite subset and the measures are entirely determined by their marginals, we conclude that $\mu = \nu$. \qed

\begin{remark}
The proof crucially relies on the condition \eqref{exponR}, corresponding to short-range interactions, in obtaining the bound \eqref{rightingupbd}.  If the interactions are finite-range, the condition \eqref{exponR} is superfluous. 
\end{remark}

\begin{remark}
The proof easily extends to tempered Gibbs measures. Alternatively, the exponential decay in \eqref{exponR} can be replaced by a sufficiently fast polynomial decay provided that  the distributions satisfy the condition in \eqref{schcond}.  
\end{remark}
	
\section{Applications - Part I: Uniqueness of Gibbs measures at high temperature.}
\label{App}

\subsection{A general result with application to the classical Heisenberg model.}
\label{app11}

At high temperature, there is in general no phase transition. Dobrushin's uniqueness theorem gives a simple explicit upper bound for the inverse temperature

\begin{proposition}
\label{highT}
Assume that $\mathcal{X}$ is either a compact convex subset of a normed space with metric $\rho$, or a compact Riemannian manifold. In the latter case, assume that the metric on $\mathcal{X}$ is given by $\rho(x,y) = \int_{0}^{1} \vert \zeta(s) \vert \,ds$, where $\zeta$ is a shortest geodesic from x to y. Consider the following formal Hamiltonian with nearest-neighbour interactions defined on $\mathcal{X}^{\mathbb{Z}^{d}}$ as
\begin{equation}
\label{famHlda}
H_{inv}(\underline{\xi}) := \sum_{j \in \mathbb{Z}^{d}} \mathcal{V}(\xi_{j}) + \sum_{j \in \mathbb{Z}^{d}} \sum_{l \in N_{1}(j)} \mathcal{V}_{l-j}(\xi_{j},\xi_{l}),
\end{equation}
where $N_{1}(j) := \{ j' \in \mathbb{Z}^{d} : \vert j'- j \vert = 1\}$ denotes the set of nearest-neighbours of $j \in \mathbb{Z}^{d}$, and where,\\
$(i)$. $\mathcal{V}: \mathcal{X} \rightarrow \mathbb{R}$ is assumed to be continuous; \\
$(ii)$. $\mathcal{V}_{k}: \mathcal{X}\times \mathcal{X} \rightarrow \mathbb{R}$, $\vert k \vert =1$ are supposed to be jointly continuous and Lipschitz-continuous in the second variable, i.e. there exists a constant $C_{0} >0$ such that, for all $(\xi, \eta,\eta') \in \mathcal{X}\times \mathcal{X}\times \mathcal{X}$,
\begin{equation*}
\vert \mathcal{V}_{k}(\xi,\eta) - \mathcal{V}_{k}(\xi,\eta')\vert \leq C_{0} \rho(\eta,\eta').
\end{equation*}
Further, let $\mu_{0} \in \mathcal{P}(\mathcal{X})$ be a  given a priori probability measure and, for a given $\beta>0$, let $\mu_{j}^{\beta}=\mu^{\beta}$ denote the associated 1-point Gibbs distribution generated by \eqref{famHlda}. Let $C_{1} > 0$ be defined as
\begin{equation*}
C_{1}(\beta) := \sup_{\xi' \in \mathcal{X}} \sup_{\underline{\varphi} \in \mathcal{X}^{N_{1}(0)}}\int_{\mathcal{X}} \rho(\xi,\xi') \mu^{\beta}(d\xi \vert \underline{\varphi}) < \infty.
\end{equation*}
Then, provided that $12 d C_{0} C_{1}(\beta) \beta < 1$, there exists a unique limit Gibbs distribution at inverse temperature $\beta>0$ in $\mathcal{P}(\mathcal{X}^{\mathbb{Z}^{d}})$, associated with $H_{inv}$. 
\end{proposition}

As a direct application of Proposition \ref{highT}, consider the classical Heisenberg model.
In this model, the single-spin space is the unit sphere $\mathbb{S}^{r}$, $r \geq 1$ with normalized Lebesgue measure, and the pair-interaction potential is
\begin{equation*}
\mathcal{V}_{l-j}(s_{j},s_{l}) = - J s_{j} \cdot s_{l},\quad s_{j},s_{l} \in \mathbb{S}^{r},
\end{equation*}
for some real coupling constant $J>0$. Proposition \ref{highT} guarantees that there is no phase transition for $\beta < (12d \pi J)^{-1}$. Indeed, the Riemann metric is bounded by $\rho(s,s') \leq \pi$ so that $C_1 \leq \pi$, and if $\theta$ denotes the angle between $s$ and $s'$,
\begin{equation*}
\vert \mathcal{V}_{k}(s_0,s)-\mathcal{V}_{k}(s_0,s')\vert = J \vert s_0 \cdot (s-s')\vert \leq J \vert s-s'\vert = 2 J \sin\left(\frac{\theta}{2}\right) \leq J\theta = J \rho(s,s').
\end{equation*}
In fact, considering $\mathbb{S}^{r}$ as a subset of $\mathbb{R}^{r+1}$, the estimate can be slightly improved to $\beta < (24d J)^{-1}$. Using the symmetry as in the Ising model in Sec. \ref{Isingdim} below, the estimate can be improved further.\\

\noindent \textbf{Proof of Proposition \ref{highT}.} Given $(\underline{\eta},\underline{\eta}') \in \mathcal{X}^{N_{1}(0)} \times \mathcal{X}^{N_{1}(0)}$ and $M \in \mathbb{N}$, we set
\begin{equation}
\label{etak}
\underline{\eta}_{k} := \left(1-\frac{k}{M}\right)\underline{\eta} + \frac{k}{M} \underline{\eta}' = \underline{\eta} + \frac{k}{M}(\underline{\eta}' - \underline{\eta}), \quad k=0,\dots, M,
\end{equation}
if $\mathcal{X}$ is a convex subset of a normed space, and we set
\begin{equation}
\label{etakl}
\eta_{k,l} := \zeta_{l}\left(\frac{k}{M}\right), \quad k=0,\dots, M,
\end{equation}
where, for $l \in N_{1}(0)$, $\zeta_{l}: [0,1] \rightarrow \mathcal{X}$ is a minimal-length geodesic from $\eta_{l}$ to $\eta_{l}'$ if $\mathcal{X}$ is a compact manifold. Note that, since $\mathcal{X}$ is assumed to be compact, there exist $C_{*}>0$ such that $\sum_{l \in N_{1}(0)} \rho(\xi_{l},\xi_{l}') \leq  C_{*}$ uniformly in $\underline{\xi},\underline{\xi}' \in \mathcal{X}^{N_{1}(0)}\times \mathcal{X}^{N_{1}(0)}$. Our assumptions on $\mathcal{X}$ then imply
\begin{equation}
\label{kid}
\sum_{l \in N_{1}(0)} \rho(\eta_{k,l},\eta_{k-1,l}) = \frac{1}{M} \sum_{l \in N_{1}(0)} \rho(\eta_{l},\eta_{l}') < \frac{C_{*}}{M}.
\end{equation}
In the following, we choose $M$ in \eqref{etak}-\eqref{etakl} large enough so that $C_{*} < M$.\\
Now suppose that the following inequality holds for some $\lambda>0$
\begin{equation}
\label{Dobrcd}
\rho_{W}\left(\mu^{\beta}(\cdot\,\vert \underline{\eta}_{k}),\mu^{\beta}(\cdot\,\vert \underline{\eta}_{k-1})\right) \leq \lambda \sum_{l \in N_{1}(0)} \rho(\eta_{k,l},\eta_{k-1,l}),\quad k=0,\dots,M.
\end{equation}
Then, by virtue of the the first equality in \eqref{kid}, we have,
\begin{equation}
\label{kift}
\rho_{W}\left(\mu^{\beta}(\cdot\,\vert \underline{\eta}),\mu^{\beta}(\cdot\,\vert \underline{\eta}')\right) \leq  \sum_{k=1}^{M} \rho_{W}\left(\mu^{\beta}(\cdot\,\vert \underline{\eta}_{k}),\mu^{\beta}(\cdot\,\vert \underline{\eta}_{k-1})\right) \leq \lambda \sum_{l \in N_{1}(0)} \rho(\eta_{l},\eta_{l}').
\end{equation}
In the rest of the proof, we prove an inequality of type \eqref{Dobrcd}. To do that, we first derive a series of key-estimates for general $\underline{\eta}, \underline{\eta}' \in \mathcal{X}^{N_{1}(0)}$. For this part of the proof, it is in fact enough that $(\mathcal{X}, \rho)$ is a compact metric space for instance. Given $\xi_{0} \in \mathcal{X}$ and $\underline{\eta} \in \mathcal{X}^{N_{1}(0)}$, define for $\beta>0$,
\begin{equation*}
\mathpzc{f}^{\beta}(\xi_{0}\vert \underline{\eta}) := \exp\left(-\beta\left(\mathcal{V}(\xi_{0}) + \sum_{l \in N_{1}(0)} \mathcal{V}_{l}(\xi_{0},\eta_{l})\right)\right),
\end{equation*}
so that (see \eqref{condma})
\begin{equation*}
\mu^{\beta}(A \vert \underline{\eta}) := \frac{\int_{A} f^{\beta}(\xi_{0} \vert \underline{\eta}) \mu_{0}(d\xi_{0})}{\int_{\mathcal{X}} f^{\beta}(\xi_{0}\vert \underline{\eta}) \mu_{0}(d\xi_{0})},\quad A \in \mathcal{B}(\mathcal{X}).
\end{equation*}
Let $(\underline{\eta},\underline{\eta}') \in \mathcal{X}^{N_{1}(0)} \times \mathcal{X}^{N_{1}(0)}$ be fixed. We now show that for all $\beta>0$ satisfying
\begin{equation}
\label{cond1beta}
\theta_{\beta} := C_{0} \beta  \sum_{l \in N_{1}(0)} \rho(\eta_{l},\eta_{l}') < 1,
\end{equation}
and for all $A \in \mathcal{B}(\mathcal{X})$, the following holds
\begin{equation}
\label{intr11}
\left \vert \mu^{\beta}(A\vert \underline{\eta}') - \mu^{\beta}(A\vert \underline{\eta})\right \vert \leq 2 \theta_{\beta} (1+ \theta_{\beta}).
\end{equation}
By using the inequality
\begin{equation*}
\mathrm{e}^{\theta} - 1 \leq \theta(1 + \theta), \quad  0 \leq \theta < 1,
\end{equation*}
followed by assumption $(ii)$, we obtain, under the conditions \eqref{cond1beta}, that
\begin{equation*}
\left\vert  \mathpzc{f}^{\beta}(\xi_{0}\vert \underline{\eta}') - \mathpzc{f}^{\beta}(\xi_{0}\vert \underline{\eta}) \right \vert = \left \vert e^{\beta \sum_{l \in N_{1}(0)} (\mathcal{V}_{l}(\xi_{0},\eta_{l}) - \mathcal{V}_{l}(\xi_{0},\eta_{l}'))}-1 \right \vert \mathpzc{f}^{\beta}(\xi_{0}\vert \underline{\eta})
\leq \theta_{\beta} (1+\theta_{\beta})\mathpzc{f}^{\beta}(\xi_{0}\vert \underline{\eta}).
\end{equation*}
Denoting
\begin{equation*}
Z^{\beta}(\underline{\eta}) := \int_{\mathcal{X}} \mathpzc{f}^{\beta}(\xi_{0}\vert \underline{\eta})  \mu_{0}(d\xi_{0}),
\end{equation*}
write
\begin{multline*}
\mu^{\beta}(A\vert \underline{\eta}') - \mu^{\beta}(A\vert \underline{\eta}) = \\ \frac{1}{Z^{\beta}(\underline{\eta})}\int_{A} \left(\mathpzc{f}^{\beta}(\xi_{0}\vert \underline{\eta}') - \mathpzc{f}^{\beta}(\xi_{0}\vert \underline{\eta})\right) \mu_{0}(d \xi_{0}) - \frac{\mu^{\beta}(A\vert \underline{\eta}')}{Z^{\beta}(\underline{\eta})} \int_{\mathcal{X}} \left(\mathpzc{f}^{\beta}(\xi_{0}\vert \underline{\eta}') - \mathpzc{f}^{\beta}(\xi_{0} \vert \underline{\eta})\right)\mu_{0}(d \xi_{0}).
\end{multline*}
Under the conditions \eqref{cond1beta}, it then follows from the above that
\begin{equation}
\label{folim}
\left \vert \mu^{\beta}(A\vert \underline{\eta}') - \mu^{\beta}(A\vert \underline{\eta}) \right \vert \leq  (1+\theta_{\beta})\theta_{\beta} \left(\mu^{\beta}(A \vert \underline{\eta}) + \mu^{\beta}(A \vert \underline{\eta}')\right).
\end{equation}
This proves \eqref{intr11}. Subsequently assume that $\theta_{\beta}$ in \eqref{cond1beta} is small enough so that
\begin{equation}
\label{cond2beta}
\kappa_{\beta} := \theta_{\beta} (1+\theta_{\beta}) < 1.
\end{equation}
Under this condition, we prove the following upper bound for the Wasserstein distance
\begin{equation}
\label{rWdtc}
\rho_{W}\left(\mu^{\beta}(\cdot\,\vert \underline{\eta}), \mu^{\beta}(\cdot\,\vert \underline{\eta}')\right) \leq  6 C_{1}(\beta)\kappa_{\beta}.
\end{equation}
We follow a strategy similar to the one used in the proof of Proposition \ref{prop1}, see Section \ref{equivcvg} below. We cover $\mathcal{X}$ by a number $N$ of open balls $B_{i} \subset \mathcal{X}$, $i=1,\dots,N$ of radius $\frac{1}{2}\epsilon$ with $0<\epsilon<1$, and define
\begin{equation*}
A_{i} := B_{i} \setminus \bigcup_{j=1}^{i-1} B_{j},\quad i=1,\dots,N.
\end{equation*}
Note that the $A_{i}$'s form a partition of $\mathcal{X}$. We may assume that $\mu^{\beta}(A_{i}\vert \underline{\eta}')>0$ for all $i=1,\dots,N$, and define, for any $E \in \mathcal{B}(\mathcal{X})$, the measures
\begin{gather}
\label{nuetetp}
\nu_{\underline{\eta},\underline{\eta}'}^{\beta}(E) := \mu^{\beta}(E\vert \underline{\eta}) - \frac{1 - \kappa_{\beta}}{1+\kappa_{\beta}} \sum_{i=1}^{N} \mu^{\beta}(E \cap A_{i}\vert \underline{\eta}) = \frac{2 \kappa_\beta}{1 + \kappa_\beta} \mu^{\beta}(E\vert \underline{\eta}),\\
\label{tnuetetp}
\tilde{\nu}_{\underline{\eta},\underline{\eta}'}^{\beta}(E) := \mu^{\beta}(E\vert \underline{\eta}') - \frac{1 - \kappa_{\beta}}{1+\kappa_{\beta}} \sum_{i=1}^{N} \mu^{\beta}(A_{i}\vert \underline{\eta}) \frac{\mu^{\beta}(E \cap A_{i}\vert \underline{\eta}')}{\mu^{\beta}(A_{i}\vert \underline{\eta}')}.
\end{gather}
Note that \eqref{nuetetp} and \eqref{tnuetetp} are both positive measures since \eqref{folim} yields, for all $i=1,\dots,N$,
\begin{equation*}
\frac{1 + \kappa_{\beta}}{1 - \kappa_{\beta}} \mu^{\beta}(A_{i}\vert \underline{\eta}) \geq \mu^{\beta}(A_{i}\vert \underline{\eta}') \geq \frac{1 - \kappa_{\beta}}{1 + \kappa_{\beta}} \mu^{\beta}(A_{i}\vert \underline{\eta}).
\end{equation*}
Then define in $\mathcal{P}(\mathcal{X}\times\mathcal{X})$
\begin{equation*}
\sigma_{\underline{\eta},\underline{\eta}'}^{\beta}(E \times F) = \frac{1 - \kappa_{\beta}}{1 + \kappa_{\beta}} \sum_{i=1}^{N} \mu^{\beta}(E \cap A_{i}\vert \underline{\eta}) \frac{\mu^{\beta}(F \cap A_{i}\vert \underline{\eta}')}{\mu^{\beta}(A_{i}\vert \underline{\eta}')} + \frac{\nu_{\underline{\eta},\underline{\eta}'}^{\beta}(E) \tilde{\nu}_{\underline{\eta},\underline{\eta}'}^{\beta}(F)}{\nu_{\underline{\eta},\underline{\eta}'}^{\beta}(\mathcal{X})}.
\end{equation*}
Note that $\sigma_{\underline{\eta},\underline{\eta}'}^{\beta}(\mathcal{X} \times F)= \mu^{\beta}(F\vert \underline{\eta}')$ and $\sigma_{\underline{\eta},\underline{\eta}'}^{\beta}(E\times \mathcal{X})= \mu^{\beta}(E\vert \underline{\eta})$ so that $\sigma_{\underline{\eta},\underline{\eta}'}^{\beta} \in \Xi_{\mathcal{X}}(\mu^{\beta}(\cdot\, \vert \underline{\eta}),\mu^{\beta}(\cdot\,\vert \underline{\eta}'))$. This follows from the fact that $\nu_{\underline{\eta},\underline{\eta}'}^{\beta}(\mathcal{X}) = \tilde{\nu}_{\underline{\eta},\underline{\eta}'}^{\beta}(\mathcal{X}) = \frac{2\kappa_\beta}{1 + \kappa_\beta}$. Now,
\begin{equation*}
\begin{split}
\int_{\mathcal{X} \times \mathcal{X}} \rho(\xi,\xi') \sigma_{\underline{\eta},\underline{\eta}'}^{\beta}(d\xi,d\xi') =
&\frac{1 - \kappa_{\beta}}{1 + \kappa_{\beta}}  \sum_{i=1}^{N} \frac{1}{\mu^{\beta}(A_{i} \vert \underline{\eta}')} \int_{A_{i}} \int_{A_{i}} \rho(\xi,\xi') \mu^{\beta}(d\xi \vert \underline{\eta}) \mu^{\beta}(d\xi' \vert \underline{\eta}') \\
&+ \frac{1}{\nu_{\underline{\eta},\underline{\eta}'}^{\beta}(\mathcal{X})} \int_{\mathcal{X}} \int_{\mathcal{X}} \rho(\xi,\xi') \nu_{\underline{\eta},\underline{\eta}'}^{\beta} (d\xi) \tilde{\nu}_{\underline{\eta},\underline{\eta}'}^{\beta} (d\xi').
\end{split}
\end{equation*}
Using the fact that $\mathrm{diam}(A_{i}) \leq \epsilon$ in the first term and the triangle inequality in the second term, we have,
\begin{equation*}
\int_{\mathcal{X} \times \mathcal{X}} \rho(\xi,\xi') \sigma_{\underline{\eta},\underline{\eta}'}^{\beta}(d\xi,d\xi') \leq \epsilon + \int_{\mathcal{X}} \rho(\xi,\tilde{\xi}) \nu_{\underline{\eta},\underline{\eta}'}^{\beta} (d\xi) + \int_{\mathcal{X}} \rho(\tilde{\xi},\xi') \tilde{\nu}_{\underline{\eta},\underline{\eta}'}^{\beta} (d\xi'),
\end{equation*}
for some $\tilde{\xi} \in \mathcal{X}$. Since the $A_{i}$'s form a partition of $\mathcal{X}$, the last two terms can be bounded as follows
\begin{gather*}
\int_{\mathcal{X}} \rho(\xi,\tilde{\xi}) \nu_{\underline{\eta},\underline{\eta}'}^{\beta} (d\xi) \leq 2 \kappa_{\beta} \int_{\mathcal{X}} \rho(\xi,\tilde{\xi}) \mu^{\beta}(d\xi\vert \underline{\eta}) \leq 2 C_{1}(\beta) \kappa_{\beta}, \\
\int_{\mathcal{X}} \rho(\tilde{\xi},\xi') \tilde{\nu}_{\underline{\eta},\underline{\eta}'}^{\beta} (d\xi') \leq \left(1 - \left(\frac{1-\kappa_{\beta}}{1+ \kappa_{\beta}}\right)^{2} \right) \int_{\mathcal{X}} \rho(\xi, \tilde{\xi}) \mu^{\beta}(d\xi \vert \underline{\eta}') \leq 4 C_{1}(\beta) \kappa_{\beta}.
\end{gather*}
Gathering the above estimates, the upper bound \eqref{rWdtc} follows after taking the limit $\epsilon \rightarrow 0$.\\
To conclude the proof of the Proposition, substitute in the right-hand side of \eqref{rWdtc} $\kappa_{\beta}$  with its definition in \eqref{cond2beta} and the explicit expression of $\theta_{\beta}$ in \eqref{cond1beta}, and then replace $(\underline{\eta},\underline{\eta}')$ by $(\underline{\eta}_{k},\underline{\eta}_{k-1})$ (see \eqref{etak}-\eqref{etakl}) in both the left-hand side and right-hand side of \eqref{rWdtc}. This gives
\begin{equation*}
\rho_{W}\left(\mu^{\beta}(\cdot\,\vert \underline{\eta}_{k}), \mu^{\beta}(\cdot\,\vert \underline{\eta}_{k-1})\right) \leq  6 C_{1}(\beta) C_{0} \beta  \sum_{l \in N_{1}(0)} \rho(\eta_{k,l},\eta_{k-1,l}) \left( 1 + C_{0} \beta  \sum_{l \in N_{1}(0)} \rho(\eta_{k,l},\eta_{k-1,l})\right).
\end{equation*}
By virtue of the upper bound in \eqref{kid}, $M$ can be chosen large enough so that the condition in \eqref{cond2beta}, implying the one in \eqref{cond1beta}, is satisfied. Moreover, in view of \eqref{kift}, the following upper bound
\begin{equation*}
\rho_{W}\left(\mu^{\beta}(\cdot\,\vert \underline{\eta}), \mu^{\beta}(\cdot\,\vert \underline{\eta}')\right) \leq  6 C_{0} C_{1}(\beta) \beta \left(1 + C_{0}\beta \frac{C_{*}}{M}\right) \sum_{l \in N_{1}(0)} \rho(\eta_{l},\eta_{l}'),
\end{equation*}
holds for any $(\underline{\eta},\underline{\eta}') \in \mathcal{X}^{N_{1}(0)} \times \mathcal{X}^{N_{1}(0)}$. It remains to take the limit $M \rightarrow \infty$ and the upper bound in Proposition \ref{highT} follows from the condition \eqref{nniupbd} in Remark \ref{nearestn}.\qed

\begin{remark}
The uniqueness result for sufficiently high temperatures in Proposition \ref{highT} can be extended to the case of more general interactions, as well as unbounded spins (the conditions of Proposition \ref{highT} have to be modified accordingly). We refer to \cite{Lev,Roy} for an improved estimate for the case of  large $N$ vector models. 
\end{remark}

\subsection{The Potts and Ising models.}
\label{IsPo}

\subsubsection{The $q$-state Potts model in dimension $d \geq 2$.}
\label{secPott}

The $q$-state Potts model is a generalization of the Ising model in that the spin values can be any arbitrary integer in $\{1,\ldots,q\}$,  see, e.g., \cite{Potts}. We refer to, e.g.,  \cite{Bax,HKW,VEF2S}, for some exact results in dimension $d=2$.\\
The single-spin space is a finite set $\mathcal{X}=\{1,\dots,q \}$ with $q \in \mathbb{N}$, $q >1$ and the pair-interaction potential is
\begin{equation*}
\mathcal{V}_{k}(s_{j},s_{k}) = - J(j-k) \delta_{s_{j},s_{k}},\quad s_{j},s_{k} \in \{1,\dots,q \},
\end{equation*}
for some real coupling constants $J>0$. In the following, we consider only the nearest-neighbour case, where $J(j-k)=J$ if $\vert j-k\vert = 1$, $J(j-k)=0$ otherwise. The discrete metric is given by
\begin{equation*}
\rho(s,s') := 1 -\delta_{s,s'},\quad s,s' \in \{1,\dots,q\},
\end{equation*}
where $\delta_{s,s'}$ is the Kronecker delta, i.e., $\delta_{s,s'}=1$ when $s=s'$ and $\delta_{s,s'}=0$ otherwise.
Given two probability measures $\mu,\mu' \in \mathcal{P}(\mathcal{X})$, the minimizing measure $\sigma \in \mathcal{P}(\mathcal{X} \times \mathcal{X})$ for $\rho_{W}(\mu,\mu')$  satisfies
\begin{gather}
\sigma(s,s') = p_{s,s'}, \quad s,s' \in \{1,\dots,q\}; \nonumber\\
\label{consss}
\sum_{s'=1}^{q} p_{s,s'} = \mu(\{s\}),\quad \sum_{s = 1}^{q} p_{s,s'} = \mu'(\{s'\});
\end{gather}
and it minimizes
\begin{equation}
\label{tominimz}
\sum_{s,s'=1}^{q} \rho(s,s') \sigma(s,s') = \sum_{\substack{s, s'=1 \\s \neq s'}}^{q} p_{s,s'} = 1 - \sum_{s=1}^{q} p_{s,s},
\end{equation}
where, to derive the last identity in the above right-hand side, we used that
\begin{equation}
\label{simplif}
\sum_{s,s'=1}^{q} p_{s,s'} = \sum_{s=1}^{q} \mu(\{s\}) = \sum_{s'=1}^{q} \mu'(\{s'\}) = 1.
\end{equation}
We now claim that the following holds
\begin{lema}
\label{Pottlem}
For two probability measures $\mu,\mu' \in \mathcal{P}(\{1,\ldots,q\})$, the Wasserstein metric is given by the following expression
\begin{equation}
\label{idroWpot}
\rho_{W}\left(\mu,\mu'\right) = \frac{1}{2} \sum_{s=1}^{q} \left\vert \mu(\{s\}) - \mu'(\{s\})\right\vert.
\end{equation}
\end{lema}
For convenience's sake, the proof of Lemma \ref{Pottlem} is deferred to the next subsection.\\
Denoting
\begin{equation*}
E_{s_{j}}(\underline{s}) := - J \sum_{l \in N_{1}(j)} \delta_{s_{j},s_{l}},
\end{equation*}
we have, using \eqref{idroWpot},
\begin{equation}
\label{rhoWPotts}
\rho_{W}\left(\mu_{j}^{\beta}(\cdot\, \vert \underline{s}),\mu_{j}^{\beta}(\cdot\, \vert \underline{s}')\right) = \frac{1}{2} \sum_{s_{j}=1}^{q} \left \vert \frac{\mathrm{e}^{-\beta E_{s_{j}}(\underline{s})}}{\sum_{s_{j}=1}^{q} \mathrm{e}^{-\beta E_{s_{j}}(\underline{s})}} - \frac{\mathrm{e}^{-\beta E_{s_{j}}(\underline{s}')}}{\sum_{s_{j}=1}^{q} \mathrm{e}^{-\beta E_{s_{j}}(\underline{s}')}}\right\vert.
\end{equation}
Below, this expression is evaluated numerically in dimensions $d=2$ and $d=3$. When $d=2$, there are in general only 9 distinct cases to be considered. Namely, since the interaction is invariant under renaming of spins, the possible distinct configurations for $\underline{s}$ are
\begin{equation*}
	(1,1,1,1),\quad (1,1,1,2), \quad (1,1,2,2),\quad (1,1,2,3),\quad (1,2,3,4).
\end{equation*}
Changing one spin in the first configuration, it can only combine with the second, i.e. $\underline{s}' = (1,1,1,2)$. If $\underline{s} = (1,1,1,2)$ then, changing $2$ back to $1$, results in nothing new. Changing $2$ to $3$, however, we obtain $\underline{s} = (1,1,1,3)$, which, while equivalent thermodynamically, has a non-zero Wasserstein distance to $(1,1,1,2)$. Changing one of the spins $1$ to $2$ leads to $\underline{s}'=(1,1,2,2)$, and changing it to a different value leads to $\underline{s}'= (1,1,2,3)$. If $\underline{s} = (1,1,2,2)$ then the spins equal 1 and 2 are equivalent. We therefore only need to consider changing a 2-spin. Changing it back to a 1, leads to a previous case (with $\underline{s}$ and $\underline{s}'$ exchanged), whereas changing it to a 3 gives $\underline{s}' = (1,1,2,3)$. 
If $\underline{s} = (1,1,2,3)$ then the spins equal 2 and 3 are equivalent. Changing the 3 to a 1 or 2 leads to a previous combination. Changing 3 to a 4 leads to $\underline{s}' = (1,1,2,4)$, which as in the case $(1,1,1,2)$ with $(1,1,1,3)$ has a non-zero Wasserstein distance to $\underline{s}$. Alternatively, we can change a 1 to a 2, leading to $\underline{s}' = (1,2,2,3)$ or we can change a 1 to another spin value, leading to $\underline{s}' = (1,2,3,4)$. 
Finally, if $\underline{s} = (1,2,3,4)$, we can replace one of these spins by another value, say 5, leading to $\underline{s}' = (1,2,3,5)$. In all, we have the following possible pairings:
\begin{eqnarray*} \underline{s} &=& (1,1,1,1), \quad \underline{s}' = (1,1,1,2) \\
	\underline{s} &=& (1,1,1,2), \quad \underline{s}' = (1,1,1,3) \\
	\underline{s} &=& (1,1,1,2), \quad \underline{s}' = (1,1,2,2) \\
	\underline{s} &=& (1,1,1,2), \quad \underline{s}' = (1,1,2,3) \\
	\underline{s} &=& (1,1,2,2), \quad \underline{s}' = (1,1,2,3) \\
	\underline{s} &=& (1,1,2,3), \quad \underline{s}' = (1,1,2,4) \\
	\underline{s} &=& (1,1,2,3), \quad \underline{s}' = (1,2,2,3) \\
	\underline{s} &=& (1,1,2,3), \quad \underline{s}' = (1,2,3,4) \\
	\underline{s} &=& (1,2,3,4), \quad \underline{s}' = (1,2,3,5).
\end{eqnarray*}
Of course, if $q<5$ then some of these combinations are excluded. Figures \ref{fig:figure1} and  \ref{fig:figure2} below show the graphs of $\rho_W$ as a function of $\beta$ for these cases with $q=4$ and $q=30$ respectively.

\begin{figure}[H]
	\centering
	\includegraphics[width=12cm,keepaspectratio]{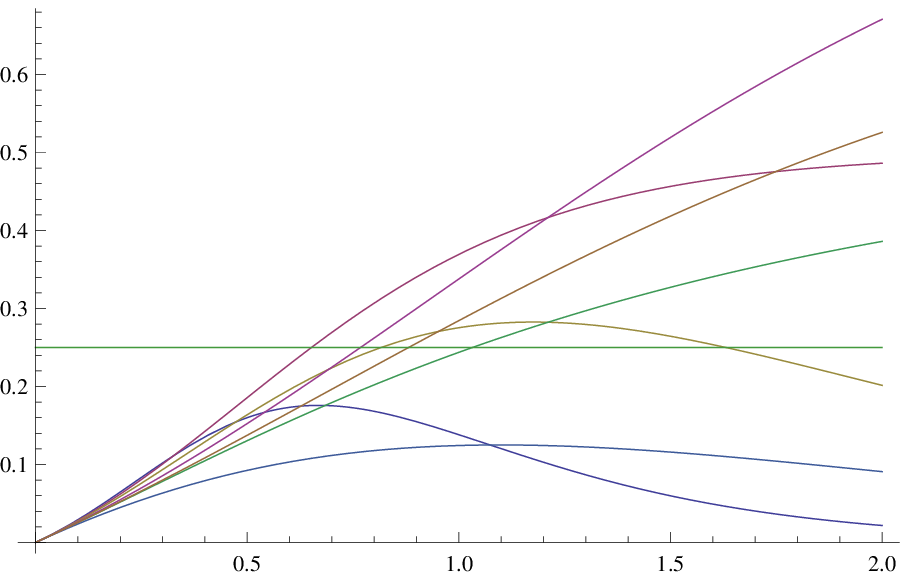}
	\caption{Graphs of $\rho_{W}$ versus $\beta$ for $q=4$.}
	\label{fig:figure1}
\end{figure}
\begin{figure}[H]
	\centering
	\includegraphics[width=12cm,keepaspectratio]{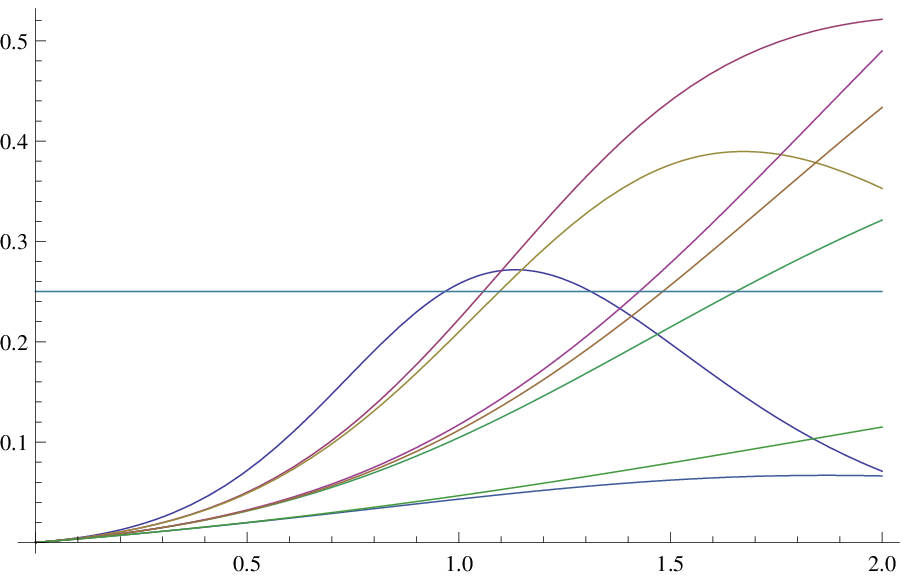}
	\caption{Graphs of $\rho_{W}$ versus $\beta$ for $q=30$.}
	\label{fig:figure2}
\end{figure}

\noindent In Figures \ref{fig:figure1} and \ref{fig:figure2}, the line $y=\frac{1}{4}$ has also been drawn, which is the upper bound on the right-hand side of the Dobrushin inequality \eqref{nniupbd}. For $q=4$, the curve which intersects the horizontal line in the left-most point corresponds to the case $\underline{s} = (1,1,1,2)$ and $\underline{s}' = (1,1,2,2)$. For $q=30$, this curve is overtaken by the case $\underline{s} = (1,1,1,1)$ and $\underline{s}' = (1,1,1,2)$. Note that the corresponding curve for $q=4$ is the blue curve with a clear maximum. The change-over occurs at $q=20$, as is clearly visible in Figure \ref{fig:figure3} below showing the Dobrushin lower bound for $\beta_{c}J$ ($\beta_{c}$ is the critical inverse temperature) as a function of $q$.
\begin{figure}[H]
	\centering
	\includegraphics[width=12cm,keepaspectratio]{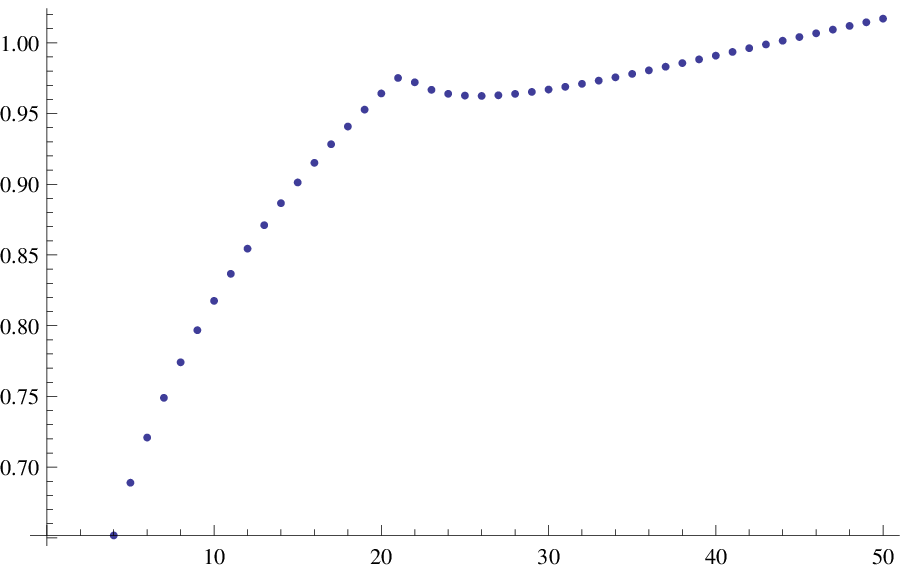}
	\captionsetup{justification=centering}
	\caption{The Dobrushin lower bound for $\beta_{c}J$ of the Potts model  in dimension $d=2$ as a function of $q$.}
	\label{fig:figure3}
\end{figure}

\begin{figure}[H]
	\centering
	\includegraphics[width=12cm,keepaspectratio]{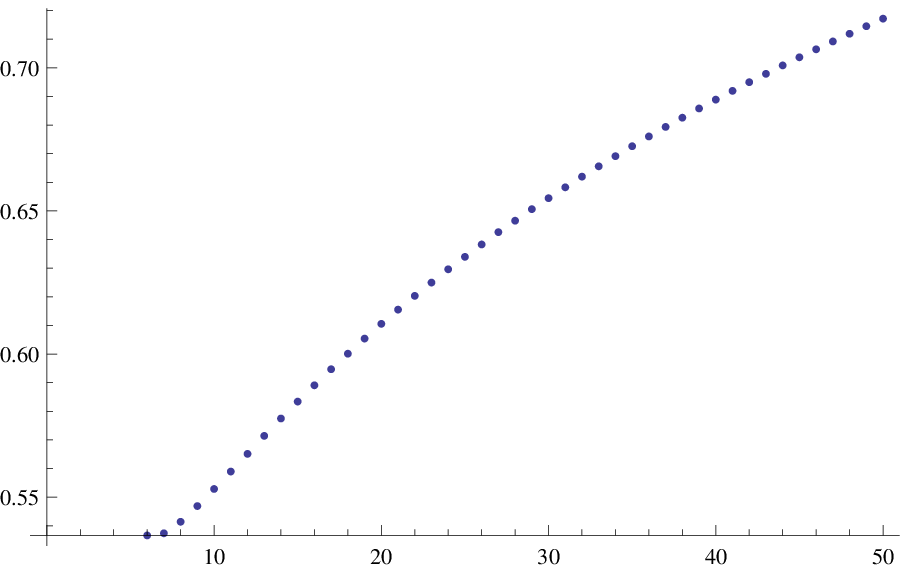}
	\captionsetup{justification=centering}
	\caption{The Dobrushin lower bound for $\beta_{c}J$ of the Potts model  in dimension $d=3$ as a function of $q$.}
	\label{fig:figure4}
\end{figure}

\noindent To obtain the asymptotics for large $q$, we only need to consider the first case, i.e. $\underline{s} = (1,1,1,1)$ and $\underline{s}' = (1,1,1,2)$. The explicit form of equation \eqref{rhoWPotts} is then
\begin{equation*} 
\begin{split}
\rho_{W}\left(\mu_{j}^{\beta}(\cdot\, \vert \underline{s}), \mu_{j}^{\beta}(\cdot\, \vert \underline{s}')\right) = &\frac{1}{2} \left\vert \frac{\mathrm{e}^{4\beta J}}{Z_{\beta}(1,1,1,1)} - \frac{\mathrm{e}^{3\beta J}}{Z_{\beta}(1,1,1,2)} \right\vert  \\
	&+ \frac{1}{2} \left\vert \frac{1}{Z_{\beta}(1,1,1,1)} - \frac{\mathrm{e}^{\beta J}}{Z_{\beta}(1,1,1,2)} \right\vert + \frac{q-2}{2} \left\vert \frac{1}{Z_{\beta}(1,1,1,1)} - \frac{1}{Z_{\beta}(1,1,1,2)} \right\vert,
\end{split}
\end{equation*} 
where, 
\begin{equation*} 
Z_{\beta}(1,1,1,1) := \mathrm{e}^{4\beta J} + q-1 \quad \textrm{and} \quad Z_{\beta}(1,1,1,2) := \mathrm{e}^{3\beta J} + \mathrm{e}^{\beta J} + q-2. 
\end{equation*}
It is easily seen that the first term between absolute values is positive, the other two negative, so that 
\begin{equation*} 
\rho_{W}\left(\mu_{j}^{\beta}(\cdot\, \vert \underline{s}), \mu_{j}^{\beta}(\cdot\, \vert \underline{s}')\right) = 
\frac{1}{2} \left\{ \frac{\mathrm{e}^{4\beta J}-q+1}{\mathrm{e}^{4\beta J} + q-1} - \frac{\mathrm{e}^{3\beta J} - \mathrm{e}^{\beta J} - q+2}{\mathrm{e}^{3\beta J} + \mathrm{e}^{\beta J} + q-2} \right\}. 
\end{equation*}
Similarly, for $q<20$, 
\begin{equation*} 
\rho_{W}\left(\mu_{j}^{\beta}(\cdot\, \vert \underline{s}), \mu_{j}^{\beta}(\cdot\, \vert \underline{s}')\right) = 
\frac{1}{2} \left\{ \frac{\mathrm{e}^{3\beta J} - \mathrm{e}^{\beta J} - q+2}{\mathrm{e}^{3\beta J} + \mathrm{e}^{\beta J} + q-2} + \frac{q-2}{2 \mathrm{e}^{2\beta J} + q-2} \right\}. 
\end{equation*}
Note the plus-sign in the second term, which is the reason these boundary conditions dominate for small $q$. 
For large $q$, we expect $\beta J$ to be large as well so that the term with 
$\mathrm{e}^{4\beta J}$ dominates and we have $q \sim \alpha \mathrm{e}^{4 \beta J}$ for some constant $\alpha$. Setting $\rho_{W}=\frac{1}{4}$, we then get asymptotically,
\begin{equation*}
\left( 1-\frac{2q}{\mathrm{e}^{4\beta J} + q} \right) - \left(1- \frac{2(q+\mathrm{e}^{\beta J})}{\mathrm{e}^{3\beta J} + \mathrm{e}^{\beta J} + q} \right) \approx \frac{1}{2},
\end{equation*} 
and inserting $q =\alpha \mathrm{e}^{4 \beta J}$, $\alpha = 3$. Therefore 
\begin{equation}
\label{Doblob} 
\beta_{c} J > \frac{1}{4} \ln \frac{q}{3}. 
\end{equation} 
In comparison, an explicit expression with exact value for $\beta_{c}J$ is known, see, e.g., \cite{Potts,Bax}: $\beta_{c}J = \ln(1+\sqrt{q})$. This is roughly twice the Dobrushin lower bound in \eqref{Doblob}.\\
\indent In dimension $d=3$, there are 23 combinations of boundary conditions to be considered. We simply show the Dobrushin lower bound in Figure \ref{fig:figure4} below analogous to Figure \ref{fig:figure3}.

The asymptotics is entirely analogously, 
\begin{equation}
\label{Doblob2} 
\beta_c J > \frac{1}{6} \ln \frac{q}{3}. 
\end{equation} 

\subsubsection{Proof of Lemma \ref{Pottlem}.}

In view of \eqref{tominimz}, we need to maximize the diagonal elements of the matrix $(p_{s,s'})_{1\leq s,s' \leq q}$ where the $s$-th row and $s'$-th column correspond to $\mu(\{s\})$ and $\mu'(\{s'\})$ respectively. If $\mu'(\{s\}) \geq \mu(\{s\})$ for some $s \in \{1,\dots,q\}$, $p_{s,s}$ is maximized when $p_{s,s} = \mu(\{s\})$. Conversely, if $\mu'(\{s\}) \leq \mu(\{s\})$ for some $s \in \{1,\ldots,q\}$, $p_{s,s}$ is maximized when $p_{s,s} = \mu'(\{s\})$. Assume for the moment that, for such a choice of $p_{s,s}$'s, we can always find some off-diagonal elements $p_{s,s'} \geq 0$, $s \neq s'$ so that the conditions in \eqref{consss} and \eqref{simplif} are fulfilled. Let $r:=\vert \{s \in \{1,\ldots, q\}: \mu'(\{s\}) \geq \mu(\{s\})\}\vert$ with $1 \leq r < q$ (as a result of \eqref{simplif}). Even if it means renaming the $\mu(\{s\})$'s, we can always assume that $\mu'(\{s\}) \geq \mu(\{s\})$, $s \in \{1,\ldots,r\}$. Then, by using \eqref{simplif}, we can write,
\begin{align*}
1 - \sum_{s=1}^{q} p_{s,s} &=  \frac{1}{2}\left(1 - \sum_{s=r+1}^{q} \mu'(\{s\}) -
\sum_{s=1}^{r} \mu(\{s\})\right) + \frac{1}{2} \left(1 - \sum_{s=1}^{r} \mu(\{s\}) - \sum_{s=r+1}^{q} \mu'(\{s\}) \right) \nonumber \\
&= \frac{1}{2} \sum_{s=1}^{r} \left(\mu'(\{s\}) - \mu(\{s\})\right) + \frac{1}{2} \sum_{s=r+1}^{q} \left(\mu(\{s\}) - \mu'(\{s\})\right) = \frac{1}{2} \sum_{s=1}^{r} \left\vert \mu'(\{s\}) - \mu(\{s\})\right\vert.
\end{align*}
This proves \eqref{idroWpot}. We now show that we can always construct a $q\times q$ lower-triangular matrix $(p_{s,s'})_{1 \leq s,s' \leq q}$ with diagonal elements $p_{s,s} = \mu(\{s\})$ for $s \in \{1,\ldots,r\}$, $p_{s,s} = \mu'(\{s\})$ for $s \in \{r+1,\ldots,q\}$ and off-diagonal elements $p_{s,s'} \geq 0$, $s \neq s'$ such that \eqref{consss} and \eqref{simplif} are fulfilled.\\
Set $p_{s} := \mu(\{s\})$ and $p_{s}':= \mu'(\{s\})$, $s \in \{1,\ldots,q\}$. Hereafter, we use $(i,j)$ in place of $(s,s')$ for the matrix indices. Let $r:=\vert \{i \in \{1,\ldots, q\}: p_{i}' \geq p_{i}\}\vert$ with $1 \leq r < q$. As before, even if it means renaming the $p_{i}$'s, we can always assume that $p_{i}' \geq p_{i}$, $i \in \{1,\ldots,r\}$. Recall the conditions
\begin{equation}
\label{condrefr}
\begin{split}
\sum_{j=1}^{r} p_{k,j} &= p_{k}-p_{k}',\quad k=r+1,\dots,q,\\
\sum_{k=r+1}^{q} p_{k,j} &= p_{j}'-p_{j},\quad j=1,\dots,r.
\end{split}
\end{equation}
We start by reordering the rows and columns so that $p_{j}'-p_j$ is in increasing order for $j=1,\dots,r$, and $p_{k}-p_{k}'$ is in decreasing order for $k=r+1,\dots,q$. We then proceed by induction on $q$, eliminating the $q$-th row and column.
We want to determine $p_{q,j}$ for $1 \leq j \leq r$ such that
\begin{equation*}
\sum_{j=1}^{r} p_{q,j} = p_{q}-p_{q}' \qquad \textrm{and} \qquad  p_{q,j} + \sum_{k=r+1}^{q-1} p_{k,j} = p_{j}' - p_{j}.
\end{equation*}
Now since $p_{1} + \dotsb + p_{q} = p_{1}'+ \dotsb + p_{q}'$,
\begin{equation*}
\sum_{j=1}^{r} (p_{j}'-p_{j}) = \sum_{k=r+1}^{q} (p_{k} - p_{k}') \geq p_{q} - p_{q}',
\end{equation*}
so that in principle this is possible. In fact, if $r=q-1$, then we can simply put $p_{q,j} = p'_j-p_j$ and no induction is needed. Otherwise, let $j_{0}$ be the maximal value such that $\sum_{j=j_{0}}^{r} (p_{j}'-p_{j}) \geq p_{q}-p_{q}'$. Put $p_{q,j} = 0$ for $1 \leq j < j_{0}$, $p_{q,j} = p_{j}'-p_{j}$ for $j_{0} < j \leq r$ and
$p_{q,j_{0}} = p_{q}-p_{q}' - \sum_{j=j_{0}+1}^{r} (p_{j}'-p_{j}) \geq 0$. To prove the claim by induction, we then need to show that we can continue this procedure for $k=q-1$. From \eqref{condrefr}, the reduced set of equations reads
\begin{align*}
\sum_{j=1}^{j_{0}} p_{k,j} &= p_{k}-p_{k}', \quad k=r+1,\dots,q-1; \\
p_{k,j} &= 0, \quad j=j_{0}+1,\dots,r;\,\,\, k=r+1,\dots,q-1; \\
\sum_{k=r+1}^{q-1} p_{k,j} &= p_{j}'-p_{j},\quad  j=1,\dots,j_{0}-1; \\
\sum_{k=r+1}^{q-1} p_{k,j_{0}} + p_{q}-p_{q}'&= p_{j_{0}}'-p_{j_{0}} + \sum_{j=j_{0}+1}^{r} (p_{j}'-p_{j}).
\end{align*}
We therefore need that
\begin{equation*}
p_{j_{0}}'-p_{j_{0}} +  \sum_{j=j_{0}+1}^{r} (p_{j}'-p_{j}) \geq p_{q}-p_{q}'.
\end{equation*}
This holds by definition of $j_{0}$. Therefore, the procedure can be continued.

\subsubsection{The Ising model in dimension $d \geq 2$.}
\label{Isingdim}

The single-spin space is $\mathcal{X}=\{-1,1\}$ and the pair-interaction potential is
\begin{equation}
\label{Vlj}
\mathcal{V}_{l-j}(s_{j},s_{l}) = - J(\vert j-l \vert) s_{j} s_{l},\,\,\,\, s_{j},s_{l} \in \{-1,1\},\,\,\,\, l \neq j,
\end{equation}
where $J$ is a sufficiently fast decreasing positive function satisfying $\sum_{\substack{k \in \mathbb{Z}^{d} \\ k\neq 0}} J(\vert k \vert) < \infty$.\\
For $\mathcal{X}=\{-1,1\}$, the discrete metric is given by
\begin{equation*}
\rho(s,s') := 1 -\delta_{s,s'}.
\end{equation*}
By virtue of \eqref{idroWpot} with $q=2$, given two probability measures $\mu,\mu' \in \mathcal{P}(\mathcal{X})$, we have,
\begin{equation}
\label{rhoIsing}
\rho_{W}(\mu,\mu') = \frac{1}{2} \sum_{s \in \{-,+\}} \vert \mu(\{s\}) - \mu'(\{s\}) \vert = \vert \mu(\{+\}) - \mu'(\{+\})\vert.
\end{equation}
Denoting
\begin{equation*}
S_{0}(\underline{s}) := \sum_{\substack{j \in \mathbb{Z}^{d} \\ j \neq 0}} J(\vert j\vert) s_{j},
\end{equation*}
we have, using \eqref{rhoIsing}, 
\begin{align}
\label{Isingid}
\rho_{W}\left(\mu_{0}^{\beta}(\cdot\,\vert \underline{s}), \mu_{0}^{\beta}(\cdot\,\vert \underline{s}')\right) &= \left\vert  \frac{\mathrm{e}^{\beta S_{0}(\underline{s})}}{\mathrm{e}^{\beta S_{0}(\underline{s})} + \mathrm{e}^{-\beta S_{0}(\underline{s})}} - \frac{\mathrm{e}^{\beta S_{0}(\underline{s}')}}{\mathrm{e}^{\beta S_{0}(\underline{s}')} + \mathrm{e}^{-\beta S_{0}(\underline{s}')}} \right\vert  \nonumber \\
&= \frac{1}{2} \left\vert \tanh(\beta S_{0}(\underline{s})) - \tanh(\beta S_{0}(\underline{s}')) \right\vert.
\end{align}
This implies  that
\begin{equation*}
\rho_{W}\left(\mu_{0}^{\beta}\left(\cdot\,\vert \underline{s}\right), \mu_{0}^{\beta}\left(\cdot\,\vert \underline{s}'\right)\right) \leq \frac{1}{2} \beta \left\vert S_{0}(\underline{s}) - S_{0}(\underline{s}')\right\vert \leq \frac{1}{2} \beta \sum_{\substack{j \in \mathbb{Z}^{d} \\ j \neq 0}} J(\vert j \vert)\vert s_{j} - s_{j}'\vert \leq \beta \sum_{\substack{j \in \mathbb{Z}^{d} \\ j \neq 0}} J(\vert j \vert) \rho(s_{j}, s_{j}').
\end{equation*}
By \eqref{consumf}, we conclude that there is no phase transition if $\beta (\sum_{\substack{j \in \mathbb{Z}^{d} \\ j\neq 0}} J(\vert j \vert)) < 1$.

For the usual $d$-dimensional Ising model with nearest-neighbour interactions, we can be more precise. In that case, 
\begin{equation*}
S_{0}(\underline{s}) = J \sum_{j \in N_{1}(0)} s_{j}, \quad J>0.
\end{equation*}
When $d=1$, it follows from \eqref{Isingid} that there is no phase transition for any $\beta >0$. Indeed, \eqref{Isingid} is at most equal to $\frac{1}{2}\tanh(2\beta J)< \frac{1}{2}$. When $d\geq 2$, changing a single boundary spin we have
\begin{equation*}
\rho_{W}\left(\mu_{0}^{\beta}\left(\cdot\,\vert \underline{s}\right), \mu_{0}^{\beta}\left(\cdot\,\vert \underline{s}'\right)\right) = \frac{1}{2}\max_{k\in \{0,\dots,d-1\}}\left\vert \tanh\left(\beta J (2d-2k)\right) - \tanh\left(\beta J (2d-2k -2)\right)\right\vert.
\end{equation*}
The maximum is attained at $k=d-1$ so that
\begin{equation*}
\rho_{W}\left(\mu_{0}^{\beta}\left(\cdot\,\vert \underline{s}\right), \mu_{0}^{\beta}\left(\cdot\,\vert \underline{s}'\right)\right) = \frac{1}{2} \tanh(2\beta J).
\end{equation*}
The Dobrushin condition is therefore
\begin{equation*}
\tanh(2\beta J) < \frac{1}{d}\quad \textrm{or} \quad \beta J < \frac{1}{4} \ln \left(\frac{d+1}{d-1}\right).	
\end{equation*}	
Numerically, for $d=2$ this yields $\beta_{c}J > 0.27465$  and for $d=3$, $\beta_{c}J > 0.17328$. In comparison, for $d=2$, an exact value for $\beta_{c}J$ is known, see, e.g., \cite{On,KW}: $\beta_{c}J = \ln(1+\sqrt{2})/2 \simeq 0.44069$. This is the same as the expression below \eqref{Doblob} for the 2-state Potts model with $2J$ as coupling constant. \\
\indent Adding an external uniform magnetic field $h$ to the Ising model in \eqref{Vlj},  the expression of the Wasserstein distance becomes
\begin{equation*}
\rho_{W}\left(\mu_{0}^{\beta,h}\left(\cdot\,\vert \underline{s}\right),
 \mu_{0}^{\beta,h}\left(\cdot\,\vert \underline{s}'\right)\right) = \frac{1}{2}\max_{\underline{s},\, \underline{s}'}\left\vert \tanh\left(\beta \left(S_{0}(\underline{s})+h\right)\right) - \tanh\left(\beta \left(S_{0}(\underline{s}')+h\right)\right)\right\vert.
\end{equation*}
Since $S_{0}(\underline{s}) \leq \sum_{\substack{j \in \mathbb{Z}^{d} \\ j\neq 0}} J(\vert j\vert)$, it follows that there is no phase transition if $h > \sum_{\substack{j \in \mathbb{Z}^{d} \\ j\neq 0}} J(\vert j\vert)$.

\subsection{Dobrushin-Shlosman theorem and linear programming.}

In order to apply the Dobrushin-Shlosman theorem (Theorem \ref{DSthm}) to some lattice spin models, we need a method, possibly computational, to estimate the Wasserstein distance for collections of spins, see, e.g., \cite{DS1}. This problem is more generally known as linear programming, see, e.g., \cite{Gas} and references therein.

\subsubsection{Linear Programming.}
\label{linprogr}

The general problem consists in minimizing a linear expression 
$f(\underline{x}) := \langle \underline{c}, \underline{x} \rangle_{\mathbb{R}^{n}}$ where $\underline{c} \in \mathbb{R}^{n}$ is a given constant vector, subject to the conditions $\underline{x} \geq \underline{0}$, i.e., $x_{j} \geq 0$ for all $j=1,\ldots,n$, and $A \underline{x} = \underline{b}$, where $\underline{b} \in \mathbb{R}^{m}$ is a given vector of lower dimension $m < n$ and $A$ is an $m \times n$ matrix. 

Applied to the case of computing the Wasserstein distance between two probability measures $\mu$ and $\nu$ on a set of discrete spins $\Omega_{0} = \{1,\ldots,q\}^{\Lambda_0}$, let $n=q^{\vert \Lambda_0\vert}$. As above for the Potts model, we can choose the diagonal elements of $\sigma \in \Xi(\mu,\nu)$ as
\begin{equation*}
\sigma\left(\underline{s},\underline{s}\right) = \min\left(\mu(\underline{s}), \nu(\underline{s})\right).
\end{equation*} 
This leaves $m(n-m)$ matrix elements of $\sigma$ to be determined, where $m := \vert \{ \underline{s}: \mu(\underline{s}) < \nu(\underline{s}) \} \vert$. These satisfy $m+(n-m) = n$ equations, one of which is linearly dependent on the others. These equations are
\begin{equation*} 
\sum_{\underline{s} \in \Omega_{0}^{>}} \sigma(\underline{s}, \underline{s}') = \nu(\underline{s}') - \mu(\underline{s}') \quad \textrm{and} \quad 
\sum_{\underline{s}' \in \Omega_{0}^{<}} \sigma(\underline{s}, \underline{s}') = \mu(\underline{s}) - \nu(\underline{s}), 
\end{equation*}
where $\Omega_{0}^{>} := \{\underline{s} \in \Omega_{0}:\, \mu(\underline{s}) > \nu(\underline{s})\}$ and
$\Omega_{0}^{<} := \{\underline{s} \in \Omega_{0}:\, \mu(\underline{s}) < \nu(\underline{s})\}$. Obviously, 
the variables $\sigma(\underline{s}, \underline{s}') \geq 0$ and we have to minimize the expression 
\begin{equation*}
f(\sigma) = \sum_{\underline{s} \in \Omega_{0}^{>}} \sum_{\underline{s}' \in \Omega_{0}^{<}} \sigma(\underline{s}, \underline{s}') \rho(\underline{s}, \underline{s}').
\end{equation*}

The linear programming algorithm is based on the following theorems

\begin{theorem} 
\label{linprog1}	
Consider the convex set (simplex) $K$ defined by 
\begin{equation*} 
K := \left\{\underline{x} \in \mathbb{R}^{n}\,:\, \underline{x} \geq \underline{0},\,\, A \underline{x} = \underline{b}\right\}, 
\end{equation*}
where $A:\mathbb{R}^{n} \to \mathbb{R}^{m}$, with $m < n$ is a linear map and $\underline{b} \in \mathbb{R}^{m}$. Given $\underline{c} \in \mathbb{R}^{n}$, the minimum of the linear function $f(\underline{x}) := \langle \underline{c}, \underline{x} \rangle_{\mathbb{R}^{n}}$, provided it exists, is attained at an extremal point of the set $K$. 
\end{theorem}

\begin{theorem} 
\label{linprog2}
Under the conditions of Theorem \ref{linprog1}, assume that the matrix $A$ has rank $m$. Then the extremal points of $K$ are given by 
the intersections of $m$ hyperplanes given by independent 
column vectors $P_{r(1)}, \ldots, P_{r(m)}$ of the matrix $A$, i.e., $(P_{j})_{i} = A_{i,j}$ such that 
\begin{equation*} 
\sum_{i=1}^{m} x_{r(i)} P_{r(i)} = \underline{b},
\end{equation*}
where $x_{j} = 0$ for $j \notin \{r(1),\ldots,r(m)\}$. 
\end{theorem}

\noindent \textbf{Proof of Theorem \ref{linprog2}}.
Let $\underline{x}$ be an extremal point of $K$ and assume that the non-zero coefficients are $x_{i}$, $i \in I \subset \{1,\ldots,n\}$. Suppose that the corresponding column vectors $P_{i}$, $i \in I$ are dependent. Then there exists a linear combination 
\begin{equation*}
\sum_{i \in I} \lambda_{i} P_{i} = 0,
\end{equation*}
where at least one $\lambda_{i} \neq 0$. By assumption,
\begin{equation*}
\sum_{i \in I} x_{i} P_{i} = \underline{b}.
\end{equation*}
Multiplying the first identity by $\delta > 0$ and adding and subtracting from the second, we have,
\begin{equation*}
\sum_{i \in I} \left(x_{i} \pm \delta \lambda_{i}\right) P_{i} = \underline{b}.
\end{equation*}
Taking $\delta > 0$ small enough, we then have two solutions $\underline{x}_\pm$ of the equations $A \underline{x}_\pm = \underline{b}$ which are non-negative, and moreover, $\underline{x} = \frac{1}{2}(\underline{x}_{+} + \underline{x}_{-})$ which implies that $\underline{x}$ is not an extremal point of $K$. \\
Conversely, suppose that there are column vectors $P_{r(1)}, \ldots, P_{r(m)}$ such that 
\begin{equation*}
\sum_{i=1}^{m} x_{r(i)} P_{r(i)} = \underline{b},
\end{equation*}
where all $x_{r(i)} \geq 0$ and $x_{j} = 0$ for $j \neq r(i)$. Assume that there exist other points $\underline{y},\underline{z} \in K$ such that $\underline{x} = \alpha \underline{y} + (1-\alpha) \underline{z}$ for 
$0<\alpha<1$. Since the elements of $\underline{x}$, $\underline{y}$ and $\underline{z}$ are non-negative, it follows that $y_{j} = 0 = z_{j}$ for $j \neq r(i)$. Therefore,
\begin{equation*}
\sum_{i=1}^{m} y_{r(i)} P_{r(i)} = \sum_{i=1}^{m} z_{r(i)} P_{r(i)} = \underline{b}.
\end{equation*}
By the independence of the vectors $P_{r(i)}$, it then follows that $\underline{y} = \underline{z} = \underline{x}$. \qed

\begin{theorem} 
\label{linear3}	
Given an extremal solution $\underline{x}$ with corresponding column vectors 
$P_{r(1)}, \ldots, P_{r(m)}$, let 
\begin{equation} 
\label{Pj} 
P_{j} = \sum_{i=1}^{m} \alpha_{i,j} P_{r(i)}, \quad j=1,\ldots,n; 
\end{equation} 
and let
\begin{equation} 
\label{fx} 
z = \sum_{i=1}^{m} c_{r(i)} x_{r(i)}, 
\end{equation}
be the corresponding value of $f(\underline{x})$. Define 
\begin{equation} 
\label{zj}  
z_{j} := \sum_{i=1}^{m} \alpha_{i,j} c_{r(i)}. 
\end{equation}
If, for any fixed $j \in \{1,\ldots,n\}$, $z_{j} - c_{j} > 0$ then there exists another extremal solution $\underline{x}'$ such that $f(\underline{x}') < f(\underline{x})$. On the other hand, if $z_{j} \leq c_{j}$ for all $j=1,\ldots,n$, then $z = f(\underline{x})$ is minimal. 
\end{theorem}

\noindent \textbf{Proof of Theorem \ref{linear3}.} First, suppose that there exists $j$ such that $z_{j} > c_{j}$. Then we multiply equation \eqref{Pj} by $\theta > 0$ and subtract from the basic equation $\sum_{i=1}^{m} x_{r(i)} P_{r(i)} = \underline{b}$,
to obtain
\begin{equation*} 
\sum_{i=1}^{m} \left(x_{r(i)} - \theta \alpha_{i,j}\right) P_{r(i)} + \theta P_{j} = \underline{b}.  
\end{equation*}
Similarly, from equations \eqref{fx} and  \eqref{zj}, we have
\begin{equation*} 
\sum_{i=1}^{m} \left(x_{r(i)} - \theta \alpha_{i,j}\right) c_{r(i)} + \theta c_{j} = z - \theta \left(z_{j} - c_{j}\right).
\end{equation*}
Clearly, if all $\alpha_{i,j} \leq 0$, we can take $\theta$ arbitrarily large and the resulting minimum tends to $-\infty$. Otherwise, let 
\begin{equation} 
\label{theta} 
\theta := \min_{i \in \{1,\ldots,m\}\,:\,\alpha_{i,j} > 0} \frac{x_{r(i)}}{\alpha_{i,j}}. 
\end{equation}
This eliminates $P_{r(l)}$ where $l$ is the minimizer in \eqref{theta}, and replaces it by $P_{j}$. Since $z_{j}>c_{j}$, we have,
\begin{equation*}
z' = z - \theta(z_{j}-c_{j}) < z.
\end{equation*} 
Next, suppose that $z_{j} \leq c_{j}$ for all $j=1,\ldots,n$, and let $\underline{y}$ be another solution with $z^{*} = \langle c,y\rangle_{\mathbb{R}^{n}}$. Since $z_{j} \leq c_{j}$, we have $\langle y,z\rangle_{\mathbb{R}^{n}} \leq z^{*}$. Writing $P_{j}$ in terms of $P_{r(i)}$ as in \eqref{Pj}, and inserting into 
\begin{equation*}
\sum_{j=1}^{n} y_{j} P_{j} = \underline{b},
\end{equation*}
we can write
\begin{equation*}
\sum_{i=1}^{m} \left( \sum_{j=1}^{n} y_{j} \alpha_{i,j} \right) P_{r(i)} = \underline{b}.
\end{equation*}
Similarly, using equation \eqref{zj},
\begin{equation*}
\sum_{i=1}^{m} \left(\sum_{j=1}^{n} y_{j} \alpha_{i,j} \right) c_{r(i)} \leq z^{*}.
\end{equation*}
But, since the vectors $P_{r(1)}, \ldots, P_{r(m)}$ are independent, 
\begin{equation*}
\sum_{j=1}^{n} y_{j} \alpha_{i,j} = x_{r(i)},
\end{equation*}
and hence,
\begin{equation*}
z = \sum_{i=1}^{m} x_{r(i)} c_{r(i)} \leq z^{*},
\end{equation*}
that is, $z$ is minimal. \qed

\subsubsection{Application to the square lattice Ising model.}
\label{DSapp}

In this section, we consider the square lattice Ising model in dimension $d=2,3$ with nearest-neighbour interactions, uniform coupling constant $J>0$ and no external magnetic field.\\
\indent In Section \ref{Isingdim}, we applied the Dobrushin theorem (Theorem \ref{Dobrushin}) to the $d$-dimensional Ising model and using the 1-point Dobrushin condition in \eqref{nniupbd} allowed us to derive  the following lower bounds for the critical inverse temperature:  $\beta_{c}J > 0.27465$ for $d=2$ and $\beta_{c}J > 0.17328$ for $d=3$. For comparison, we recall that an explicit expression with exact value is known for the case $d=2$:  $\beta_{c}J = \ln(1+\sqrt{2})/2 \simeq 0.44069$, see, e.g., \cite{On,KW}. For $d=3$, there is no such explicit expression but a numerical estimation is known: $\beta_{c}J \simeq 0.22165$, see, e.g., \cite{Liv,TB}.\\
\indent Improved lower bounds for $\beta_{c}J$ may be obtained by applying  Dobrushin-Shlosman' conditions  \eqref{DScond0}-\eqref{DScond} in Theorem \ref{DSthm}. For illustration purposes, we consider the following cases: a square of $2\times2$ spins and a square of $3\times3$ spins for $d=2$, and a cube of $2\times 2 \times 2$ spins for $d=3$. The linear programming technique was applied to compute the Wasserstein distance numerically for all  relevant  configurations of boundary conditions. Our results are summarised in Table \ref{tab:truthTables}  below.

\begin{table}[H]
	\centering
	\begin{tabular}{|M{3.5cm}|M{2.8cm}|M{3.5cm}|M{1.8cm}| }
		\hline
		 Model & Dobrushin's lower bound & Dobrushin-Shlosman's lower bound  & Theoretical value \\
		\hline
		square of $2 \times 2$ spins  & 0.27465 & 0.30817  &0.44069 \\
		\hline
		square of $3 \times 3$ spins & 0.27465 & 0.33021 & 0.44069\\
		\hline
		cube of $2 \times 2 \times 2$ spins & 0.17328 & 0.18727 &  0.22165 \\
		\hline
	\end{tabular}
	\caption{Lower bounds for $\beta_{c}J$ for some Ising square lattice models.}
	\label{tab:truthTables}   
\end{table}

\indent \textit{Square of $2 \times 2$ spins.} This configuration is depicted in the following figure.

\begin{center}
	\setlength{\unitlength}{1mm}
	\begin{picture}(50,50) \thinlines
		\multiput(10,10)(10,0){4}{\line(0,1){30}} \multiput(10,10)(0,10){4}{\line(1,0){30}} \multiput(20,20)(0,10){2}{\circle*{2}} \multiput(30,20)(0,10){2}{\circle*{2}} 
		\multiput(10,20)(0,10){2}{\circle{2}} \multiput(40,20)(0,10){2}{\circle{2}}
		\multiput(20,10)(10,0){2}{\circle{2}} \multiput(20,40)(10,0){2}{\circle{2}}
		\put(42,31){4a} \put(31,42){4b} \put(15,42){1a} \put(5,32){1b} 
		\put(5,18){2a} \put(15,5){2b} \put(31,5){3a} \put(42,18){3b}
	\end{picture}
\end{center}
The spins of $\Lambda_{0}$ are indicated with black dots, the boundary spins with open circles. The boundary spins divide into groups of two, interacting only with a single spin of $\Lambda_0$. Each pair of corner boundary spins $\pi_{i}$, $i\in I_{4}:=\{1,2,3,4\}$ takes values in $\{-2,0,2\}$. We expect that the maximum unweighted Wasserstein distance is obtained for the following pairs of boundary conditions: \\
$\mathrm{(i)}$ $\pi_{i_{1}} \neq 0$ and $\pi_{i}=0$ for $i \in I_{4} \setminus \{i_{1}\}$ (bc1), $\pi_{i}= 0$ for $ i \in I_{4}$ (bc2); \\
$\mathrm{(ii)}$ $\pi_{i_{1}} \neq 0$ and $\pi_{i}=0$ for $i \in I_{4} \setminus \{i_{1}\}$ (bc1), $\pi_{i_{2}} = - \pi_{i_{1}}$ and $\pi_{i}=0$ for $i \in I_{4} \setminus \{i_{1},i_{2}\}$ (bc2).\\
Each pair of boundary conditions (bc1 and bc2) differs by one single pair of corner boundary spins. For the calculation of the resulting Wasserstein distance, a weighting accounting for the number of spin changes  is needed.  
In the square of $2\times 2$ spins configuration, the maximum Wasserstein distance equals $\frac{1}{2}$ (4 spins in $\Lambda_{0}$, 8 boundary spins), see \eqref{DScond0}-\eqref{DScond}. By computing the Wasserstein distance using the linear programming algorithm and then optimizing for $\beta J$, we obtain  the lower bound $\beta_{c}J > 0.30817$ for the critical inverse temperature. While the Wasserstein distance was computed for all possible boundary conditions, only a selection is given in Table \ref{tab:truthTables1}  below.

\begin{table}[H]
		\centering
	\begin{tabular}{|M{0.2cm}|M{2.7cm}|M{2.7cm}|M{2cm}|}
		\hline
		& Pairs of corner boundary spins 1 & Pairs of corner boundary spins 2 & Unweighted Wasserstein distance  \\
		\hline
		\multirow{4}{*}{1} & (0,0,0,-2)   & (0,0,0,0)  &\multirow{4}{*}{0.4999988}\\
		 & (0,0,0,-2)   & (0,0,2,-2)  & \\
		 & (0,0,0,-2)   & (0,2,0,-2)  & \\
		 & (0,0,0,-2)   & (2,0,0,-2)   & \\
		\hline
		\multirow{3}{*}{2} & (0,-2,2,-2)   & (2,-2,2,-2) &\multirow{3}{*}{0.4859922}  \\
		& (0,-2,2,-2)   & (0,0,2,-2)  &\\
		& 	(2,0,2,-2)   & (0,0,2,-2)  & \\
		\hline
	\end{tabular}
		\caption{Computation of unweighted Wasserstein distance for some boundary configurations with $\beta J = 0.308176$.}
	\label{tab:truthTables1}   
	\end{table}

\noindent Noting that there are 8 possible changes, each corresponding to a sign flip of a different corner boundary spin, and each change results in the same unweighted Wasserstein distance, the resulting Wasserstein distance therefore equals the unweighted Wasserstein distance. \\

\textit{Square of $3 \times 3$ spins.} This configuration is depicted in the following figure.

\begin{center}
	\setlength{\unitlength}{1mm}
	\begin{picture}(60,60) \thinlines
		\multiput(10,10)(10,0){5}{\line(0,1){40}} \multiput(10,10)(0,10){5}{\line(1,0){40}} \multiput(20,20)(0,10){3}{\circle*{2}} \multiput(30,20)(0,10){3}{\circle*{2}} 
		\multiput(40,20)(0,10){3}{\circle*{2}} 
		\multiput(10,20)(0,10){3}{\circle{2}} \multiput(50,20)(0,10){3}{\circle{2}}
		\multiput(20,10)(10,0){3}{\circle{2}} \multiput(20,50)(10,0){3}{\circle{2}}
		\put(52,41){8a} \put(40,52){8b} \put(29,52){1} \put(17,52){2a} 
		\put(4,41){2b} \put(5,29){3} \put(4,18){4a} \put(17,5){4b} 
		\put(29,5){5} \put(40,5){6a} \put(52,18){6b} \put(52,29){7}
	\end{picture}
\end{center}
The spins of $\Lambda_{0}$ are indicated with black dots, the boundary spins with open circles. The boundary spins now come in two groups. The middle boundary spins are numbered as 1, 3, 5, 7, each one of those interacts  with only one spin of $\Lambda_0$. The corner boundary spins come in pairs, interacting with a corner spin of $\Lambda_{0}$. Each pair of corner boundary spins $\pi_{j}$, $j\in J_{4}:=\{2,4,6,8\}$ takes values in $\{-2,0,2\}$ while each middle boundary spin $\sigma_{j'}$, $j' \in \{1,3,5,7\}$  takes values in $\{-1,1\}$.  We expect that the maximum unweighted Wasserstein distance is obtained for the following pair of boundary conditions: \\
$\pi_{j} = 0$ for $j \in J_{4}$ and $\sum_{j'\in J_{4}} \sigma_{j'-1} = 0$ (bc1), $\pi_{j}= 0$ for $ j \in J_{4}$ and  $\vert \sum_{j'\in J_{4}} \sigma_{j'-1} \vert = 1$ (bc2).\\
The pair of boundary conditions only differs by one middle boundary spin. For the calculation of the resulting Wasserstein distance, a weighting accounting for the number of spin changes (corner spins vs middle spins) is needed.  
In the square of $3\times 3$ spins, the maximum Wasserstein distance equals $\frac{3}{4}$ (9 spins in $\Lambda_{0}$, 12 boundary spins), see \eqref{DScond0}-\eqref{DScond}. By computing the Wasserstein distance using the linear programming algorithm and then optimizing for $\beta J$, we get the lower bound $\beta_{c}J > 0.33021$ for the critical inverse temperature. Due to the number of possible combinations, the computation of the unweighted Wasserstein distance was limited to a selection of relevant boundary conditions. The main cases are presented in Table \ref{tab:truthTables2} below.

\begin{table}[H]
	\centering
	\begin{tabular}{|M{0.2cm}|M{0.3cm}|M{2.3cm}|M{1.8cm}|M{2.3cm}|M{1.8cm}|M{2.0cm}| }
		\hline
		\multicolumn{2}{|c|}{} & Pairs of corner boundary spins 1 & Middle boundary spins 1 &
		Pairs of corner boundary spins 2 & Middle boundary spins 2 & Unweighted Wasserstein distance \\
		\hline
		\multirow{4}{*}{1} & \multirow{2}{*}{1a} & (0,0,0,0)   & (1,-1,1,-1)  & (0,0,0,0)   & (1,1,1,-1) &\multirow{2}{*}{0.8904972} \\
		&  & (0,0,0,0)  & (1,-1,1,-1)  & (0,0,0,0)   & (1,-1,-1,-1) & \\ \cline{2-7} 
		
		& \multirow{2}{*}{1b} & (0,0,0,0)   & (1,-1,1,-1)  & (0,0,0,2)   & (1,-1,1,-1)
		&\multirow{2}{*}{0.6797176} \\
		& & (0,0,0,0)   & (1,-1,1,-1)  & (0,-2,0,0)   & (1,-1,1,-1) & \\
		\hline
		\multirow{10}{*}{2} & \multirow{2}{*}{2a} & (0,0,0,0)   & (1,1,-1,-1)  & (0,0,0,0)   & (1,1,1,-1) &\multirow{2}{*}{0.8904972} \\ 
		& & (0,0,0,0)   & (1,1,-1,-1)  & (0,0,0,0)   & (-1,1,-1,-1) & \\  \cline{2-7}
		& \multirow{2}{*}{2b} & (0,0,0,0)   & (1,1,-1,-1)  & (-2,0,0,0)   & (1,1,-1,-1)  &\multirow{2}{*}{0.7284996} \\ 
		& & (0,0,0,0)   & (1,1,-1,-1)  & (0,0,2,0)   & (1,1,-1,-1) & \\  \cline{2-7}
		& \multirow{2}{*}{2c}  & (0,0,0,0)   & (1,1,-1,-1)  & (0,2,0,0)   & (1,1,-1,-1) 	&\multirow{2}{*}{0.6693575} \\  
		& & (0,0,0,0)   & (1,1,-1,-1)  & (0,0,0,2)   & (1,1,-1,-1) & \\ \cline{2-7}
		& \multirow{2}{*}{2d} & (0,0,0,0)   & (1,1,-1,-1)  & (0,-2,0,0)   & (1,1,-1,-1) &\multirow{2}{*}{0.6693575} \\  
		& & (0,0,0,0)   & (1,1,-1,-1)  & (0,0,0,-2)   & (1,1,-1,-1)& \\ \cline{2-7}
		& \multirow{2}{*}{2e} & (0,0,0,0)   & (1,1,-1,-1)  & (2,0,0,0)   & (1,1,-1,-1)
		&\multirow{2}{*}{0.6060878} \\ 
		& & (0,0,0,0)   & (1,1,-1,-1)  & (0,0,-2,0)   & (1,1,-1,-1) & \\
		\hline
	\end{tabular}
	\caption{Computation of unweighted Wasserstein distance for some boundary configurations with $\beta J = 0.33021$.}
	\label{tab:truthTables2}   
\end{table}

\noindent Noting that there are 12 possible changes, each corresponding to a sign flip of a different middle boundary spin and corner boundary spin, and each change may result in a different unweighted Wasserstein distance, the resulting Wasserstein distance is computed in Table \ref{tab:truthTables3} below for the 2 sets of boundary conditions from Table \ref{tab:truthTables2}.\\

\begin{table}[H]
	\centering
	\begin{tabular}{|M{0.2cm}|M{0.3cm}|M{1cm}|M{2cm}| }
		\hline
		\multicolumn{2}{|c|}{} & Weight &  Resulting Wasserstein distance  \\
		\hline
		\multirow{2}{*}{1} & 1a & 1/3 & \multirow{2}{*}{0.74997747} \\  \cline{2-3}
		& 1b & 2/3 & \\ \hline
		\multirow{5}{*}{2} & 2a & 1/3 &	 \multirow{5}{*}{0.74238279}\\ \cline{2-3} 
		& 2b & 1/6 & \\ \cline{2-3}
		& 2c & 1/6 & \\ \cline{2-3}
		& 2d & 1/6 & \\ \cline{2-3}
		& 2e & 1/6 & \\ \cline{2-3}
		\hline
	\end{tabular}
	\caption{Computation of the resulting Wasserstein distance for the 2 sets of boundary conditions from Table \ref{tab:truthTables2} with $\beta J = 0.33021$.}
	\label{tab:truthTables3}   
\end{table}

\textit{Cube of $2 \times 2 \times 2$ spins.} This configuration is depicted in the following figure

\begin{center}
	\setlength{\unitlength}{1mm}
	\begin{picture}(60,60) \thinlines
		\multiput(25,10)(15,0){2}{\line(0,1){45}} \multiput(10,25)(0,15){2}{\line(1,0){45}}
		\multiput(35,15)(15,0){2}{\line(0,1){45}} \multiput(20,30)(0,15){2}{\line(1,0){45}}
		\multiput(15,20)(0,15){2}{\line(2,1){30}} \multiput(30,20)(0,15){2}{\line(2,1){30}}
		\multiput(25,25)(0,15){2}{\circle*{2}} \multiput(35,30)(0,15){2}{\circle*{2}} 
		\multiput(40,25)(0,15){2}{\circle*{2}} \multiput(50,30)(0,15){2}{\circle*{2}} 
		\multiput(10,25)(0,15){2}{\circle{2}} \multiput(15,20)(0,15){2}{\circle{2}}
		\multiput(20,30)(0,15){2}{\circle{2}} \multiput(25,10)(0,45){2}{\circle{2}}
		\multiput(30,20)(0,15){2}{\circle{2}} \multiput(35,15)(0,45){2}{\circle{2}}
		\multiput(40,10)(0,45){2}{\circle{2}} \multiput(45,35)(0,15){2}{\circle{2}}
		\multiput(50,15)(0,45){2}{\circle{2}} \multiput(55,25)(0,15){2}{\circle{2}}
		\multiput(60,35)(0,15){2}{\circle{2}} \multiput(65,30)(0,15){2}{\circle{2}}
		\put(26,20){1} \put(41,20){2} \put(51,26){3} \put(36,26){4} 
		\put(26,36){5} \put(41,36){6} \put(51,46){7} \put(32,46){8} 
	\end{picture}
\end{center}
The spins of $\Lambda_{0}$ are indicated with black dots, the boundary spins with open circles. The boundary spins divide into groups of three, interacting only with a single spin of $\Lambda_0$. Each triplet of corner boundary spins $\pi_{i}$, $i\in I_{8}:=\{1,\ldots,8\}$ takes values in $\{-3,-1,1,3\}$. We expect that the maximum unweighted Wasserstein distance is obtained for the following pair of boundary conditions: \\
$(\mathrm{i})$ $\pi_{4+i}= \pi_{5-i}$ for $i \in \{1,\ldots,4\}\setminus\{i_{0}\}$ and $\pi_{i+1} \pi_{i}=-1$ for $i \in \{1,2,3\}$ or $i \in \{5,6,7\}$ (bc1), $\pi_{4+i}= \pi_{5-i}$ for $i \in \{1,\ldots,4\}$ and $\pi_{i+1} \pi_{i}=-1$ for $i \in \{1,2,3\}$  (bc2);\\ $(\mathrm{ii})$ $\pi_{4+i}= \pi_{5-i}$ for $i \in \{1,\ldots,4\}\setminus\{i_{0}\}$ and $\pi_{i+1} \pi_{i}=-1$ for $i \in \{1,2,3\}$ or $i \in \{5,6,7\}$ (bc1),  $\sum_{i=1}^{8} \pi_{i} = 0$ and $\pi_{i+1}\pi_{i}=-1$ for $i \in \{1,2,3\}$ or $i \in \{5,6,7\}$, or $\sum_{i=1}^{8} \pi_{i} = 0$ and $\pi_{i_{1}} \pi_{i_{1}+1}=1$ for $i_{1} \in \{1,2,3\}$ and  $\pi_{i_{2}} \pi_{i_{2}+1}=1$ for $i_{2} \in \{5,6,7\}$ (bc2).\\
Each pair of boundary conditions (bc1 and bc2) only differs by one single triplet of corner boundary spins. For the calculation of the resulting Wasserstein distance, a weighting accounting for the number of spin changes is needed.  In the cube of $2\times 2\times 2$ spins configuration, the maximum Wasserstein distance equals $\frac{1}{3}$ (8 spins in $\Lambda_{0}$, 24 boundary spins), see \eqref{DScond0}-\eqref{DScond}. By computing the Wasserstein distance using the linear programming algorithm and then optimizing for $\beta J$, we obtain  the lower bound $\beta_{c}J > 0.18727$ for the critical inverse temperature. Due to the number of possible combinations, the computation of the Wasserstein distance was limited to a selection of relevant boundary conditions. The main cases are presented in Table \ref{tab:truthTables4}  below.

\begin{table}[H]
	\centering
	\begin{tabular}{ |M{0.2cm}|M{0.3cm}|M{3.1cm}|M{3.1cm}|M{2cm}|  }
		\hline
		\multicolumn{2}{|c|}{} & Triplets of corner boundary spins 1 & Triplets of corner boundary spins 2 & Unweighted Wasserstein distance \\
		\hline
		\multirow{9}{*}{1} & \multirow{7}{*}{1a} & (1,1,-1,1,1,-1,1,-1) &	(-1,1,-1,1,1,-1,1,-1)  &\multirow{7}{*}{0.3385898} \\
	 	& & (-1,1,-1,1,1,-1,-1,-1) &  (-1,1,-1,1,1,-1,1,-1)  &  \\ 
		& & (1,-1,1,-1,1,1,-1,1)   &  (1,-1,1,-1,-1,1,-1,1)  &  \\ 
		& & (1,-1,1,-1,-1,1,-1,-1) &  (1,-1,1,-1,-1,1,-1,1)  &  \\
		& & (1,1,1,-1,-1,1,-1,1)   &  (1,1,1,-1,-1,-1,-1,1)  &  \\
		& & (1,-1,1,-1,1,1,-1,1)   &  (1,-1,-1,-1,1,1,-1,1)  &  \\
		& & (1,-1,-1,-1,-1,1,-1,1) &  (1,-1,-1,-1,-1,1,1,1)  &  \\
		\cline{2-5} 
		&  \multirow{2}{*}{1b} & (1,-3,1,-1,-1,1,-1,1) &  (1,-1,1,-1,-1,1,-1,1)  & \multirow{2}{*}{0.3128179} \\
		& & (1,-1,1,-1,-1,1,-1,3) & (1,-1,1,-1,-1,1,-1,1)  &  \\
		\hline
		\multicolumn{2}{|c|}{2} &  (1,1,-1,-1,-1,1,1,1) &  (1,1,-1,-1,-1,1,1,-1)   & 0.3382529 \\ \hline
		\multicolumn{2}{|c|}{\multirow{2}{*}{3}}  & (-1,-1,-1,-1,1,1,1,-1) & (-1,-1,-1,-1,1,1,1,1)   &\multirow{2}{*}{0.3356344}  \\
		\multicolumn{2}{|c|}{} &  (1,1,-1,1,1,1,-1,-1)  & (1,1,-1,-1,1,1,-1,-1) &  \\ 
		\hline
		\multicolumn{2}{|c|}{\multirow{2}{*}{4}}  &   (1,1,1,1,-1,1,-1,1)  & (1,-1,1,1,-1,1,-1,1)  &\multirow{2}{*}{0.3337556}   \\
		\multicolumn{2}{|c|}{} & (-1,-1,-1,-1,-1,1,-1,1)  & (1,-1,-1,-1,-1,1,-1,1) &  \\
		\hline
	\end{tabular}
	\caption{Computation of unweighted Wasserstein distance for some boundary configurations with $\beta J = 0.18727$.}
	\label{tab:truthTables4}   
\end{table}

\noindent Noting that there are 24 possible changes, each corresponding to a sign flip of a different corner boundary spin, and each change may result in a different unweighted Wasserstein distance, the resulting Wasserstein distance is computed in Table \ref{tab:truthTables5} below  for the first set of boundary conditions from Table \ref{tab:truthTables4}.\\

\begin{table}[H]
	\centering
	\begin{tabular}{|M{0.2cm}|M{0.3cm}|M{1cm}|M{2cm}| }
		\hline
		\multicolumn{2}{|c|}{} & Weight &  Resulting Wasserstein distance  \\
		\hline
		\multirow{2}{*}{1} & 1a & 2/3 & \multirow{2}{*}{0.329999} \\  \cline{2-3}
		& 1b & 1/3 & \\ \hline
	\end{tabular}
	\caption{Computation of the resulting Wasserstein distance for the first set of boundary conditions from Table \ref{tab:truthTables4} with $\beta J = 0.18727$.}
	\label{tab:truthTables5}   
\end{table}

\section{Applications - Part II: Convex perturbation of the Gaussian free field model.}
\label{classiC}

In this section, we focus on some models of Euclidean lattice field theories, see, e.g., \cite{GRS,Sim1}.\\
\indent \textit{Notations.} The state space is $\mathbb{R}^{n}$ with its standard topology and Lebesgue measure. The standard inner product and Euclidean norm are denoted by $\langle \cdot\,,\cdot\,\rangle$ and
$\Vert \cdot\,\Vert$ respectively. Below, the norm $\Vert \cdot\,\Vert$ on the space $\mathbb{M}_{n}(\mathbb{C})$ of $n\times n$ matrices denotes the operator norm.

\subsection{The Gaussian free field model with $n$-dimensional spins.}
\label{Gaussfr}

\begin{proposition}
\label{class}
For any integer $n > 1$, let $L \in \mathbb{M}_{n}(\mathbb{R})$ be a non-negative symmetric matrix. Consider the following formal Hamiltonians with nearest-neighbour interactions defined on $(\mathbb{R}^{n})^{\mathbb{Z}^{d}}$ as
\begin{equation}
\label{Hn}
H_{n}^{\textrm{cla}}(\underline{x}) := \sum_{j \in \mathbb{Z}^{d}} \frac{1}{2} \alpha \Vert x_{j}\Vert^{2} - \sum_{j \in \mathbb{Z}^{d}} \sum_{l \in N_{1}(j)} \langle x_{j}, L x_{l} \rangle.
\end{equation}
Then, provided that $\alpha> 2d \Vert L \Vert$, there exists a unique limit Gibbs distribution at any inverse temperature $\beta>0$ in $\mathcal{P}((\mathbb{R}^{n})^{\mathbb{Z}^{d}})$, associated with $H_{n}^{\textrm{cla}}$, with marginal distributions satisfying \eqref{schcond}.

\end{proposition}

\begin{remark}
The interactions between particles are of ferromagnetic type since $L$ is non-negative.
\end{remark}

\noindent \textbf{Proof.} Fix $j \in \mathbb{Z}^{d}$ and $\underline{y}, \underline{y}' \in (\mathbb{R}^{n})^{N_{1}(j)}$ distinct. Set $y := \sum_{l \in N_{1}(j)} y_{l} \in \mathbb{R}^{n}$, $y':=\sum_{l \in N_{1}(j)} y_{l}' \in \mathbb{R}^{n}$. In view of \eqref{Hn}, the corresponding 1-point Gibbs distribution reads
\begin{equation*}
\mu_{j}^{\beta, (n)}(A\vert \underline{y}) := \frac{1}{Z_{j}^{\beta, (n)}(\underline{y})} \int_{A} \exp \left(-\beta \left(\frac{1}{2} \alpha \Vert x\Vert^{2} - \langle x, L y \rangle\right)\right) d^{n} x,\quad A \in \mathcal{B}(\mathbb{R}^{n}),
\end{equation*}
with
\begin{equation*}
Z_{j}^{\beta, (n)}(\underline{y}) := \int_{\mathbb{R}^{n}} \exp\left(- \beta \left( \frac{1}{2}\alpha \Vert x\Vert^{2} - \langle x, L y\rangle\right)\right) d^{n}x.
\end{equation*}
We now construct a coupling in $\mathcal{P}(\mathbb{R}^{n}\times \mathbb{R}^{n})$ such that the marginals coincide with the 1-point Gibbs distribution above with the different boundary conditions, see \eqref{measn} and \eqref{measnn} below.\\
Set $\theta_{y'-y}^{(n)}:= \frac{1}{\alpha} L(y' - y)$. Define
$\sigma_{j;\underline{y},\underline{y}'}^{\beta,(n)} \in \mathcal{P}(\mathbb{R}^{n}\times \mathbb{R}^{n})$ as
\begin{equation}
\label{measn}
\sigma_{j;\underline{y},\underline{y}'}^{\beta,(n)}(A \times B) := \frac{1}{Z_{j}^{\beta, (n)}(\underline{y})} \int_{A \times B} \exp\left(-\beta \left(\frac{1}{2} \alpha \Vert x\Vert^{2} - \langle x, L  y \rangle\right)\right)\, \delta\left(x'  - x - \theta_{y'-y}^{(n)}\right) d^{n}x d^{n}x',
\end{equation}
for any $A,B \in \mathcal{B}(\mathbb{R}^{n})$. Here, $\delta$ denotes the Dirac measure. It is easily seen that
\begin{equation}
\label{measnn}
\sigma_{j;\underline{y},\underline{y}'}^{\beta,(n)}(A \times \mathbb{R}^{n}) = \mu_{j}^{\beta, (n)}(A\vert \underline{y})\quad \textrm{and}\quad
\sigma_{j;\underline{y},\underline{y}'}^{\beta,(n)}(\mathbb{R}^{n}\times B) = \mu_{j}^{\beta, (n)}(B\vert \underline{y}').
\end{equation}
From \eqref{measn}, we have, by direct calculation,
\begin{equation*}
\int_{\mathbb{R}^{n} \times \mathbb{R}^{n}} \Vert x - x'\Vert\, \sigma_{j;\underline{y},\underline{y}'}^{\beta,(n)}(d^{n} x, d^{n}x') =
\frac{1}{\alpha} \Vert L(y' - y) \Vert.
\end{equation*}
The proposition now follows from the condition in \eqref{nniupbd} since the above identity yields
\begin{equation*}
\rho_{W}\left(\mu_{j}^{\beta, (n)}(\cdot\,\vert \underline{y}), \mu_{j}^{\beta, (n)}(\cdot\,\vert\underline{y}')\right)
\leq \frac{1}{\alpha} \Vert L \Vert \sum_{l \in N_{1}(j)} \Vert y_{l} - y_{l}'\Vert. \tag*{\qed}\\
\end{equation*}

\begin{remark}
\label{casen1}
In the special case $n=1$, the formal Hamiltonian on $\mathbb{R}^{\mathbb{Z}^{d}}$ commonly considered is
\begin{equation}
\label{H1}
H_{1}^{\textrm{cla}}(\underline{x}) := \sum_{j \in \mathbb{Z}^{d}} \frac{1}{2} \alpha x_{j}^{2} + \sum_{j \in \mathbb{Z}^{d}} \sum_{l \in N_{1}(j)}\frac{1}{2}(x_{j} - x_{l})^{2}.
\end{equation}
We can mimic the arguments used in the proof of Proposition \ref{class} by gathering the quadratic terms together, giving the factor $\alpha_{d}:= \alpha + 4d$. The existence and uniqueness, for all $\beta>0$, of a limit Gibbs distribution in $\mathcal{P}(\mathbb{R}^{\mathbb{Z}^{d}})$ with marginal distributions satisfying \eqref{schcond},
 is guaranteed whenever $\alpha>0$.
\end{remark}

\begin{remark}
\label{remmmm}
The method used in the proof given above allows us to extend the uniqueness result of Proposition \ref{class} to long-range pair-interaction potentials, typically of the form
\begin{equation}
\label{forpot}
\sum_{j \in \mathbb{Z}^{d}} \sum_{\substack{l \in \mathbb{Z}^{d} \\ l\neq j}} J(\vert j - l \vert)(x_{j} - x_{l})^{2},
\end{equation}
where $J$ is a sufficiently fast decreasing function satisfying $\sum_{\substack{j \in \mathbb{Z}^{d} \\ j \neq 0}} J(\vert j \vert) < \infty$.
\end{remark}

\subsection{Convex perturbation of the Gaussian free field model.}
\label{convxx}

We extend the uniqueness result to the Gaussian free field model perturbed by a convex self-interaction potential. Proposition \ref{convex} covers the case of 1-dimensional spins while Proposition \ref{convex2} covers the case of $n$-dimensional spins for a particular class of convex potentials.

\subsubsection{The case of 1-dimensional spins.}

\begin{proposition}
\label{convex}
Consider the following formal Hamiltonian with nearest-neighbour interactions defined on $\mathbb{R}^{\mathbb{Z}^{d}}$, which is a perturbation of \eqref{H1} in the sense
\begin{equation}
\label{Hconv}
\tilde{H}_{1}^{\textrm{cla}}(\underline{x}) := \sum_{j \in \mathbb{Z}^{d}} \left(\frac{1}{2} \alpha x_{j}^{2} + g(x_{j})\right) + \sum_{j \in \mathbb{Z}^{d}} \sum_{l \in N_{1}(j)} \frac{1}{2}(x_{j} - x_{l})^{2}.
\end{equation}
Assume that $g: \mathbb{R} \rightarrow \mathbb{R}$ is a convex function. Then provided that $\alpha>0$, there exists a unique limit Gibbs distribution at any inverse temperature $\beta>0$ in $\mathcal{P}(\mathbb{R}^{\mathbb{Z}^{d}})$, associated with $\tilde{H}_{1}^{\textrm{cla}}$, with marginal distributions satisfying \eqref{schcond}.
\end{proposition}

\begin{remark}
In the lattice approximation of quantum field theory, the $\phi^{4}$ model corresponds to $g(x)= c x^{4}$, $c >0$ in \eqref{Hconv}. Further models are covered by our assumptions, see, e.g., \cite{GRS, Sim1}.
\end{remark}

\begin{remark}
The uniqueness result of Proposition \ref{convex} can be extended to long-range pair interaction potentials of the form \eqref{forpot}, see Remark \ref{remmmm}.
\end{remark}

To prove Proposition \ref{convex}, we need the following lemma.

\begin{lema}
\label{incres}
Let $\nu \in \mathcal{P}(\mathbb{R})$ be a probability measure. Then, for any $\nu$-integrable and non-decreasing function $f$ on $\mathbb{R}$, the function
\begin{equation*}
z \mapsto w_{\nu}(z) := \frac{1}{\nu((-\infty,z])} \int_{-\infty}^{z} f(x) \nu(dx)
\end{equation*}
is non-decreasing on $\mathbb{R}$.
\end{lema}

\noindent \textbf{Proof.} Suppose that $z < z'$. We have
\begin{equation*}
w_{\nu}(z') - w_{\nu}(z) = \frac{1}{\nu((-\infty,z'])}
\left(\int_{z}^{z'} f(x)\nu(dx) \\
- \frac{\nu((z,z'])}{\nu((-\infty,z])} \int_{-\infty}^{z} f(x) \nu(dx)\right).
\end{equation*}
Since $f$ is non-decreasing by assumption, $f(x) \geq f(z)$ on $(z,z']$ and $f(x)\leq f(z)$ on $(-\infty,z]$.\\ As a result,
\begin{equation*}
\nu((-\infty,z'])\left(w_{\nu}(z') - w_{\nu}(z)\right)  \geq  f(z)\left(\nu((z,z']) - \frac{\nu((z,z'])\nu((-\infty,z])}{\nu((-\infty,z])}\right) =0. \tag*{\qed}
\end{equation*}

We now turn to\\

\noindent \textbf{Proof of Proposition \ref{convex}.} Fix $j \in \mathbb{Z}^{d}$ and $\underline{y}, \underline{y}' \in \mathbb{R}^{N_{1}(j)}$ distinct. Set $y := \sum_{l \in N_{1}(j)} y_{l} \in \mathbb{R}$ and $y':=\sum_{l \in N_{1}(j)} y_{l}' \in \mathbb{R}$.  From \eqref{Hconv}, the corresponding 1-point Gibbs distribution reads
\begin{equation*}
\mu_{j}^{\beta}(A \vert \underline{y}) := \frac{1}{Z_{j}^{\beta}(\underline{y})} \int_{A} \exp\left(- \beta \left(\frac{1}{2} \alpha_{d} x^{2} + g(x) - xy\right)\right) dx,\quad A \in \mathcal{B}(\mathbb{R}),
\end{equation*}
where $\alpha_{d} := \alpha + 4d$, and with
\begin{equation*}
Z_{j}^{\beta}(\underline{y}) := \int_{\mathbb{R}} \exp\left(-\beta\left(\frac{1}{2}\alpha_{d} x^{2} + g(x) - xy\right)\right) dx.
\end{equation*}
We now construct a coupling in $\mathcal{P}(\mathbb{R}\times \mathbb{R})$ such that the marginals coincide with the 1-point Gibbs distribution above with the different boundary conditions, see \eqref{sigmyy'} and \eqref{coinn} below.
Let $(z_{m})_{m \in \mathbb{Z}}$ be the increasing sequence of points $z_{m}:= m \delta$ with $0<\delta<1$. Define the sequence $(z_{m}')_{m \in \mathbb{Z}}$ such that
\begin{equation}
\label{eqq1}
\mu_{j}^{\beta}\left((-\infty,z_{m}']\vert \underline{y}'\right) = \mu_{j}^{\beta}\left((-\infty,z_{m}]\vert \underline{y}\right).
\end{equation}
This implies
\begin{equation*}
\mu_{j}^{\beta}\left((z_{m-1}',z_{m}']\vert \underline{y}'\right) = \mu_{j}^{\beta}\left((z_{m-1},z_{m}]\vert \underline{y}\right).
\end{equation*}
From the foregoing, define the coupling $\sigma_{j;\underline{y},\underline{y}'}^{\beta} \in \mathcal{P}(\mathbb{R}\times \mathbb{R})$ as
\begin{equation}
\label{sigmyy'}
\sigma_{j;\underline{y},\underline{y}'}^{\beta}(A \times B) := \sum_{m \in \mathbb{Z}} \frac{\mu_{j}^{\beta}\left(A \cap (z_{m-1},z_{m}]\vert \underline{y}\right) \mu_{j}^{\beta}\left(B \cap (z_{m-1}',z_{m}']\vert \underline{y}'\right)}{\mu_{j}^{\beta}\left((z_{m-1},z_{m}]\vert \underline{y}\right)},
\end{equation}
for any $A,B \in \mathcal{B}(\mathbb{R})$. It is easily seen that
\begin{equation}
\label{coinn}
\sigma_{j;\underline{y},\underline{y}'}^{\beta}(A \times \mathbb{R})= \mu_{j}^{\beta}(A\vert \underline{y}) \quad \textrm{and}\quad \sigma_{j;\underline{y},\underline{y}'}^{\beta}(\mathbb{R}\times B)= \mu_{j}^{\beta}(B\vert \underline{y}').
\end{equation}Furthermore,
\begin{equation}
\label{temp}
\begin{split}
\int_{\mathbb{R}\times \mathbb{R}} \vert x - x'\vert \sigma_{j;\underline{y},\underline{y}'}^{\beta}(dx, dx') &= \sum_{m \in \mathbb{Z}} \int_{(z_{m-1},z_{m}] \times (z_{m-1}', z_{m}']} \vert x - x'\vert \frac{\mu^{\beta}_{j}(dx\vert \underline{y}) \mu^{\beta}_{j}(dx'\vert \underline{y}')}{\mu^{\beta}_{j}\left((z_{m-1},z_{m}]\vert \underline{y}\right)} \\
&\leq \sum_{m \in \mathbb{Z}} \left(\sup_{(z,z') \in (z_{m-1},z_{m}] \times (z_{m-1}', z_{m}']} \vert z - z'\vert \right)  \mu^{\beta}_{j}\left((z_{m-1}',z_{m}']\vert \underline{y}'\right).
\end{split}
\end{equation}
Next, we claim that under our conditions (a proof is given below)
\begin{equation}
\label{claim1}
\sup_{(z,z') \in (z_{m-1},z_{m}] \times (z_{m-1}', z_{m}']} \vert z - z'\vert \leq \delta + \frac{1}{\alpha_{d}} \vert y - y' \vert.
\end{equation}
Inserting \eqref{claim1} into the right-hand side of \eqref{temp} and letting $\delta \rightarrow 0$, we get,
\begin{equation*}
\rho_{W}\left(\mu_{j}^{\beta}(\cdot\,\vert \underline{y}), \mu_{j}^{\beta}(\cdot\,\vert \underline{y}')\right) \leq \frac{1}{\alpha_{d}} \vert y - y' \vert \leq \frac{1}{\alpha_{d}} \sum_{l \in N_{1}(j)} \vert y_{l} - y_{l}'\vert,
\end{equation*}
and the Proposition follows from the condition in \eqref{nniupbd}. To complete the proof of the Proposition, we now prove \eqref{claim1}. Suppose that $y < y'$ (the case $y>y'$ can be treated by similar arguments). First, we show that $z_{m} \leq z_{m}'$ for all $m \in \mathbb{Z}$. It suffices to remark that
\begin{equation*}
\mu_{j}^{\beta}\left((-\infty,z] \vert \underline{y}'\right) = \frac{\int_{-\infty}^{z} \mathrm{e}^{\beta x (y'-y)} \mu_{j}^{\beta}(dx \vert \underline{y})}{\int_{-\infty}^{+\infty} \mathrm{e}^{\beta x (y'-y)} \mu_{j}^{\beta}(dx \vert \underline{y})} \leq \mu_{j}^{\beta}\left((-\infty,z] \vert \underline{y}\right),
\end{equation*}
as a result of Lemma \ref{incres} since $x \mapsto \mathrm{e}^{\beta x (y'-y)}$ is increasing.
Set $z=z_{m}$ and then use \eqref{eqq1}. Secondly, we show that $z_{m}' \leq z_{m} + \theta_{y'-y}$ with $\theta_{y' - y} := \frac{1}{\alpha_{d}} (y' - y) >0$ proving \eqref{claim1} when $y<y'$. Due to \eqref{eqq1}, it suffices to prove that
\begin{equation}
\label{inh}
\mu_{j}^{\beta}\left((-\infty,z+\theta_{y'-y}] \vert \underline{y}'\right) \geq  \mu_{j}^{\beta}\left((-\infty,z] \vert \underline{y}\right).
\end{equation}
By a change of variables, we have the rewriting
\begin{equation*}
\mu_{j}^{\beta}\left((-\infty,z+\theta_{y'-y}] \vert \underline{y}'\right) = \frac{\int_{-\infty}^{z} e^{-\frac{1}{2}\beta \alpha_{d}(x- \alpha_{d}^{-1}y)^{2} - \beta g(x+\theta_{y'-y})} dx}{\int_{-\infty}^{+\infty} e^{-\frac{1}{2}\beta \alpha_{d}(x- \alpha_{d}^{-1}y)^{2} - \beta g(x+\theta_{y'-y})} dx}.
\end{equation*}
This yields the following identity
\begin{multline}
\label{cleq}
\frac{\int_{-\infty}^{+\infty} e^{-\frac{1}{2}\beta \alpha_{d}(x- \alpha_{d}^{-1}y)^{2} - \beta g(x+\theta_{y'-y})} dx}{\int_{-\infty}^{z} e^{-\frac{1}{2}\beta \alpha_{d}(x- \alpha_{d}^{-1}y)^{2} - \beta g(x)} dx} \left(\mu_{j}^{\beta}\left((-\infty,z+\theta_{y'-y}] \vert \underline{y}'\right) - \mu_{j}^{\beta}\left((-\infty,z] \vert \underline{y}\right)\right) \\ = \frac{1}{\mu_{j}^{\beta}\left((-\infty,z]\vert \underline{y}\right)} \int_{-\infty}^{z} \mathrm{e}^{-\beta(g(x+\theta_{y'-y}) - g(x))} \mu_{j}^{\beta}(dx\vert \underline{y}) - \int_{-\infty}^{\infty} \mathrm{e}^{-\beta(g(x+\theta_{y'-y}) - g(x))} \mu_{j}^{\beta}(dx\vert \underline{y}).
\end{multline}
Note that $x \mapsto g(x+\theta) - g(x)$, $\theta>0$ is non-decreasing since $g$ is convex.
By multiplying the right-hand side of \eqref{cleq} by $(-1)$ then applying Lemma \ref{incres} with $f(x)= - e^{-\beta[g(x+\theta_{y'-y}) - g(x)]}$ and $\nu = \mu_{j}^{\beta}(\cdot\,\vert \underline{y})$, we conclude that \eqref{cleq} is non-negative. Therefore, \eqref{inh} follows. \qed

\subsubsection{The case of $n$-dimensional spins.}

\begin{proposition}
\label{convex2}
For any integer $n > 1$, let $L \in \mathbb{M}_{n}(\mathbb{R})$ be a non-negative symmetric matrix. Consider the following formal Hamiltonians with nearest-neighbour interactions defined on $(\mathbb{R}^{n})^{\mathbb{Z}^{d}}$, which are a perturbations of \eqref{Hn} given by
\begin{equation}
\label{Hconvn}
\tilde{H}_{n}^{\textrm{cla}}(\underline{x}) := \sum_{j \in \mathbb{Z}^{d}} \left(\frac{1}{2} \alpha \Vert x_{j}\Vert^{2} + \sum_{r=1}^{n} g_{r}(x_{j,r}) \right)- \sum_{j \in \mathbb{Z}^{d}} \sum_{l \in N_{1}(j)} \langle x_{j}, L x_{l} \rangle.
\end{equation}
Assume that $g_{r}: \mathbb{R} \rightarrow \mathbb{R}$, $r=1\ldots,n$ are convex functions. Then, provided that $\alpha> 2d \Vert L \Vert$, there exists a unique limit Gibbs distribution at any inverse temperature $\beta>0$ in $\mathcal{P}((\mathbb{R}^{n})^{\mathbb{Z}^{d}})$, associated with $\tilde{H}_{n}^{\textrm{cla}}$, with marginal distributions satisfying \eqref{schcond}.
\end{proposition}

\begin{remark}
The uniqueness result of Proposition \ref{convex2} can be extended to long-range pair interaction potentials of the form \eqref{forpot}, see Remark \ref{remmmm}.
\end{remark}

\noindent \textbf{Proof of Proposition \ref{convex2}}. Fix $j \in \mathbb{Z}^{d}$ and $\underline{y}, \underline{y}' \in (\mathbb{R}^{n})^{N_{1}(j)}$ distinct. Set $y := \sum_{l \in N_{1}(j)} y_{l} \in \mathbb{R}^{n}$, $y':=\sum_{l \in N_{1}(j)} y_{l}' \in \mathbb{R}^{n}$. In view of \eqref{Hconvn}, the corresponding 1-point Gibbs distribution reads
\begin{equation*}
\mu_{j}^{\beta, (n)}(A\vert \underline{y}) := \frac{1}{Z_{j}^{\beta, (n)}(\underline{y})} \int_{A} \prod_{r=1}^{n} \exp \left(-\beta \left(\frac{1}{2} \alpha x_{r}^2 + g_{r}(x_{r}) - x_{r} (L y)_{r}  \right)\right) dx_{1} \dotsb dx_{n},\quad A \in \mathcal{B}(\mathbb{R}^{n}),
\end{equation*}
with
\begin{equation*}
Z_{j}^{\beta, (n)}(\underline{y}) := \int_{\mathbb{R}^{n}} \prod_{r=1}^{n} \exp \left(-\beta \left(\frac{1}{2} \alpha x_{r}^2 + g_{r}(x_{r}) - x_{r} (L y)_{r}  \right)\right) dx_{1} \dotsb dx_{n}.
\end{equation*}
Pick $0<\delta<1$. Let $(z_{1,m_{1}})_{m_{1} \in \mathbb{Z}}$ be the  sequence  $z_{1,m_{1}} := m_{1} \kappa_{n} \delta^{2}$ with
\begin{equation*}
\kappa_{n}:= \frac{1}{3 \max(1,\alpha^{-1}\Vert L(y-y')\Vert)\sqrt{n}}.
\end{equation*}
Define the sequence $(z_{1,m_{1}}')_{m_{1} \in \mathbb{Z}}$ such that
\begin{equation*}
\mu_{j}^{\beta, (n)}(\{z \in \mathbb{R}^{n}\,:\, z_{1} \in (-\infty,z_{1,m_{1}}]\}\vert \underline{y})
= \mu_{j}^{\beta,(n)}(\{z' \in \mathbb{R}^{n}\,:\, z_{1}' \in (-\infty,z_{1,m_{1}}']\}\vert \underline{y}').
\end{equation*}
Set $A_{m_{1}}^{(1)} := \{z \in \mathbb{R}^{n}\,:\, z_{1} \in (z_{1,m_{1}-1},z_{1,m_{1}}]\}$ and  $A_{m_{1}}'^{(1)} := \{z' \in \mathbb{R}^{n}\,:\, z'_{1} \in (z'_{1,m_{1}-1},z'_{1,m_{1}}]\}$. Then, for $r=2,\ldots,n$, let $(z_{r,m_{r}})_{m_{r} \in \mathbb{Z}}$ be the same increasing sequences $z_{r,m_{r}} := m_{r} \kappa_{n} \delta^{2}$. Define recursively the sets $A_{m_{1},\ldots,m_{r}}^{(r)} :=\{z \in A_{m_{1},\ldots,m_{r-1}}^{(r- 1)}\, :\, z_{r} \in (z_{r,m_{r}-1},z_{r,m_{r}}]\}$, $2\leq r \leq n$. Define the sequences $(z_{r,m_{r}}')_{m_{r} \in \mathbb{Z}}$, $r=2,\ldots,n$ such that
\begin{equation*}
\mu_{j}^{\beta,(n)}(A_{m_{1},\ldots,m_{r}}^{(r)}\vert \underline{y}) = \mu_{j}^{\beta,(n)}(A_{m_{1},\ldots,m_{r}}'^{(r)}\vert \underline{y}'),\quad 2\leq r \leq n,
\end{equation*}
where the sets $A_{m_{1},\ldots,m_{r}}'^{(r)}$ are defined similarly to $A_{m_{1},\ldots,m_{r}}^{(r)}$ but with $(z_{r}')_{r}$ and $(z_{r,m_{r}}')_{r}$. Define
\begin{equation*}
\sigma_{j;\underline{y},\underline{y}'}^{\beta,(n)}(A \times B) := \sum_{m_{1},\ldots,m_{n} \in \mathbb{Z}} \frac{\mu_{j}^{\beta,(n)}(A \cap A_{m_{1},\ldots,m_{n}}^{(n)}\vert \underline{y}) \mu_{j}^{\beta,(n)}(B \cap A_{m_{1},\ldots,m_{n}}'^{(n)}\vert \underline{y}')}{\mu_{j}^{\beta,(n)}(A_{m_{1},\ldots,m_{n}}^{(n)}\vert \underline{y})},
\end{equation*}
for any $A,B \in \mathcal{B}(\mathbb{R}^{n})$. It is easily seen that
\begin{equation*}
\sigma_{j;\underline{y},\underline{y}'}^{\beta,(n)}(A \times \mathbb{R}^{n}) = \mu_{j}^{\beta,(n)}(A\vert \underline{y}) \quad \textrm{and} \quad \sigma_{j;\underline{y},\underline{y}'}^{\beta,(n)}(\mathbb{R}^{n} \times B) = \mu_{j}^{\beta,(n)}(B\vert \underline{y}').
\end{equation*}
Furthermore,
\begin{multline}
\label{togty}
\int_{\mathbb{R}^{n} \times \mathbb{R}^{n}} \Vert x - x'\Vert \sigma_{j;\underline{y},\underline{y}'}^{\beta,(n)}(d^n x, d^n x')
\leq \sum_{m_{1},\ldots, m_{n} \in \mathbb{Z}}  \left(\sup_{(z,z') \in A_{m_{1},\ldots,m_{n}}^{(n)} \times A_{m_{1},\ldots,m_{n}}'^{(n)}} \Vert z - z'\Vert \right) \\ \times \frac{\mu_{j}^{\beta,(n)}(A_{m_{1},\ldots,m_{n}}^{(n)}\vert \underline{y}) \mu_{j}^{\beta,(n)}(A_{m_{1},\ldots,m_{n}}'^{(n)}\vert \underline{y}')}{\mu_{j}^{\beta,(n)}(A_{m_{1},\ldots,m_{n}}^{(n)}\vert \underline{y})}.
\end{multline}
We now claim that, under our conditions, the following holds
\begin{equation}
\label{claimn}
\sup_{(z,z') \in A_{m_{1},\ldots,m_{n}}^{(n)} \times A_{m_{1},\ldots,m_{n}}'^{(n)}} \Vert z - z'\Vert \leq \delta + \frac{1}{\alpha} \Vert L (y - y') \Vert.
\end{equation}
Inserting \eqref{claimn} into the right-hand side of \eqref{togty} and letting $\delta \rightarrow 0$, we get,
\begin{equation*}
\rho_{W}\left(\mu_{j}^{\beta,(n)}(\cdot\,\vert \underline{y}), \mu_{j}^{\beta,(n)}(\cdot\,\vert \underline{y}')\right) \leq \frac{1}{\alpha} \Vert L(y-y')\Vert \leq \frac{1}{\alpha} \Vert L \Vert \sum_{l \in N_{1}(j)} \Vert y_{l} - y_{l}'\Vert,
\end{equation*}
and the Proposition follows from \eqref{nniupbd}. To prove \eqref{claimn}, it is enough to show that
\begin{equation*}
\sup_{(x_{r},x_{r}') \in (z_{r,m_{r}-1}, z_{r,m_{r}}] \times (z_{r,m_{r}-1}', z_{r,m_{r}}']} \vert x_{r} - x_{r}' \vert \leq \kappa_{n} \delta^{2} + \frac{1}{\alpha} \vert (L (y'-y))_{r} \vert, \quad r=1,\ldots,n.
\end{equation*}
Since this is now similar to the case of 1-dimensional spins, it suffices to repeat the arguments used to prove \eqref{claim1} in the proof of Proposition \ref{convex}. \qed

\begin{remark}
Some translation-invariant Hamiltonians with infinite-range pair-interactions are considered in \cite{LP,COPP,BK}. A typical pair-interaction potential is of the form \eqref{forpot}. In \cite{LP,COPP,BK}, the self-interaction potentials diverge at least quadratically and the conditions set on the pair-interaction potentials assure the superstability of the Hamiltonians. For a wide class of boundary conditions, existence of 'superstable' limit Gibbs distributions (see \cite{Ru0,Ru1}) is proven in \cite{LP} (the state space is $\mathbb{R}^{n}$) and existence of tempered limit Gibbs distributions is proven in \cite{COPP,BK} (the state space is $\mathbb{R}$). Conditions for uniqueness are also discussed in \cite{LP,COPP,BK}. In \cite{COPP}, sufficient conditions are derived from Dobrushin's uniqueness theorem using an explicit  expression for the Wasserstein distance between probability measures on $\mathbb{R}$, see \cite[Theorem]{Val} and \cite[Thm. 2.2]{COPP}.
\end{remark}

\section{Proof of Theorem \ref{complete}.}
\label{compl}

\subsection{Some technical results.}

We start with two lemmas:

\begin{lema}
\label{lem1}
Let $\mathcal{X}$ be a completely regular Hausdorff space. Let $K \subset \mathcal{X}$ be a non-empty compact set and $O \subset \mathcal{X}$ an open set such that $K\subset O$. Let
$(\mu_{n})_{n \in \mathbb{N}} \subset \mathcal{P}(\mathcal{X})$ be a sequence of probability measures converging weakly to $\mu \in \mathcal{P}(\mathcal{X})$. Then, given $\epsilon >0$, there exists an open set $V \subset O$ such that $K \subset V$, and for all $n$ large enough, $\mu_{n}(V) < \mu(K) + \epsilon$.
\end{lema}

\noindent \textbf{Proof.} Given $0<\delta<1$, we may replace $O$ by an open set $U$ such that $K \subset U$ and $\mu(U\setminus K) < \delta$. By Urysohn's lemma, there exists a continuous function $f: \mathcal{X} \rightarrow [0,1]$ such that $\mathbf{1}_{K} \leq f \leq \mathbf{1}_{U}$. Since $(\mu_{n})_{n}$ converges weakly to $\mu$, then it follows that, for $n$ sufficiently large,
\begin{equation*}
\left \vert \int_{\mathcal{X}} f(x)  \mu (dx) - \int_{\mathcal{X}} f(x)  \mu_{n} (dx) \right\vert < \delta.
\end{equation*}
Now define $V := \{x \in \mathcal{X}: f(x) > 1 - \delta\}$. Clearly, $K \subset V \subset U$. From the above, we have, for $n$ large enough,
\begin{equation*}
\mu_{n}(V) < \frac{1}{1-\delta} \int_{\mathcal{X}} f(x)  \mu_{n} (dx) < \frac{1}{1-\delta} \left(\int_{\mathcal{X}} f(x)  \mu(dx) + \delta \right) < \frac{1}{1-\delta} (\mu(K) + 2\delta ),
\end{equation*}
where we used that $\mu(U) < \mu(K) + \delta$. Set $\epsilon = 3\delta/(1 - \delta)$ and the lemma follows. \qed \\

We recall the following well-known result

\begin{lema}
\label{lemm2}
Let $\mathcal{X}$ be a completely regular Hausdorff space. Let
$(\mu_{n})_{n \in \mathbb{N}} \subset \mathcal{P}(\mathcal{X})$ be a sequence of probability measures converging weakly to $\mu \in \mathcal{P}(\mathcal{X})$. Then, for all open sets $O \subset \mathcal{X}$,
\begin{equation*}
\liminf_{n \rightarrow \infty} \mu_{n}(O) \geq \mu(O).
\end{equation*}
\end{lema}

We continue with the following two technical lemmas

\begin{lema}
\label{lem3}
Let $\mathcal{X}$ be a completely regular Hausdorff space. Let $K \subset \mathcal{X}$ be a non-empty compact set and $O_{1},O_{2} \subset \mathcal{X}$ be two open sets such that $K\subset (O_{1}\cup O_{2})$. Let
$(\mu_{n})_{n \in \mathbb{N}} \subset \mathcal{P}(\mathcal{X})$ be a sequence of probability measures converging weakly to $\mu \in \mathcal{P}(\mathcal{X})$. Then, given $0<\delta<1$ and $0<\epsilon<1$, there exist two open sets $V_{1} \subset O_{1}$ and $V_{2} \subset O_{2}$ such that $K \subset (V_{1} \cup V_{2})$, with $(K \setminus O_{2}) \subset V_{1}$ and $(K \setminus V_{1}) \subset V_{2}$, and such that, for all $n$ large enough, all the following hold\\
$(i)$. Either $\mu(K \setminus O_{2}) =0$ and $\mu_{n}(V_{1}) < \delta$, or $\mu(K \setminus O_{2}) > 0$ and $\mu_{n}(V_{1}) < (1+\epsilon) \mu(K \setminus O_{2})$;\\
$(ii)$. Either $\mu(K \setminus V_{1}) =0$ and $\mu_{n}(V_{2}) < \delta$, or $\mu(K \setminus V_{1}) > 0$ and $\mu_{n}(V_{2}) < (1+\epsilon) \mu(K \setminus V_{1})$ ; \\
$(iii)$. $\mu_{n}(V_{1}\cap V_{2}) < \frac{1}{2} \epsilon \mu(K\setminus V_{1})$ whenever $\mu(K \setminus V_{1}) > 0$. 
\end{lema}

\noindent \textbf{Proof.} Set $K_{1}:= K \setminus O_{2}$. Assume first that $\mu(K)=0$.  Clearly, there exists an open set $V_{1} \subset O_{1}$ such that $K_{1}\subset V_{1}$ and $\mu_{n}(V_{1}) < \delta$ for $n$ large enough. Set $K_{2}:= K \setminus V_{1}$. Similarly, there exists an open set $V_{2} \subset O_{2}$ such that $K_{2}\subset V_{2}$ and $\mu_{n}(V_{2}) < \delta$ for $n$ large enough. So we can now assume that $\mu(K)>0$.
If $\mu(K_{1})=0$, then as above, there exists an open set $V_{1} \subset O_{1}$ such that $K_{1}\subset V_{1}$ and $\mu_{n}(V_{1}) < \delta$ for $n$ sufficiently large. If $\mu(K_{1})>0$ then by Lemma \ref{lem1}, there exists an open set $V_{1}\subset O_{1}$ such that $K_{1} \subset V_{1}$ and, for $n$ large enough, $\mu_{n}(V_{1}) - \mu(K_{1}) < \epsilon \mu(K_{1})$. This proves $(i)$. We point out that, if $\mu(K\setminus V_{1}) >0$, we may assume that, for $n$ large enough, we also have,
\begin{equation}
\label{im2}
\mu_{n}(V_{1}) - \mu(K_{1}) < \frac{1}{4} \epsilon \mu(K\setminus V_{1}).
\end{equation}
Indeed, we can replace $V_{1}$ by an open set $V_{1}' \subset V_{1}$ such that $K_{1} \subset V_{1}'$ and for $n$ sufficiently large, $\mu_{n}(V_{1}') < \mu(K_{1}) + \frac{1}{4} \epsilon \mu(K \setminus V_{1}) < \mu(K_{1}) + \frac{1}{4} \epsilon \mu(K \setminus V_{1}')$. Write $V_{1}$ instead of $V_{1}'$ and \eqref{im2} follows.
We next turn to $(ii)$. Set $K_{2}:= K \setminus V_{1}$. Notice that $K_{1} \cap K_{2} = \emptyset$. Let $U_{1}\subset V_{1}$ and $U_{2} \subset O_{2}$ be two open sets such that $U_{1}\cap U_{2} = \emptyset$ and $K_{1} \subset U_{1}$ and $K_{2}\subset U_{2}$. By the same arguments used to prove $(i)$, we may choose an open set $V_{2}\subset U_{2}$ such that $K_{2}\subset V_{2}$ and for $n$ sufficiently large, $\mu_{n}(V_{2}) < \delta$ if $\mu(K_{2})=0$ and $\mu_{n}(V_{2}) - \mu(K_{2}) < \epsilon \mu(K_{2})$ if $\mu(K_{2})>0$. To prove $(iii)$, first note that $\mu_{n}(V_{1}\cap V_{2}) \leq \mu_{n}(V_{1})-\mu_{n}(U_{1})$ because $V_{2}\subset U_{2} \subset ((O_{1}\cup O_{2})\setminus U_{1})$. Assuming $\mu(K_{2})>0$, it follows from Lemma \ref{lemm2} that $\mu_{n}(V_{1}\cap V_{2}) \leq \mu_{n}(V_{1}) - \mu(U_{1}) + \frac{1}{4}\epsilon\mu(K_{2})$ for $n$ sufficiently large. It remains to use \eqref{im2} noting that $K_{1}\subset U_{1}$. 
\qed \\

\begin{lema}
\label{t3}
Let $\mathcal{X}$ be a completely regular Hausdorff space. Let $K \subset \mathcal{X}$ be a non-empty compact set and $B_{1},\dotsb, B_{N} \subset \mathcal{X}$ $N \geq 2$ open sets such that $K\subset \bigcup_{i=1}^{N} B_{i}$. Let
$(\mu_{n})_{n \in \mathbb{N}} \subset \mathcal{P}(\mathcal{X})$ be a sequence of probability measures converging weakly to $\mu \in \mathcal{P}(\mathcal{X})$. Then, given $0<\epsilon<1$, there exist $N$ open sets $V_{1},\dots,V_{N}$ with $V_{i} \subset B_{i}$, $i=1,\dots,N$ such that $K \subset \bigcup_{j=1}^{N} V_{j}$, and, setting
\begin{equation}
\label{Ki}
\begin{split}
K_{1} &:= K\setminus \bigcup_{j=2}^{N} B_{j};\\
K_{i} &:= K \setminus \left(\bigcup_{j=1}^{i-1} V_{j} \cup \bigcup_{j=i+1}^{N} B_{j}\right),\quad i=2,\dots, N-1;\\
K_{N} &:= K\setminus \bigcup_{j=1}^{N-1} V_{j},
\end{split}
\end{equation}
such that $K_{1} \subset V_{1}$ and $K_{i} \subset V_{i} \setminus \bigcup_{j=1}^{i-1} V_{j}$, $i=2,\ldots,N$, and, for all $n$ sufficiently large, either $\mu(K_{i})=0$ and $\mu_{n}(V_{i}) < \frac{\epsilon}{N}$, or $\mu(K_{i})>0$ and $\mu_{n}(V_{i}) < (1+\epsilon) \mu(K_{i})$ for all $i=1,\ldots,N$, and moreover,
\begin{equation}
\label{intermaj}
\mu_{n}\left(V_{i} \cap \bigcup_{j=1}^{i-1} V_{j}\right) < \frac{1}{2} \epsilon \mu(K_{i}).
\end{equation}
\end{lema}

\noindent \textbf{Proof.} We start by applying Lemma \ref{lem3} to $K^{(2)}:= K\setminus \bigcup_{j=3}^{N} B_{j} \subset (B_{1} \cup B_{2})$. Hence, there exist two open sets $V_{1}'\subset B_{1}$ and $V_{2}'\subset B_{2}$ such that $K^{(2)} \subset (V_{1}' \cup V_{2}')$, and such that, for $n$ large enough, denoting $K_{1}':= K^{(2)}\setminus B_{2}\subset V_{1}'$ and $K_{2}' := K^{(2)}\setminus V_{1}' \subset V_{2}'$, either $\mu(K_{i}')=0$ and $\mu_{n}(V_{i}') < \frac{\epsilon}{N}$, or $\mu(K_{i}')> 0$ and $\mu_{n}(V_{i}') <(1+\epsilon) \mu(K_{i}')$, $i=1,2$, and, moreover, $\mu_{n}(V_{1}'\cap V_{2}') < \frac{1}{2}\epsilon \mu(K_{2}')$.
We now proceed by induction on $N$, assuming that the statement holds for $N-1$ and modifying the previous sets $V_{i}'$, $i=1,\dots, N-2$ in the process (the sets $V_{i}'$, $i=1,\ldots,j$ constructed at the induction step $j$ are different from the sets $V_{i}'$, $i=1,\ldots,j$ constructed at the step $j+1$).
Assume thus that we have constructed $V_{i}'$, $i=1,\dots,N-1$ for the set $K^{(N-1)}:= K\setminus B_{N} \subset \bigcup_{i=1}^{N-1}B_{i}$. We then apply Lemma \ref{lem3} to $K\subset (\bigcup_{i=1}^{N-1} V_{i}' \cup B_{N})$. Hence, there exist two open sets $U_{1}\subset \bigcup_{i=1}^{N-1} V_{i}'$ and $U_{2}\subset B_{N}$ such that $K \subset (U_{1} \cup U_{2})$ with $K^{(N-1)} \subset U_{1}$ and $K \setminus U_{1} \subset U_{2}$, and such that, for $n$ large enough, either $\mu(K^{(N-1)})=0$ and $\mu_{n}(U_{1}) < \frac{\epsilon}{N}$  or $\mu(K^{(N-1)})> 0$ and $\mu_{n}(U_{1}) <(1+\epsilon) \mu(K^{(N-1)})$, and also, either $\mu(K\setminus U_{1})=0$ and $\mu_{n}(U_{2}) < \frac{\epsilon}{N}$  or $\mu(K\setminus U_{1})> 0$ and $\mu_{n}(U_{2}) <(1+\epsilon) \mu(K\setminus U_{1})$, and moreover, $\mu_{n}(U_{1}\cap U_{2}) < \frac{1}{2}\epsilon \mu(K \setminus U_{1})$.
Now let us set $V_{N} := U_{2}$ and $V_{i} := V_{i}' \cap U_{1}$, $i=1,\ldots, N-1$. With this definition, $V_{i} \subset V_{i}'$ and $U_{1} = \bigcup_{i=1}^{N-1} V_{i}$. Set $K_{N} := K \setminus U_{1}$,
\begin{equation*}
K_{i}:= K \setminus \left(\bigcup_{j=1}^{i-1} V_{j} \cup \bigcup_{j=i+1}^{N} B_{j}\right),\quad i=2,\dots, N-1,
\end{equation*}
and $K_{1} := K\setminus \bigcup_{j=2}^{N} B_{j}$. Set also
\begin{equation*}
K^{(i)} := K \setminus \bigcup_{j=i+1}^{N} B_{j},\quad i=2,\ldots, N-1.
\end{equation*}
By construction of the sets $V_{i}'$, we have $K^{(i)} \setminus \bigcup_{j=1}^{i-1} V_{j}' \subset V_{i}'$. Since $K^{(N-1)} \subset U_{1} = \bigcup_{i=1}^{N-1} V_{i}$, we have $K^{(i)} \setminus \bigcup_{j=1}^{i-1} V_{j}' = K^{(i)} \setminus \bigcup_{j=1}^{i-1} V_{j}=K_{i}$. It remains to use that $K_{i} \subset K^{(N-1)}$ to conclude that $K_{1} \subset V_{1}$ and $K_{i} \subset V_{i} \setminus \bigcup_{j=1}^{i-1} V_{j} \subset V_{i}$, $i=2,\ldots,N-1$. Hence, if $\mu(K_{i})=0$ then $\mu_{n}(V_{i}) \leq \mu_{n}(V_{i}') < \frac{\epsilon}{N}$, while if $\mu(K_{i})>0$ then $\mu_{n}(V_{i}) \leq \mu_{n}(V_{i}') < (1+\epsilon) \mu(K_{i})$. Moreover, $\mu_{n}(V_{i} \cap \bigcup_{j=1}^{i-1} V_{j}) < \mu_{n}(V_{i}' \cap \bigcup_{j=1}^{i-1} V_{j}') < \frac{1}{2} \epsilon \mu(K_{i})$. When $i=N$, $K_{N} \subset V_{N}$ since $K\setminus U_{1} \subset U_{2}$. Hence, if $\mu(K_{N})=0$ then $\mu_{n}(V_{N}) = \mu_{n}(U_{2}) < \frac{\epsilon}{N}$, otherwise $\mu_{n}(V_{N}) = \mu_{n}(U_{2}) < (1+\epsilon) \mu(K \setminus U_{1}) = (1+\epsilon) \mu(K_{N})$. Moreover, $\mu_{n}(V_{N} \cap \bigcup_{j=1}^{N-1} V_{j}) = \mu_{n}(U_{2} \cap U_{1}) < \frac{1}{2} \epsilon \mu(K_{N})$. This proves the lemma. \qed

\subsection{Equivalence between convergence in Wasserstein metric and weak convergence with modified Prokhorov condition.}
\label{equivcvg}

The following Proposition contains the key-results for the proof of Theorem \ref{complete} $(ii)$.

\begin{proposition}
\label{prop1}
Let $(\mathcal{X},\rho)$ be a metric space. Let $\mu \in \mathcal{P}_{1}(\mathcal{X})$ and $(\mu_{n})_{n \in \mathbb{N}}\subset \mathcal{P}(\mathcal{X})$ be a sequence of probability measures such that $\mu_{n} \in \mathcal{P}_{1}(\mathcal{X})$ for $n$ large enough.\\
$(i)$. Suppose that $\rho_{W}(\mu_{n},\mu) \rightarrow 0$ when $n \rightarrow \infty$. Then $(\mu_{n})$ converges weakly to $\mu$.\\
$(ii)$. Suppose that $(\mu_{n})$ converges weakly to $\mu$. Further, assume that a Prokhorov-like condition holds, i.e. for every $\epsilon>0$, there exists a non-empty compact set $K \subset \mathcal{X}$ such that for all $n$ large enough and for some $x_{0} \in \mathcal{X}$,
\begin{equation}
\label{unifti}
\int_{K^{c}} \rho(x,x_{0}) \mu_{n}(dx) < \epsilon.
\end{equation}
Then $\rho_{W}(\mu_{n},\mu) \rightarrow 0$ when $n \rightarrow \infty$.
\end{proposition}

\begin{remark}
\label{unifWass}
The proof of Proposition \ref{prop1} relies on Lemma \ref{t3} which holds for general completely regular Hausdorff spaces $\mathcal{X}$. This enables the extension to uniform spaces, replacing the Wasserstein metric by a Wasserstein uniformity.
\end{remark}

\noindent \textbf{Proof.}\,$(i)$. Fix $\epsilon >0$. Let $f:\mathcal{X} \rightarrow \mathbb{R}$ be a bounded \textit{uniformly} continuous function. By definition, there exists $\delta >0$ such that for every $x,y \in \mathcal{X}$ with $\rho(x,y)<\delta$, $\vert f(x) - f(y)\vert < \frac{\epsilon}{2}$. Besides, in view of the definition \eqref{rhoW}, there exists a sequence of probability measures $\sigma_{n} \in \Xi_{\mathcal{X}}(\mu_{n},\mu)$ on $\mathcal{X}\times \mathcal{X}$ such that for all $n$ large enough,
\begin{equation}
\label{funde}
\int_{\mathcal{X}\times\mathcal{X}} \rho(x,y) \sigma_{n}(dx,dy) < \rho_{W}(\mu_{n},\mu) + \frac{\delta^{2}}{2}.
\end{equation}
Let $\Delta_{\delta} := \{(x,y) \in \mathcal{X}\times\mathcal{X}\,:\,\rho(x,y) < \delta\}$ be a  $\delta$-neighbourhood of the diagonal. Then \eqref{funde} implies that, for all $n$ large enough,
\begin{equation}
\label{funde2}
\sigma_{n}(\Delta_{\delta}^{c}) < \frac{1}{\delta} \int_{\Delta_{\delta}^{c}} \rho(x,y) \sigma_{n}(dx,dy) < \frac{1}{\delta}\left(\rho_{W}(\mu_{n},\mu) + \frac{\delta^{2}}{2}\right).
\end{equation}
Since by assumption $\rho_{W}(\mu_{n},\mu) \rightarrow 0$, \eqref{funde2} yields $\sigma_{n}(\Delta_{\delta}^{c}) < \delta$ for $n$ large enough. This implies
\begin{equation}
\label{funde3}
\left\vert \int_{\mathcal{X}} f(x) \mu_{n}(dx) - \int_{\mathcal{X}} f(x) \mu(dx)\right\vert
\leq \int_{\Delta_{\delta} \cup \Delta_{\delta}^{c}} \vert f(x) - f(y)\vert \sigma_{n}(dx,dy)
< \frac{\epsilon}{2} + 2 \Vert f \Vert_{\infty} \delta.
\end{equation}
It remains to replace $\delta$ above by $\min(\delta, \frac{\epsilon}{4 \Vert f \Vert_{\infty}})$. As \eqref{funde3} holds for all bounded uniformly continuous functions $f:\mathcal{X} \rightarrow \mathbb{R}$, the weak convergence follows, see, e.g., \cite[Thm 2.1]{Billi}.\\
We now prove $(ii)$. Fix $x_{0} \in \mathcal{X}$ and $0<\epsilon<1$. By assumption, there exists a  non-empty compact set $K \subset \mathcal{X}$ such that, for $n$ sufficiently large,
\begin{equation*}
\int_{K^{c}} \rho(x,x_{0}) \mu(dx) + \int_{K^{c}} \rho(x,x_{0}) \mu_{n}(dx) < \frac{\epsilon}{6}.
\end{equation*}
We cover $K$ with a number $N$ of open balls $B_{i} \subset \mathcal{X}$ of radius $0<r<1$ to be chosen hereafter. Applying Lemma \ref{t3}, given $0<\kappa < 1$, there exist $N$ open sets $V_{1},\dots,V_{N}$ with $V_{i} \subset B_{i}$ and also $N$ compact sets $K_{1},\dots,K_{N}$ which are defined in \eqref{Ki} such that, setting
\begin{equation}
\label{Ai}
\begin{split}
A_{1} :=& V_{1},\\
A_{i} :=& V_{i} \setminus \bigcup_{j=1}^{i-1} V_{j},\quad i=2,\dots,N,
\end{split}
\end{equation}
we have $K_{i} \subset A_{i} \subset V_{i}$, $i=1,\dots, N$ and, for all $n$ sufficiently large, either $\mu(K_{i})=0$ and $\mu_{n}(A_{i}) \leq \mu_{n}(V_{i}) < \frac{\kappa}{N}$, or $\mu(K_{i})>0$ and
\begin{equation}
\label{upbbnd}
\mu_{n}(A_{i}) \leq \mu_{n}(V_{i}) < (1+\kappa) \mu(K_{i}),\quad i=1,\ldots,N.
\end{equation}
Next, set $I:= \{i\in \{1,\dots,N\}: \mu(K_{i}) > 0\}$ and define a measure $\nu \in \mathcal{M}^{+}(\mathcal{X})$ by
\begin{equation}
\label{nu}
\nu(E) := \mu(E) - (1-\kappa) \sum_{i \in I} \mu(E \cap A_{i}), \quad E \in \mathcal{B}(\mathcal{X}),
\end{equation}
and, also, introduce the sequence of measures $(\nu_{n})_{n \in \mathbb{N}}$ by
\begin{equation}
\label{nun}
\nu_{n}(E) := \mu_{n}(E) - (1 - \kappa) \sum_{i \in I} \mu(A_{i}) \frac{\mu_{n}(E\cap A_{i})}{\mu_{n}(A_{i})},\quad E \in \mathcal{B}(\mathcal{X}).
\end{equation}
Note that $\nu_{n} \in \mathcal{M}^{+}(\mathcal{X})$ for all $n$ large enough since for $i \in I$,
\begin{equation*}
\mu_{n}(A_{i}) = \mu_{n}(V_{i}) - \mu_{n}\left(V_{i} \cap \bigcup_{j=1}^{i-1} V_{j}\right) > \left(\mu(V_{i}) - \frac{1}{2}\kappa \mu(K_{i})\right) - \frac{1}{2} \kappa\mu(K_{i}) > (1 -\kappa) \mu(A_{i}).
\end{equation*}
Here, we used Lemma \ref{lemm2} combined with the upper bound \eqref{intermaj}.  In view of \eqref{nu} and \eqref{nun}, define for all $n$ large enough $\sigma_{n} \in \mathcal{P}(\mathcal{X}\times \mathcal{X})$ as follows
\begin{equation*}
\sigma_{n}(E\times F):= (1-\kappa) \sum_{i \in I} \mu(E \cap A_{i}) \frac{\mu_{n}(F \cap A_{i})}{\mu_{n}(A_{i})} + \frac{\nu(E) \nu_{n}(F)}{\nu(\mathcal{X})},\quad E,F \in \mathcal{B}(\mathcal{X}).
\end{equation*}
It follows from the fact
\begin{equation*}
\nu(\mathcal{X}) = 1 - (1 - \kappa) \sum_{i \in I} \mu(A_{i}) = \nu_{n}(\mathcal{X}),
\end{equation*}
that we have $\sigma_{n} \in \Xi_{\mathcal{X}}(\mu,\mu_{n})$. Moreover, 
\begin{multline*}
\int_{\mathcal{X}\times \mathcal{X}} \rho(x,y) \sigma_{n}(dx,dy)
\\ = (1 - \kappa) \sum_{i \in I}\frac{1}{\mu_{n}(A_{i})} \int_{A_{i} \times A_{i}} \rho(x,y) \mu(dx)\mu_{n}(dy)
+ \frac{1}{\nu(\mathcal{X})} \int_{\mathcal{X} \times \mathcal{X}} \rho(x,y) \nu(dx) \nu_{n}(dy).
\end{multline*}
To conclude the proof of the proposition, it suffices to show that, for all $n$ large enough,
\begin{equation}
\label{2sho}
\int_{\mathcal{X}\times \mathcal{X}} \rho(x,y) \sigma_{n}(dx,dy) < \epsilon.
\end{equation}
On the one hand, since $\mathrm{diam}(A_{i}) \leq \mathrm{diam}(B_{i})= 2r$, we have,
\begin{equation*}
(1 - \kappa) \sum_{i\in I} \frac{1}{\mu_{n}(A_{i})} \int_{A_{i} \times A_{i}} \rho(x,y) \mu(dx)\mu_{n}(dy) \leq 2r (1-\kappa) \sum_{i \in I} \mu(A_{i}) \leq 2r.
\end{equation*}
On the other hand, by the triangle inequality, we have, for all $n$ large enough,
\begin{equation}
\label{spllit}
\frac{1}{\nu(\mathcal{X})} \int_{\mathcal{X} \times \mathcal{X}} \rho(x,y) \nu(dx) \nu_{n}(dy) \leq
\int_{\mathcal{X}} \rho(x,x_{0}) \nu(dx) + \int_{\mathcal{X}} \rho(x_{0},y) \nu_{n}(dy).
\end{equation}
In view of \eqref{nu}, the first term in the right-hand side of \eqref{spllit} can be rewritten as
\begin{equation*}
\int_{\mathcal{X}} \rho(x,x_{0}) \nu(dx) = \int_{\mathcal{X} \setminus \cup_{i \in I} A_{i}} \rho(x,x_{0}) \mu(dx) + \kappa \sum_{i\in I} \int_{A_{i}} \rho(x,x_{0}) \mu(dx),
\end{equation*}
and can be bounded as follows
\begin{multline*}
\int_{\mathcal{X}} \rho(x,x_{0}) \nu(dx)\\
\begin{split}
&\leq \int_{\mathcal{X} \setminus \cup_{i \in I} A_{i}} \rho(x,x_{0}) \mu(dx) + \kappa \left(\int_{(\cup_{i \in I} A_{i})\setminus K} \rho(x,x_{0}) \mu(dx) + \sum_{i \in I} \int_{A_{i} \cap K} \rho(x,x_{0}) \mu(dx)\right) \\
&\leq \int_{K^{c}} \rho(x,x_{0}) \mu(dx) + \kappa \int_{K} \rho(x,x_{0}) \mu(dx).
\end{split}
\end{multline*}
For the second term in the right-hand side of \eqref{spllit}, in view of the definition \eqref{nun}, we have,
\begin{equation*}
\int_{\mathcal{X}} \rho(x,x_{0}) \nu_{n}(dx) \leq  \int_{\mathcal{X}} \rho(x,x_{0}) \mu_{n}(dx) - \frac{1-\kappa}{1 + \kappa} \sum_{i\in I} \int_{A_{i}} \rho(x,x_{0}) \mu_{n}(dx),
\end{equation*}
where we used the bound in \eqref{upbbnd}. It then follows that,
\begin{equation*}
\begin{split}
\int_{\mathcal{X}} \rho(x,x_{0}) \nu_{n}(dx)
\leq  & \int_{\mathcal{X} \setminus \cup_{i \in I} A_{i}} \rho(x,x_{0}) \mu_{n}(dx) \\ &+  2\kappa \left(\int_{(\cup_{i \in I} A_{i})\setminus K} \rho(x,x_{0}) \mu_{n}(dx) + \sum_{i \in I} \int_{A_{i} \cap K} \rho(x,x_{0}) \mu_{n}(dx)\right).
\end{split}
\end{equation*}
Gathering the above estimates together, we eventually get, for all $n$ large enough,
\begin{equation*}
\int_{\mathcal{X}\times \mathcal{X}} \rho(x,y) \sigma_{n}(dx,dy) < 2r + 2 \left(\int_{K^{c}} \rho(x,x_{0})  \mu(dx) + \int_{K^{c}} \rho(x,x_{0}) \mu_{n}(dx)\right) + 3 \kappa \sup_{x \in K} \rho(x,x_{0}).
\end{equation*}
\eqref{2sho} follows by taking $r=\frac{1}{6}\epsilon$ and $\kappa = \frac{1}{9 \max(1,\sup_{x \in K} \rho(x,x_{0}))} \epsilon$. \qed

\subsection{Proof of Theorem \ref{complete} $(i)$.}

\textit{Non-degeneracy}. Let us show that $\rho_{W}(\mu,\nu)=0 \Longrightarrow \mu=\nu$ (the converse is obvious). Pick $\epsilon>0$. From \eqref{rhoW}, there exists a probability measure $\sigma \in \Xi_{\mathcal{X}}(\mu,\nu)$ such that,
\begin{equation}
\label{funde2'}
\int_{\mathcal{X}\times\mathcal{X}} \rho(x,y) \sigma(dx,dy) < \epsilon.
\end{equation}
Denote $\Delta_{\sqrt{\epsilon}} := \{(x,y) \in \mathcal{X}\times\mathcal{X}\,:\,\rho(x,y) < \sqrt{\epsilon}\}$. From \eqref{funde2'}, we have,
\begin{equation}
\label{fundeq3'}
\sigma(\Delta_{\sqrt{\epsilon}}^{c}) < \frac{1}{\sqrt{\epsilon}} \int_{\Delta_{\sqrt{\epsilon}}^{c}} \rho(x,y) \sigma(dx,dy) < \sqrt{\epsilon}.
\end{equation}
Suppose now that $\mu \neq \nu$. Then there exists a non-empty compact set $K \subset \mathcal{X}$ such that $\mu(K)> \nu(K)$ (without loss of generality). Setting $\delta := \frac{1}{2}(\mu(K) - \nu(K))>0$, there exists an open set $O \subset \mathcal{X}$ such that $K \subset O$ and $\mu(K) > \nu(O) + \delta$. We may assume that $O$ is the $\delta$-neighbourhood of $K$. Indeed, if $K_{\delta} \subset O$ then $\mu(K)> \nu(K_{\delta}) + \delta$, and if $\rho(K,O^{c}) < \delta$, then we can replace $\delta$ by $\rho(K,O^{c})$. For the set $(K \times O^{c}) \subset \Delta_{\delta}^{c}$, we have,
\begin{equation*}
\sigma(K \times O^{c}) = \sigma(K \times \mathcal{X}) - \sigma(K \times O) > \sigma(K \times \mathcal{X}) - \sigma(\mathcal{X} \times O) = \mu(K) - \nu(O) > \delta.
\end{equation*}
It follows that $\sigma(\Delta_{\delta}^{c})> \delta$ which contradicts \eqref{fundeq3'} if we set $\delta = \sqrt{\epsilon}$.\\
\textit{Triangle inequality.} Let $\mu_{l} \in \mathcal{P}_{1}(\mathcal{X})$, $l=1,2,3$. We need to show that
\begin{equation}
\label{trine}
\rho_{W}(\mu_{1},\mu_{3}) \leq \rho_{W}(\mu_{1},\mu_{2}) + \rho_{W}(\mu_{2},\mu_{3}).
\end{equation}
Given $\epsilon>0$, there exist $\sigma_{1,2} \in \Xi_{\mathcal{X}}(\mu_{1},\mu_{2})$ and $\sigma_{2,3} \in \Xi_{\mathcal{X}}(\mu_{2},\mu_{3})$ such that,
\begin{equation}
\label{rhoWtwo}
\int_{\mathcal{X}\times \mathcal{X}} \rho(x_{l},x_{l+1}) \sigma_{l,l+1}(dx_{l}, dx_{l+1}) < \rho_{W}(\mu_{l},\mu_{l+1}) + \epsilon,\quad l=1,2.
\end{equation}
Denote by $\mu_{1,2}$ the conditional probability measure defined by (see Section \ref{appX})
\begin{equation}
\label{sigma12}
\sigma_{1,2}(A_{1} \times A_{2}) = \int_{A_{2}} \mu_{1,2}(A_{1}\vert x_{2}) \mu_{2}(dx_{2}), \quad \textrm{$A_{1}, A_{2} \in \mathcal{B}(\mathcal{X})$}.
\end{equation}
Subsequently, we put
\begin{equation}
\label{sigma13}
\sigma_{1,3}(A_{1}\times A_{3}) := \int_{\mathcal{X} \times A_{3}} \mu_{1,2}(A_{1}\vert x_{2}) \sigma_{2,3}(dx_{2},dx_{3}), \quad \textrm{$A_{1}, A_{3} \in \mathcal{B}(\mathcal{X})$}.
\end{equation}
Using the triangle inequality and the definition in \eqref{sigma13}, we have,
\begin{multline*}
\int_{\mathcal{X}\times \mathcal{X}} \rho(x_{1},x_{3}) \sigma_{1,3}(dx_{1},dx_{3})\\
\begin{split}
&\leq \int_{\mathcal{X}\times \mathcal{X} \times \mathcal{X}} \rho(x_{1},x_{2}) \mu_{1,2}(dx_{1}\vert x_{2}) \sigma_{2,3}(dx_{2},dx_{3}) + \int_{\mathcal{X}\times \mathcal{X}} \rho(x_{2},x_{3}) \mu_{1,2}(\mathcal{X}\vert x_{2}) \sigma_{2,3}(dx_{2},dx_{3}) \\
&= \int_{\mathcal{X}\times \mathcal{X}} \rho(x_{1},x_{2}) \mu_{1,2}(dx_{1}\vert x_{2}) \mu_{2}(dx_{2}) + \int_{\mathcal{X}\times \mathcal{X}} \rho(x_{2},x_{3}) \sigma_{2,3}(dx_{2},dx_{3})
\end{split}
\end{multline*}
Using \eqref{sigma12} in the first term of the above right-hand side, this yields
\begin{equation*}
\int_{\mathcal{X}\times \mathcal{X}} \rho(x_{1},x_{3}) \sigma_{1,3}(dx_{1},dx_{3})
\leq \int_{\mathcal{X}\times \mathcal{X}} \rho(x_{1},x_{2}) \sigma_{1,2}(dx_{1},dx_{2}) + \int_{\mathcal{X}\times \mathcal{X}} \rho(x_{2},x_{3})\sigma_{2,3}(dx_{2},dx_{3}).
\end{equation*}
To obtain \eqref{trine}, it remains to use \eqref{rhoWtwo} and take the limit $\epsilon \rightarrow 0$.

\subsection{Proof of Theorem \ref{complete} $(ii)$.}

Given a non-empty compact set $Q$ and a real $\alpha >0$, in the following we denote
\begin{equation}
\label{Keta}
Q_{\alpha} := \{x \in \mathcal{X}: \mathrm{dist}(x,Q) < \alpha\}.
\end{equation}
Let $(\mu_{n})_{n \in \mathbb{N}}$ be a Cauchy sequence in $\mathcal{P}_{1}(\mathcal{X})$. This means that, given $\delta >0$, there exists $N \in \mathbb{N}$ such that $\rho_{W}(\mu_{n},\mu_{m}) < \frac{\delta^2}{2}$ for all $n,m \geq N$. As a result, there exists  a coupling $\sigma_{n,m} \in \Xi_{\mathcal{X}}(\mu_{n},\mu_{m})$ for each $n,m \geq N$ such that,
\begin{equation}
\label{goo1}
\int_{\mathcal{X} \times \mathcal{X}} \rho(x,y) \sigma_{n,m}(dx,dy) \leq \delta^{2}.
\end{equation}
By setting $\Delta_{\delta} := \{(x,y) \in \mathcal{X}\times \mathcal{X}: \rho(x,y) < \delta\}$, \eqref{goo1} implies that $\sigma_{n,m}(\Delta_{\delta}^{c}) < \delta$, see \eqref{fundeq3'}. \\
Further, for any uniformly continuous function $f: \mathcal{X} \rightarrow \mathbb{R}$ satisfying
\begin{equation}
\label{Lipt}
\vert f(x) - f(y)\vert < C \rho(x,y),
\end{equation}
for some constant $C>0$, we have, for all $n,m \geq N$,
\begin{equation}
\label{unifcobd}
\left \vert \int_{\mathcal{X}} f(x) \mu_{n}(dx) - \int_{\mathcal{X}} f(y) \mu_{m}(dy)\right\vert = \left \vert \int_{\Delta_{\delta} \cup \Delta_{\delta}^{c}} (f(x) - f(y)) \sigma_{n,m}(dx,dy)  \right\vert< (C + 2 \Vert f \Vert_{\infty}) \delta.
\end{equation}
\indent Fix $0 < \epsilon < 1$. Let $\delta_{0}:= \frac{\epsilon^{2}}{12}>0$ and $n_{0} \in \mathbb{N}$ such that $\rho_{W}(\mu_{n},\mu_{m}) < \frac{\delta_{0}^{2}}{2}$ for all $n,m \geq n_{0}$.  Choose a compact set $K_{0}$ such that $\mu_{n_{0}}(K_{0}^{c}) < \frac{\epsilon}{2}$. Put $O_{0} := K_{0,\epsilon}$ (see \eqref{Keta}) and introduce
\begin{equation*}
f_{0}(x) := \left\{\begin{array}{ll}
1 - \frac{1}{\epsilon} \mathrm{dist}(x,K_{0}),\,\, &\textrm{if $\mathrm{dist}(x,K_{0}) \leq \epsilon$}\\
0, &\textrm{if $\mathrm{dist}(x,K_{0}) > \epsilon$}
\end{array}\right.,
\end{equation*}
where it is understood that $f_{0} = 1$ on $K_{0}$. Clearly, $f_{0}$ is uniformly continuous and satisfies \eqref{Lipt} with $C = \frac{1}{\epsilon}$. It follows from \eqref{unifcobd} that, for all $n \geq n_{0}$,
\begin{equation*}
\mu_{n}(O_{0}) > \int_{\mathcal{X}} f_{0}(x) \mu_{n}(dx) > \int_{\mathcal{X}} f_{0}(x) \mu_{n_{0}}(dx) - \left(2 + \frac{1}{\epsilon}\right)\delta_{0} > \mu_{n_{0}}(K_{0}) - \left(2 + \frac{1}{\epsilon}\right)\delta_{0}.
\end{equation*}
Since $\mu_{n_{0}}(K_{0}) > 1 -\frac{\epsilon}{2}$, we find that $\mu_{n}(O_{0}^{c}) < \frac{3}{4} \epsilon$ for all $n \geq n_{0}$.\\
Next, we proceed by induction to construct a non-decreasing subsequence $(n_{i})_{i \in \mathbb{N}}$ and compact sets $K_{i}$, $i \in \mathbb{N}$ such that, denoting $O_{i}:= K_{i,2^{-i} \epsilon}$ (see \eqref{Keta}), we have, for all $n \geq n_{i}$,
\begin{equation*}
\mu_{n_{i}}(K_{i}^{c}) < \left(1 - \frac{1}{2^{i+1}}\right)\epsilon \quad \textrm{and} \quad \mu_{n}(O_{i}^{c})< \left(1 - \frac{1}{2^{i+2}}\right)\epsilon.
\end{equation*}
Suppose that we have found $n_{i-1} \in \mathbb{N}$ and we have constructed a compact set $K_{i-1}$ such that $\mu_{n_{i-1}}(K_{i-1}^{c}) < (1 - \frac{1}{2^{i}})\epsilon$ and $\mu_{n}(O_{i-1}^{c}) < (1 - \frac{1}{2^{i+1}})\epsilon$ for all $n \geq n_{i-1}$. Let $\delta_{i}:= \frac{\epsilon^{2}}{2^{i+3}(1 + 2^{i-1})}>0$ and $n_{i} \in \mathbb{N}$ such that $\rho_{W}(\mu_{n},\mu_{m}) < \frac{\delta_{i}^{2}}{2}$ for all $n,m \geq n_{i}$. We choose $K_{i} \subset O_{i-1}$ compact such that $K_{i-1} \subset K_{i}$ and $\mu_{n_{i}}(K_{i}^{c}) < (1 - \frac{1}{2^{i+1}})\epsilon$, and we put $O_{i} = K_{i,2^{-i}\epsilon}$. Then, setting
\begin{equation*}
f_{i}(x) := \left\{\begin{array}{ll}
1 - \frac{2^{i}}{\epsilon} \mathrm{dist}(x,K_{i}),\,\, &\textrm{if $\mathrm{dist}(x,K_{i}) \leq \frac{\epsilon}{2^{i}}$}\\
0, &\textrm{if $\mathrm{dist}(x,K_{i}) > \frac{\epsilon}{2^{i}}$}
\end{array}\right.,
\end{equation*}
it follows from \eqref{unifcobd} that, for all $n \geq n_{i}$,
\begin{equation*}
\mu_{n}(O_{i}) > \int_{\mathcal{X}} f_{i}(x) \mu_{n}(dx) > \int_{\mathcal{X}} f_{i}(x) \mu_{n_{i}}(dx) - \left(2 + \frac{2^{i} }{\epsilon}\right)\delta_{i} > \mu_{n_{i}}(K_{i}) - \left(2 + \frac{2^{i}}{\epsilon}\right)\delta_{i}.
\end{equation*}
It remains to use that $\mu_{n}(K_{i}) > 1 - (1 - \frac{1}{2^{i+1}})\epsilon$ to conclude this induction step.\\
\indent Define $K := \overline{\bigcup_{i \in \mathbb{N}} K_{i}}$ and let us show that $K$ is compact. Fix  $0<\eta<1$ and let $i_{0} \in \mathbb{N}$ such that $\frac{1}{2^{i_{0}-3}} < \eta$. We can cover $K_{i_{0}}$ by finitely many balls of radius $\frac{\eta}{2}$ denoted by $B_{\frac{\eta}{2}}^{(l)}$, $l=1,\dots,M$. Since $K_{j} \subset O_{j-1}$, $\mathrm{dist}(K_{j},K_{j-1}) < \frac{1}{2^{j-1}}$. It follows by induction that $\mathrm{dist}(K_{j},K_{i}) < \frac{1}{2^{i-1}}$ for all $j>i$. Now, if $x' \in K$, there exists $j \geq i_{0}$ such that $\mathrm{dist}(x',K_{j})<\frac{\eta}{4}$ and hence $\mathrm{dist}(x',K_{i_{0}})< \frac{\eta}{4} + \frac{1}{2^{i_{0}-1}} < \frac{\eta}{2}$. Therefore, there exists $x \in K_{i_{0}}$ with $\rho(x,x')< \frac{\eta}{2}$, and there is a ball $B_{\frac{\eta}{2}}^{(l_{0})}$ with $x \in B_{\frac{\eta}{2}}^{(l_{0})}$ so that $x' \in B_{\eta}^{(l_{0})}$. We conclude that $K$ is covered by the balls $B_{\eta}^{(l)}$ with double the radius. This means that $K$ is totally bounded, and since $\mathcal{X}$ is complete, $K$ is compact.\\
\indent Consider now the sequence of probability measures $(\mu_{n_{i}})_{i \in \mathbb{N}}$. From the foregoing, we have found a compact set $K$ such that $\mu_{n_{i}}(K^{c}) < \epsilon$ for all $i \in \mathbb{N}$. By Prokhorov's theorem, see e.g. \cite[Sec. IX.5.5]{Bou}, there exists a subsequence $(\mu_{m_{k}})_{k \in \mathbb{N}}$, with $\mu_{m_{k}} = \mu_{n_{i_{k}}}$ converging weakly to a probability measure $\mu \in \mathcal{P}(\mathcal{X})$. Let us show that $\mu \in \mathcal{P}_{1}(\mathcal{X})$. Note that  $\mu_{m_{k}} \in \mathcal{P}_{1}(\mathcal{X})$ for all $k \in \mathbb{N}$ by assumption.
Further, since $\rho_{W}(\mu_{m},\mu_{m_{1}}) < \frac{\delta_{1}^{2}}{2}$ for all $m \geq m_{1} \geq n_{1}$, by mimicking the arguments leading to \eqref{goo1}, there exists, for each $m$, a coupling $\sigma_{m,m_{1}} \in \Xi_{\mathcal{X}}(\mu_{m},\mu_{m_{1}})$ such that,
\begin{equation*}
\int_{\mathcal{X} \times \mathcal{X}} \rho(x,y) \sigma_{m,m_{1}}(dx,dy) \leq \delta_{1}^{2} < 1.
\end{equation*}
As a result, given $x_{0} \in \mathcal{X}$, there exists a constant $c_{m_{1}}>0$ such that, for $k \in \mathbb{N}$, $k \geq 2$,
\begin{equation}
\label{goo2}
\int_{\mathcal{X}} \rho(x,x_{0}) \mu_{m_{k}}(dx) \leq \int_{\mathcal{X} \times \mathcal{X}} \rho(x,y) \sigma_{m_{k},m_{1}}(dx,dy) + \int_{\mathcal{X}} \rho(y,x_{0}) \mu_{m_{1}}(dy) < 1 + c_{m_{1}}.
\end{equation}
This means, in particular, that the left-hand side of \eqref{goo2} is uniformly bounded for all $k \in \mathbb{N}$. By virtue of the monotone convergence theorem, we conclude that
\begin{equation*}
\int_{\mathcal{X}} \rho(x,x_{0}) \mu(dx) \leq 1+ c_{m_{1}},
\end{equation*}
and hence $\mu \in \mathcal{P}_{1}(\mathcal{X})$. We can now apply Proposition \ref{prop1} $(ii)$ giving that $\rho_{W}(\mu_{m_{k}},\mu) \rightarrow 0$ when $k \rightarrow \infty$. Using the triangle inequality, we conclude that $\rho_{W}(\mu_{m},\mu) \rightarrow 0$ when $m \rightarrow \infty$. \qed

\section{Acknowledgement.} 
The authors are grateful to the referees, whose remarks led us to make significant improvements in our original manuscript. 

\section{Appendix.}
\label{appX}

\subsection{Disintegration theorem.}
\label{disintg}

We recall the disintegration theorem for the existence of conditional probability measures, see, e.g.,
\cite[Sec. IX.2.7]{Bou}. Given a topological space $(\mathcal{X},\tau)$, let $\mathcal{B}(\mathcal{X})$ denote the Borel $\sigma$-algebra of subsets of $\mathcal{X}$.

\begin{theorem}
\label{existence}
Let $\mathcal{X}$ be a completely regular Hausdorff space and assume that the compact subspaces of $\mathcal{X}$ are metrizable. Let $\mathcal{Y}$ be a Hausdorff space. Let $\mu \in \mathcal{P}(\mathcal{X})$ be a Radon probability measure and $\pi: \mathcal{X} \rightarrow \mathcal{Y}$ be a $\mu$-measurable function. Let $\nu \in \mathcal{P}(\mathcal{Y})$ be the image measure $\nu = \pi_{*}(\mu) = \mu \circ \pi^{-1}$. Then there exists a map $\mu_{y} : \mathcal{Y} \rightarrow \mathcal{P}(\mathcal{X})$,
$y \mapsto \mu_{y}$ such that for all $A \in \mathcal{B}(\mathcal{Y})$ and all bounded Borel-measurable functions $f: \mathcal{X} \rightarrow \mathbb{R}$,
\begin{equation}
\label{iden0}
\int_{A} \int_{\mathcal{X}} f(x) \mu_{y}(dx) \nu(dy) = \int_{\pi^{-1}(A)} f(x) \mu(dx).
\end{equation}
The above map is a.e. unique and $\mu_{y}$ is concentrated on $\pi^{-1}(\{y\})$ for a.e. $y$. One usually writes $\mu_{y}(\cdot\,) = \mu(\cdot\,\vert y)$. Further, if $f \in \mathcal{L}^{1}(\mu)$ then $f \in \mathcal{L}^{1}(\mu_{y})$ for a.e. $y$, and the identity \eqref{iden0} still holds.
\end{theorem}

\begin{remark}
Taking the indicator function $f = \mathbf{1}_{E}$, $E \subseteq \mathcal{X}$ and choosing $A=\mathcal{Y}$ in \eqref{iden0} yields
\begin{equation*}
\int_{\mathcal{Y}} \mu_{y}(E) \nu(dy) = \mu(E).
\end{equation*}
\end{remark}

The special case where $\mathcal{X}$ is a product space is particularly important. Considering $\mathcal{X} = \mathcal{X}_{1} \times \mathcal{X}_{2}$ and considering the projection map onto one of the factors, for instance $\pi_{2}: \mathcal{X} \mapsto \mathcal{X}_{2}$, all spaces $\pi_{2}^{-1}(\{y\})$ are equivalent to $\mathcal{X}_{1}$. Given $\mu \in \mathcal{P}(\mathcal{X})$ a Radon probability measure,  we can thus define $\mu_{y}'(A) := \mu_{y}(A \times \{y\})$ for any $A \in \mathcal{B}(\mathcal{X}_{1})$. In the case of probability measures, we have in fact:

\begin{corollary}
Let $\mathcal{X}=\mathcal{X}_{1} \times \mathcal{X}_{2}$ be a product of completely regular Hausdorff spaces. Assume in addition that the compact subspaces of $\mathcal{X}_{1}$ are metrizable. Let $\mu \in \mathcal{P}(\mathcal{X})$ be a Radon probability measure and $\pi_{2}: \mathcal{X} \rightarrow \mathcal{X}_{2}$ be the projection map. Let $\nu \in \mathcal{P}(\mathcal{X}_{2})$ be the image measure $\nu = \mu \circ \pi_{2}^{-1}$. Then there exists a map $\mu_{y} : \mathcal{X}_{2} \rightarrow \mathcal{P}(\mathcal{X}_{1})$,
$y \mapsto \mu_{y}$ such that for all $B \in \mathcal{B}(\mathcal{X}_{2})$ and all Borel-measurable functions $f \in \mathcal{L}^{1}(\mu)$,
\begin{equation}
\label{desintr}
\int_{B} \nu(dy) \int_{\mathcal{X}_{1}} f(x,y) d\mu_{y}(dx) = \int_{\mathcal{X}_{1} \times B} f(x,y) d\mu(dx,dy).
\end{equation}
\end{corollary}

The probability measure $\mu_{y}$ is called the conditional measure on $\mathcal{X}_{1}$ and is usually denoted $\mu(\cdot\,\vert y)$, i.e. $\mu(A \vert y)$ is the probability of $A \in \mathcal{B}(\mathcal{X}_{1})$ given that $\pi_{2}(\underline{x}) = y$.

\subsection{Duality.}
\label{duality}

\subsubsection{The dual problem in linear programming.}

There is a dual problem corresponding to the linear programming problem, see Section \ref{linprogr}.  
This can be formulated more generally as follows, see, e.g., \cite{BP}.\\
\indent Suppose that $F: \mathbb{R}^{n} \times \mathbb{R}^{m} \to \mathbb{R} \cup \{+\infty\}$ is a lower semi-continuous convex function, and set $f(\underline{x}) := F(\underline{x},\underline{0})$. Consider the minimization problem
$\inf_{\underline{x} \in \mathbb{R}^{n}} f(\underline{x})$. The dual problem is defined as 
\begin{equation*}
\sup_{\underline{v} \in \mathbb{R}^{m}} \left(-F^{*}(\underline{0},\underline{v})\right),
\end{equation*}
where $F^{*}:  \mathbb{R}^{n} \times \mathbb{R}^{m} \to \mathbb{R} \cup \{+\infty\}$ is the Legendre transform of $F$, see, e.g., \cite[Section 12]{Rok},  
\begin{equation} 
\label{Legendre}
F^{*}\left(\underline{u}, \underline{v}\right) := \sup_{(\underline{x},\underline{y}) \in \mathbb{R}^{n} \times \mathbb{R}^{m}}
\left(\left\langle \underline{x},\underline{u} \right\rangle_{\mathbb{R}^{n}} + \left\langle \underline{y}, \underline{v} \right\rangle_{\mathbb{R}^{m}} - F\left(\underline{x}, \underline{y}\right)\right).
\end{equation}
The equivalence of these problems follows from the Legendre inversion theorem and is defined by
\begin{equation*} 
-\infty < \sup_{\underline{v} \in \mathbb{R}^{m}} \left(-F^{*}\left(\underline{0},\underline{v}\right)\right) = \inf_{\underline{x} \in \mathbb{R}^{n}} f\left(\underline{x}\right) < +\infty. 
\end{equation*}

To prove the equivalence, we define
\begin{equation}
\label{defhLT}
h\left(\underline{y}\right) := \inf_{\underline{x} \in \mathbb{R}^{n}} F\left(\underline{x},\underline{y}\right).
\end{equation}

\begin{lema} 
\label{lemconvex}
The function $h: \mathbb{R}^{m} \to \mathbb{R}\cup\{-\infty,+\infty\}$ is convex, and $h^{*}(\underline{v}) = F^{*}(0,\underline{v})$.
\end{lema}

\noindent \textbf{Proof.} Let $(\underline{y}_{1}, \underline{y}_{2}) \in \mathbb{R}^{m}\times \mathbb{R}^{m}$ and let $\epsilon > 0$. There exist $(\underline{x}_{1}, \underline{x}_{2}) \in \mathbb{R}^{n}\times \mathbb{R}^{n}$ such that 
\begin{equation*}
h\left(\underline{y}_{i}\right) \leq F\left(\underline{x}_{i}, \underline{y}_{i}\right) \leq h\left(\underline{y}_{i}\right) + \epsilon, \quad i=1,2.
\end{equation*}
Then, if $\lambda_{1}, \lambda_{2} \geq 0$ with $\lambda_{1} + \lambda_{2} = 1$, we have,
\begin{align*} 
h\left(\lambda_{1} \underline{y}_{1} + \lambda_{2} \underline{y}_{2}\right) &=
\inf_{\underline{x} \in \mathbb{R}^{n}} F\left(\underline{x},\lambda_{1} \underline{y}_{1} + \lambda_{2} \underline{y}_{2}\right) \\
&\leq F\left(\lambda_{1} \underline{x}_{1} + \lambda_{2} \underline{x}_{2}, \lambda_{1} \underline{y}_{1} + \lambda_{2} \underline{y}_{2}\right) \\ 
&\leq \lambda_{1} F\left(\underline{x}_{1}, \underline{y}_{1}\right) + \lambda_{2} F\left(\underline{x}_{2}, \underline{y}_{2}\right) 
\leq \lambda_{1} h\left(\underline{y}_{1}\right) + \lambda_{2} h\left(\underline{y}_{2}\right) + \epsilon. 
\end{align*}
This proves the convexity since $\epsilon > 0$ is arbitrary. We now compute the Legendre transform
\begin{align*}	 
h^{*}\left(\underline{v}\right) = \sup_{\underline{y} \in \mathbb{R}^{m}} \left(\left\langle  \underline{y}, \underline{v} \right\rangle_{\mathbb{R}^{m}}
- h\left(\underline{y}\right) \right) &= \sup_{\underline{y} \in \mathbb{R}^{m}} \left(\left\langle  \underline{y}, \underline{v} \right\rangle_{\mathbb{R}^{m}} - \inf_{\underline{x} \in \mathbb{R}^{n}} F\left(\underline{x}, \underline{y}\right) \right) \\
&= \sup_{(\underline{x} ,\underline{y}) \in \mathbb{R}^{n} \times \mathbb{R}^{m}} \left(\left\langle  \underline{y}, \underline{v} \right\rangle_{\mathbb{R}^{m}} - 
F\left(\underline{x}, \underline{y}\right) \right) = F^{*}\left(\underline{0},\underline{v}\right). \tag*{\qed}
\end{align*} 

An immediate corollary is
\begin{lema} 
The dual problem is given by 
\begin{equation} 
\label{hdoubstar}	
\sup_{\underline{v} \in \mathbb{R}^{m}} \left(-F^{*}\left(\underline{0},\underline{v}\right)\right) = h^{**}(0). 
\end{equation}
\end{lema}

We now first note that

\begin{lema} We have
\begin{equation*}
\sup_{\underline{v} \in \mathbb{R}^{m}} \left(-F^{*}\left(\underline{0},\underline{v}\right)\right) \leq 
	\inf_{\underline{x} \in \mathbb{R}^{n}} f\left(\underline{x}\right). 
\end{equation*}
\end{lema}

\noindent \textbf{Proof.} This follows from $F^{*}(\underline{0},\underline{v}) \geq -F(\underline{x}, \underline{0}) = -f(\underline{x})$ for any 
$\underline{x} \in \mathbb{R}^{n}$. \qed \\

We next prove the equivalence.

\begin{theorem} 
\label{thmdualinf}
The dual problem $\sup_{\underline{v} \in \mathbb{R}^{m}} (-F^{*}(\underline{0},\underline{v}))$ is equivalent to the minimization problem 
$\inf_{\underline{x} \in \mathbb{R}^{n}} f(\underline{x})$ with $f(\underline{x}):= F(\underline{x},\underline{0})$, i.e.,
\begin{equation*}
\sup_{\underline{v} \in \mathbb{R}^{m}} \left(-F^{*}(\underline{0},\underline{v})\right) = \inf_{\underline{x} \in \mathbb{R}^{n}} f\left(\underline{x}\right),
\end{equation*}
provided that $h$ defined in \eqref{defhLT} satisfies: $h(\underline{0})$ is finite and $h$ is lower semi-continuous at $\underline{y} =\underline{0}$. 
\end{theorem} 

\noindent \textbf{Proof of Theorem \ref{thmdualinf}}. By the Legendre inversion theorem, $h^{**}$ is the closed convex hull of $h$, see, e.g., \cite[Theorem C.5]{Do}.
Since $h$ is lower semi-continuous at $\underline{y}=\underline{0}$ by assumption and $h$ is convex by Lemma \ref{lemconvex}, it follows that $h^{**}(\underline{0}) = h(\underline{0})$. The Theorem now follows from \eqref{hdoubstar}. \qed 

\subsubsection{Application to the linear programming problem.}

The original problem consists in minimizing $\langle \underline{c}, \underline{x} \rangle_{\mathbb{R}^{n}}$ where $\underline{c} \in \mathbb{R}^{n}$ is a given constant vector, subject to the conditions $\underline{x} \geq \underline{0}$, i.e., $x_{j} \geq 0$, $j=1,\ldots,n$, and $A \underline{x} = \underline{b}$, where $\underline{b} \in \mathbb{R}^{m}$ is a given vector of lower dimension $m < n$ and $A$ is an $m \times n$ matrix. This can be reformulated as follows.\\
Define two functions $f: \mathbb{R}^{n} \to \mathbb{R} \cup \{+\infty\}$ and $g: \mathbb{R}^{m} \to \mathbb{R} \cup \{+\infty\}$ by
\begin{equation*} 
f\left(\underline{x}\right) := 
\begin{cases} \left\langle \underline{x}, \underline{c} \right\rangle_{\mathbb{R}^{n}} &\text{if $\underline{x} \geq \underline{0}$;} \\
+\infty &\text{otherwise}; 
\end{cases} 
\end{equation*}
and 
\begin{equation*} 
g\left(\underline{y}\right) := 
\begin{cases} 0 &\text{if $\underline{y} = \underline{b}$;} \\ +\infty 
&\text{otherwise.} 
\end{cases} 
\end{equation*} 
Then set 
\begin{equation*} 
F\left(\underline{x}, \underline{y}\right) := f\left(\underline{x}\right) + g\left(A \underline{x} - \underline{y}\right). 
\end{equation*}
It is easy to see that $F$ is convex and lower semi-continuous. The original problem $\inf_{\underline{x} \in \mathbb{R}^{n}} F(\underline{x},\underline{0})$ is the given linear programming problem. By Theorem \ref{thmdualinf}, the dual problem is 
$\sup_{\underline{v} \in \mathbb{R}^{m}} (-F^{*}(\underline{0},\underline{v}))$. Now we compute $F^{*}$:
\begin{align*} 
F^{*}\left(\underline{u}, \underline{v}\right) &= \sup_{(\underline{x},\underline{y}) \in \mathbb{R}^{n} \times \mathbb{R}^{m}}
\left(\left\langle \underline{x}, \underline{u} \right\rangle_{\mathbb{R}^{n}} + \left\langle \underline{y}, \underline{v} \right\rangle_{\mathbb{R}^{m}} - f\left(\underline{x}\right) - g\left(A \underline{x} - \underline{y}\right) \right) \\ 
&= \sup_{(\underline{x},\underline{y}') \in \mathbb{R}^{n} \times \mathbb{R}^{m}}
\left(\left\langle \underline{x}, \underline{u} \right\rangle_{\mathbb{R}^{n}} + \left\langle A \underline{x} - \underline{y}', \underline{v} \right\rangle_{\mathbb{R}^{m}} - f\left(\underline{x}\right) - g\left(\underline{y}'\right)\right)\\ 
&= \sup_{\underline{x} \in \mathbb{R}^{n}} \left(\left\langle \underline{x}, \underline{u} + A^T \underline{v} \right\rangle_{\mathbb{R}^{n}} - f\left(\underline{x}\right)\right) + \sup_{\underline{y}' \in \mathbb{R}^{m}} \left(-\left\langle \underline{y}', \underline{v} \right\rangle_{\mathbb{R}^{m}} - g\left(\underline{y}'\right)\right) \\ 
&= \sup_{\underline{x} \geq \underline{0}} \left(\left\langle \underline{x}, \underline{u} + A^T \underline{v} \right\rangle_{\mathbb{R}^{n}} - \left\langle \underline{x}, \underline{c} \right\rangle_{\mathbb{R}^{n}}\right) - \left\langle \underline{b}, \underline{v} \right\rangle_{\mathbb{R}^{m}}  \\
&= \begin{cases} 
-\left\langle \underline{b}, \underline{v} \right\rangle_{\mathbb{R}^{m}} &\text{if $\underline{u} + A^T \underline{v} \leq 
\underline{c} $; } \\ +\infty &\text{otherwise.} 
\end{cases} 
\end{align*}
The dual problem  therefore consists in maximizing $\langle \underline{b}, \underline{v} \rangle_{\mathbb{R}^{m}}$ for $\underline{v} \in \mathbb{R}^{m}$
subject to the condition that $A^{T} \underline{v} \leq \underline{c}$, i.e., $(A^{T} \underline{v})_{j} \leq c_{j}$, $j=1,\ldots,n$. Note that there is no positivity condition.\\
\indent As a special case, consider the case of the Wasserstein metric, see, e.g., proof of Lemma \ref{Pottlem}. The problem consists in minimizing 
\begin{equation*}
\sum_{k=m+1}^{n} \sum_{l=1}^{m} \rho_{k,l} p_{k,l},
\end{equation*}
for the $\sigma$-probabilities $p_{k,l}$ subject to the conditions 
\begin{equation*}
\sum_{l=1}^{m} p_{k,l} = p_{k}'-p_{k} \quad \textrm{and} \quad \sum_{k=m+1}^{n} p_{k,l} = p_{l}- p_{l}'.
\end{equation*}
The corresponding matrix $A$ is an $n \times (n-m)m$ matrix and has the following form 
\begin{equation*}
A_{j,(k,l)} := \begin{cases} 
1 &\textrm{if $j=l$;} \\ 
1 &\textrm{if $j=k$;} \\ 
0 &\textrm{otherwise,}
\end{cases}
\end{equation*}
where $j \in \{1,\ldots,n\}$, $l \in \{1,\ldots,m\}$ and $k \in \{m+1,\ldots,n\}$. Here, we assume that $p_{j} < p_{j}'$ for all $1 \leq j\leq m$ and $p_{j}' < p_{j}$ for all $m+1 \leq j \leq n$. The corresponding vector $\underline{b}\in \mathbb{R}^{n}$ is given by 
\begin{equation*}
b_{j} = 
\begin{cases} 
p_{l}' - p_{l} &\textrm{if $j = l \leq m$}; \\ 
p_{k} - p_{k}' &\textrm{if $j=k \geq m+1$}.
\end{cases}
\end{equation*}
It is in fact more convenient to multiply the first $m$ rows of $A$ by $-1$ to obtain
\begin{equation*}
\tilde{A}_{j,(k,l)} = 
\begin{cases} 
-1 &\textrm{if $j=l\leq m$}; \\ 
1  &\textrm{if $j=k \geq m+1$}; \\ 
 0 &\textrm{otherwise}; 
\end{cases}
\end{equation*}
with $b_{j} = p_{j}-p_{j}'$ for all $j=1,\ldots,n$.\\
The dual problem then consists in maximizing
\begin{equation}
\label{2maximKR}
\sum_{k=1}^{n} \left(p_{k} - p_{k}'\right)f_{k},
\end{equation}
with respect to the coefficients $f_{k}$ subject to the conditions 
\begin{equation} 
\label{2maximcKR}
f_{k} - f_{l}\leq \rho_{k,l}, \quad k=m+1,\ldots,n; \quad  l=1,\ldots,m.
\end{equation}
This is in fact equivalent to the Kantorovich-Rubinstein formulation which states that the Wasserstein distance equals 
the maximum of \eqref{2maximKR} with respect to the coefficients $f_{k}$ subject to the conditions $\vert f_{k} - f_{l}\vert \leq \rho_{k,l}$ for all $k,l \in \{1,\ldots,n\}$. To prove this, we show that maximising \eqref{2maximKR} subject to the conditions \eqref{2maximcKR} is obtained when $\mathrm{(i)}$ and $\mathrm{(ii)}$ below hold\\
$\mathrm{(i)}$
\begin{equation*}
\vert f_{k}-f_{k'} \vert \leq \rho_{k,k'} \,\,\, \textrm{for $k,k' \in  \{m+1,\ldots,n\}$} \quad \textrm{and} \quad  \vert f_{l}-f_{l'}\vert \leq \rho_{l,l'} \,\,\, \textrm{for $l,l' \in \{1,\ldots, m\}$}; 
\end{equation*}
$\mathrm{(ii)}$
\begin{equation*}
\vert f_{k} - f_{l}\vert \leq \rho_{k,l}.
\end{equation*}
To prove (i), we arrange the  $f_{k}$ in decreasing order, i.e.,
$f_{k_{1}} \geq f_{k_{2}} \geq \cdots \geq f_{k_{n-m}}$. 
Suppose that $f_{k_{1}} - f_{k_{2}} > \rho_{k_{1},k_{2}}$ and let $f_{l}+\rho_{k_{2},l}=\min_{1 \leq j \leq m}(f_{j}+\rho_{k_{2},j})$. Then we can increase $f_{k_{2}}$ until it equals $f_{l}+\rho_{k_{2},l}$. It then follows that 
\begin{equation*}
f_{k_{1}} - f_{k_{2}} \leq \left(f_{k_{1}} - f_{l}\right) - \left(f_{k_2}-f_{l}\right) \leq \rho_{k_{1},l} - \rho_{k_{2},l} \leq \rho_{k_{1},k_{2}}. 
\end{equation*}
Note that, if $f_{k_{1}} < f_{l} + \rho_{k_{2},l}$ then we need to invert the order of $f_{k_{1}}$ and $f_{k_{2}}$. We can continue this process. If $f_{k_{2}} - f_{k_{3}} > \rho_{k_{2},k_{3}}$ then we can increase $f_{k_{3}}$ 
until it equals $\min_{j \geq 1}(f_{j} + \rho_{k_{3},j})$, etc. In the same way we can order the $f_{l}$, $1 \leq l \leq m$ in increasing order and argue that $f_{l_{r}} - f_{l_{r-1}} \leq \rho_{l_{r},l_{r-1}}$.\\
To prove (ii), suppose that $f_{k_{i}} \geq f_{l_{m}} > f_{k_{i-1}}$ for some $i>m+1$. Decreasing $f_{l_{m}}$ until it equals $\min_{j\leq i} (f_{k_{j}} - \rho_{k_{j},l_{m}})$ increases the expression
$\sum_{k=1}^{n} (p_{k}-p_{k}')f_{k}$ while decreasing the differences $f_{l_{m}}-f_{k_{r}}$ ($r < i$). Then $f_{l_{m}} - f_{k_{r}} \leq (f_{k_{j}} - f_{k_{r}}) - (f_{k_{j}} - f_{l_{m}}) \leq \rho_{k_{j},k_{r}} - \rho_{k_{j},l_{m}} \leq \rho_{k_{r},l_{m}}$. We can repeat this process again with $f_{l_{m-1}}$, etc. 

\subsubsection{Kantorovich-Rubinstein duality theorem.} 
\label{Kanto}

The above analysis can be generalized to the general case (for an alternative proof, see, e.g., \cite{Ed}).

\begin{theorem}
\label{KanRub} 
Let $\mu$ and $\nu$ be two probability measures on a metric space $(\mathcal{X},\rho)$ such that
\begin{equation*}
\int_{\mathcal{X}} \rho(x,x_{0})\,\mu(dx) <+\infty \quad \textrm{and} \quad \int_{\mathcal{X}} \rho(x,x_{0})\,\nu(dx) < +\infty, \quad \textrm{for some $x_{0} \in \mathcal{X}$}.
\end{equation*} 
Then the Wasserstein metric equals the Kantorovich-Rubinstein metric, i.e., 
\begin{equation} 
\rho_{W}(\mu,\nu) = \sup_{f \in {\mathrm{Lip}}_{1}(\mathcal{X})} \left(\int_{\mathcal{X}} f(x)\,\mu(dx) - \int_{\mathcal{X}} f(x)\,\nu(dx)\right),
\end{equation}
where $\mathrm{Lip}_{1}(\mathcal{X})$ denotes the space of  1-Lipschitz functions on $\mathcal{X}$.
\end{theorem}

\noindent \textbf{Proof of Theorem \ref{KanRub}.} Consider the signed measure $\mu-\nu$. Its Hahn decomposition, see, e.g., \cite[Thm 32.1]{Billi2}, can be written as 
\begin{equation*}  
\mu - \nu = (\mu-\nu)_{+} - (\mu-\nu)_{-},
\end{equation*}
where
\begin{equation*}
(\mu-\nu)_{+} := (\mu-\nu)\vert_{\mathcal{X}_{+}}, \quad (\mu-\nu)_{-} := (\nu-\mu)\vert_{\mathcal{X}_{-}},
\end{equation*}
with
\begin{equation*}
\mathcal{X}_{+} \cup \mathcal{X}_{-} = \mathcal{X} \quad \textrm{and} \quad  \mathcal{X}_{+} \cap \mathcal{X}_{-} = \emptyset. 
\end{equation*}
Now suppose that $\sigma \in \Xi(\mu,\nu)$. Let $\sigma_{\pm \pm}$ denote the restrictions of $\sigma$ to $\mathcal{X}_{\pm} \times \mathcal{X}_{\pm}$.\\ We first show that, on $(\mathcal{X}_{+} \times \mathcal{X}_{+}) \cup (\mathcal{X}_{-} \times \mathcal{X}_{+}) \cup (\mathcal{X}_{-} \times\mathcal{X}_{-})$, 
we can replace $\sigma$ by $\tilde{\sigma}$ defined by 
\begin{equation*}
\tilde{\sigma}_{-+} := 0;\quad \tilde{\sigma}_{--} := \delta_{\Delta} \mu;\quad \tilde{\sigma}_{++} := \delta_{\Delta} \nu;
\end{equation*}
where $\delta_{\Delta}$ denotes the delta-measure on the diagonal 
$\Delta := \{(x,y) \in \mathcal{X} \times \mathcal{X}\,:\, x=y\}$. Thus, 
\begin{equation*}
\tilde{\sigma}_{++}(A \times B) = \nu(A \cap B),\quad A,B \in \mathcal{B}(\mathcal{X}).
\end{equation*}
The claim follows from the fact that 
\begin{equation*}
\int_{\mathcal{X}\times \mathcal{X}} \rho(x,y) \tilde{\sigma}_{++}(dx,dy) = 0 \quad \textrm{and} \quad \int_{\mathcal{X}\times \mathcal{X}} \rho(x,y) \tilde{\sigma}_{--}(dx,dy) = 0.
\end{equation*}
To see that, let $\epsilon > 0$ and let  $K \subset (\mathcal{X}_{+} \times \mathcal{X}_{+})$ be a compact set such that 
\begin{equation*}
\int_{(\mathcal{X}_{+} \times \mathcal{X}_{+}) \setminus K} \rho(x,y)\,\tilde{\sigma}(dx,dy) < \epsilon.
\end{equation*}
Then subdivide $K$ into a finite number $N$ of disjoint subsets $\mathcal{U}_{i}$ of diameter $<\epsilon$. Then 
\begin{equation*}
\begin{split} 
\int_{\mathcal{X}_{+} \times \mathcal{X}_{+}} \rho(x,y)\,\tilde{\sigma}(dx,dy) &= 
\int_{(\mathcal{X}_{+} \times \mathcal{X}_{+}) \setminus K} \rho(x,y)\,\tilde{\sigma}(dx,dy)  + 
\sum_{i=1}^{N} \int_{\mathcal{U}_{i} \times \mathcal{U}_{i}} \rho(x,y) \,\tilde{\sigma}(dx,dy) \\ 
&\leq \epsilon + \epsilon \sum_{i=1}^{N} \mu_{+}(\mathcal{U}_{i}) \leq 2\epsilon.
\end{split} 
\end{equation*}
We can therefore replace $\sigma$ by $\tilde{\sigma}$ on $(\mathcal{X}_{+} \times \mathcal{X}_{+}) 
\cup (\mathcal{X}_{-} \times \mathcal{X}_{+}) \cup (\mathcal{X}_{-} \times \mathcal{X}_{-})$. It remains to minimize 
\begin{equation*}
\int_{\mathcal{X}_{+}\times \mathcal{X}_{-}} \rho(x,y)\,\sigma_{+-}(dx,dy),
\end{equation*}
subject to the conditions 
\begin{equation*}
\int_{\mathcal{X}_{+} \times B} \sigma_{+-}(dx,dy) = \nu(B) - \mu(B) \quad \textrm{and} \quad \int_{A\times \mathcal{X}_{-}} \sigma_{+-}(dx,dy) = 
\mu(A) - \nu(A). 
\end{equation*} 
We now reformulate the minimization problem. Denote by $\mathcal{M}^{b}(\mathcal{X}_{\pm})$ the space of bounded measures on $\mathcal{X}_{\pm}$. Let $\mathcal{M}^{b}_{\rho}(\mathcal{X}_{+}\times \mathcal{X}_{-})$ denote the space of bounded measures $\sigma$ on 
$\mathcal{X}_{+} \times \mathcal{X}_{-}$ such that
\begin{equation*}
\int_{\mathcal{X}_{+}\times \mathcal{X}_{-}} \rho(x,y)\,\vert \sigma \vert(dx,dy) <+\infty.
\end{equation*}
Its dual space is the space of continuous functions satisfying 
$\vert f(x,y)\vert \leq C (1+\rho(x,y))$ for some constant $C>0$ and all $(x,y) \in \mathcal{X}_{+}\times \mathcal{X}_{-}$. Introduce the functions 
$F: \mathcal{M}^{b}_{\rho}(\mathcal{X}_{+}\times \mathcal{X}_{-}) \to  \mathbb{R} \cup \{+\infty\}$ and $G: \mathcal{M}^{b}(\mathcal{X}_{+}) \oplus \mathcal{M}^{b}(\mathcal{X}_{-}) \to \mathbb{R} \cup \{+\infty\}$ respectively defined as
\begin{equation} 
\label{Fsigma}
F(\sigma) := 
\begin{cases} 
\displaystyle{\int_{\mathcal{X}_{+}\times \mathcal{X}_{-}} \rho(x,y)\,\sigma(dx,dy)} &\text{if $\sigma \geq 0$;} \\ 
+\infty &\text{otherwise.}
\end{cases}
\end{equation}
\begin{equation} 
\label{Galbe}
G((\alpha,\beta)) := \begin{cases} 
0 &\textrm{if $\alpha = (\mu - \nu)_{+}$
and $\beta = (\mu-\nu)_{-}$;} \\ 
+\infty  &\text{otherwise.} 
\end{cases} 
\end{equation}
Further, introduce the linear operator $A: \mathcal{M}^{b}_{\rho}(\mathcal{X}_{+}\times \mathcal{X}_{-}) \to \mathcal{M}^{b}(\mathcal{X}_{+}) \oplus \mathcal{M}^{b}(\mathcal{X}_{-})$ defined by 
\begin{equation*} 
A \sigma := \left(\int_{\mathcal{X}_{-}} \sigma(\cdot\,, dy), \int_{\mathcal{X}_{+}} \sigma(dx, \cdot\,)\right). 
\end{equation*}
In view of \eqref{Fsigma}-\eqref{Galbe}, let  $H: \mathcal{M}^{b}_{\rho}(\mathcal{X}_{+}\times \mathcal{X}_{-}) \times (\mathcal{M}^{b}(\mathcal{X}_{+}) \oplus \mathcal{M}^{b}(\mathcal{X}_{-})) \rightarrow \mathbb{R}\cup \{+\infty\}$ defined as
\begin{equation}
\label{Hsigalbe}
H\left(\sigma, (\alpha,\beta)\right) := F(\sigma) + G\left(A \sigma - (\alpha,\beta)\right).
\end{equation}
The minimization problem then becomes 
\begin{equation*}
\inf_{\sigma \in \mathcal{M}^{b}_{\rho}(\mathcal{X}_{+}\times \mathcal{X}_{-})} H\left(\sigma,(0,0)\right).
\end{equation*}
Denoting by $C^{b}(\mathcal{X}_{\pm})$ the space of bounded continuous functions on $\mathcal{X}_{\pm}$, the dual problem is
\begin{equation*}
\sup_{(f,g)\in C^{b}(\mathcal{X}_{+}) \oplus C^{b}(\mathcal{X}_{-})} \left(-H^{*}(0,(f,g))\right).
\end{equation*} 
This is equivalent to the original problem because the Legendre inversion theorem is still valid for infinite-dimensional spaces as it only relies on the Hahn-Banach theorem. The same proof applies, see also \cite[Thm 1.4, Chapt. 2]{BP}). Let us now compute $H^{*}$. In view of \eqref{Hsigalbe}, we have,
\begin{multline*} 
H^{*}\left(\phi, (f,g)\right) \\
\begin{split}
&= \sup_{\sigma \in \mathcal{M}^{b}_{\rho}(\mathcal{X}_{+}\times \mathcal{X}_{-})} 
\sup_{(\alpha,\beta) \in \mathcal{M}^{b}(\mathcal{X}_{+}) \oplus \mathcal{M}^{b}(\mathcal{X}_{-})}
\left(\langle \phi, \sigma \rangle + \langle f,\alpha \rangle + \langle g,\beta \rangle  - F(\sigma) -  G\left(A \sigma - (\alpha,\beta)\right) \right) \\ 
&= \sup_{\sigma \in \mathcal{M}^{b}_{\rho}(\mathcal{X}_{+}\times \mathcal{X}_{-})}  
\sup_{(\alpha',\beta') \in \mathcal{M}^{b}(\mathcal{X}_{+}) \oplus \mathcal{M}^{b}(\mathcal{X}_{-})}
\left( \left\langle \phi, \sigma \right\rangle + \left\langle (f,g), A \sigma - (\alpha',\beta')  \right\rangle - 
F(\sigma) - G(\alpha',\beta') \right).
\end{split}
\end{multline*}
Inserting \eqref{Fsigma} and \eqref{Galbe}, we have,
\begin{align}
\label{Hstarph}
\begin{split}
H^{*}\left(\phi, (f,g)\right) &=\,\, \sup_{\sigma \in \mathcal{M}^{b}_{\rho}(\mathcal{X}_{+}\times \mathcal{X}_{-})}  \left(\left\langle \phi + A^T (f,g), \sigma \right\rangle - F(\sigma)\right) \\
&\,\, \qquad + \sup_{(\alpha',\beta') \in \mathcal{M}^{b}(\mathcal{X}_{+}) \oplus \mathcal{M}^{b}(\mathcal{X}_{-})}  
\left(-\langle (f,g), (\alpha',\beta') \rangle - G(\alpha',\beta')\right), 
\end{split} \nonumber
\\
&=\,\,\sup_{\sigma \in \mathcal{M}^{b}_{\rho}(\mathcal{X}_{+}\times \mathcal{X}_{-})}  \left(\left\langle \phi + A^T (f,g), \sigma \right \rangle 
- F(\sigma)\right) - \left\langle (f,g) ,  ((\mu-\nu)_{+},(\mu-\nu)_{-}) \right\rangle  \nonumber
\\
&= \,\,\begin{cases} -\langle (f,g) ,  ((\mu-\nu)_+,(\mu-\nu)_-) \rangle 
&\text{if $\phi + A^T (f,g) \leq  \rho $; } \\ +\infty &\text{otherwise.} \end{cases} 
\end{align}
Here, $A^{T} (f,g)$ is defined by 
\begin{equation*}
\left\langle A^T (f,g), \sigma \right\rangle = \left\langle (f,g), A \sigma \right\rangle =  \int_{\mathcal{X}_{+}\times \mathcal{X}_{-}} (f(x) + g(y))\, \sigma(dx,dy). 
\end{equation*}
As a result of \eqref{Hstarph}, the dual problem therefore consists in maximizing
\begin{equation*}
\int_{\mathcal{X}_{+}} f(x) (\mu-\nu)(dx) + \int_{\mathcal{X}_{-}} g(y) (\nu-\mu)(dy),
\end{equation*} 
subject to the condition that $f(x) + g(y) \leq \rho(x,y)$.\\
Defining $f(y) = -g(y)$ for $y \in \mathcal{X}_{-}$, we have a function $f$ defined on $\mathcal{X}$ and we must maximize 
\begin{equation*}
\int_{\mathcal{X}} f(x) (\mu-\nu)(dx),
\end{equation*} 
subject to the condition that $f(x) - f(y) \leq \rho(x,y)$ for $(x,y) \in \mathcal{X}_{+} \times \mathcal{X}_{-}$.\\ As in the discrete case, it remains to show that the maximum is attained when\\
(i) $\vert f(x) - f(x')\vert \leq \rho(x,x')$ for $(x, x') \in \mathcal{X}_{+}\times \mathcal{X}_{+}$ and for $(x, x') \in \mathcal{X}_{-}\times \mathcal{X}_{-}$; \\
(ii) $\vert f(x)-f(y)\vert \leq \rho(x,y)$ for $(x,y) \in \mathcal{X}_{+} \times\mathcal{X}_{-}$. \\
To prove (i), suppose that there exist $(x_{0}, x') \in \mathcal{X}_{+}\times \mathcal{X}_{+}$ such that $f(x_{0}) - f(x') > \rho(x_{0},x')$.\\ Then, we redefine $f$ on $\mathcal{X}_{+}$ as follows
\begin{equation*}
\tilde{f}(x) := \begin{cases} 
\inf_{y \in \mathcal{X}_{-}} \left(f(y) + \rho(x,y)\right)
&,\textrm{if $f(x) < f(x_{0})$;} \\ 
f(x) &,\text{otherwise.} 
\end{cases}
\end{equation*}
Note that $\tilde{f}(x) \geq f(x)$ since $f(x) \leq f(y) + \rho(x,y)$ for all $y \in \mathcal{X}_{-}$. Let us show that $\tilde{f}\in \mathrm{Lip}_{1}(\mathcal{X}_{+})$. Suppose $f(x_{1}) \leq f(x_{2})$. If $f(x_{2}) \geq f(x_{0})$ then $\tilde{f}(x_{2})- \tilde{f}(x_{1}) \leq f(x_{2})-f(x_{1}) \leq \rho(x_{1},x_{2})$ by (i). If $f(x_{2}) < f(x_{0})$ and $f(x_{1}) <  f(x_{0}) $ then $\tilde{f}(x_{1}) = f(y_{1}) - \rho(x_{1},y_{1})$ for some $y_{1} \in \mathcal{X}_{-}$ and 
$\tilde{f}(x_{2}) \leq f(y_{1}) - \rho(x_{2},y_{1})$ so $\tilde{f}(x_{2}) - \tilde{f}(x_{1}) \leq \rho(x_{1},y_{1}) - \rho(x_{2},y_{1}) \leq \rho(x_{1},x_{2})$. The case of $f(x_{2}) < f(x_{0})$ and $f(x_{1}) \geq f(x_{0})$ is straightforward. To continue with the proof of $(\mathrm{i})$,  note that $\tilde{f}(x') > f(x')$. Indeed, if $\tilde{f}(x') = f(y) + \rho(x',y)$ for some $y \in \mathcal{X}_{-}$ then $f(x') < f(x_{0}) - \rho(x_{0},x') \leq f(y) + \rho(x_{0},y) - \rho(x_{0},x')$. Therefore, 
\begin{equation*}
\int_{\mathcal{X}_{+}} \tilde{f}(x)\,(\mu-\nu)_{+}(dx) - \int_{\mathcal{X}_{-}} f(x)\,(\mu-\nu)_{-}(dx) >
\int_{\mathcal{X}_{+}} f(x)\,(\mu-\nu)_{+}(dx) - \int_{\mathcal{X}_{-}} f(x)\,(\mu-\nu)_{-}(dx).
\end{equation*} 
Besides, by definition, for any $y \in \mathcal{X}_{-}$, it is still true that $\tilde{f}(x) - f(y) \leq \rho(x,y)$. Therefore, $f$ cannot be a maximizer. Analogously, one proves that $f(y') - f(y_{0}) > \rho(y_{0},y')$ leads to a contradiction.\\
To prove (ii), assume that there exist $(x_{0}, y_{0}) \in \mathcal{X}_{+}\times \mathcal{X}_{-}$ such that $f(y_{0}) - f(x_{0}) > \rho(x_{0},y_{0})$. Then, we redefine $f$ on $\mathcal{X}_{-}$ as follows
\begin{equation}
\label{tildaf2}
\tilde{f}(y) := \begin{cases} 
\sup_{x \in \mathcal{X}_{+}} \left(f(x) - \rho(x,y)\right)  &,\textrm{if $f(y) > f(x_{0}) + \rho(x_{0},y)$;} \\ f(y) &,\textrm{otherwise.} 
\end{cases} 
\end{equation}
Note that $\tilde{f}(y) \leq f(y)$ since $f(y) \geq f(x) - \rho(x,y)$ for all $x \in \mathcal{X}_{+}$. Let us show that $\tilde{f}\in \mathrm{Lip}_{1}(\mathcal{X}_{-})$. Suppose $f(y_{1}) \leq f(y_{2})$. If $f(y_{1}) \leq f(x_{0}) + \rho(x_{0},y_{1})$ then $\tilde{f}(y_{2})- \tilde{f}(y_{1}) \leq f(y_{2})-f(y_{1}) \leq \rho(y_{1},y_{2})$ by (i). If $f(y_{1}) > f(x_{0}) + \rho(x_{0},y_{1})$ and $f(y_{2}) > f(x_{0}) + \rho(x_{0},y_{2}) $ then $\tilde{f}(y_{2}) = f(x_{2}) - \rho(x_{2},y_{2})$ for some $x_{2} \in \mathcal{X}_{+}$ and 
$\tilde{f}(y_{1}) \geq f(x_{2}) - \rho(x_{2},y_{1})$ so $\tilde{f}(y_{2}) - \tilde{f}(y_{1}) \leq \rho(x_{2},y_{1}) - \rho(x_{2},y_{2}) \leq \rho(y_{1},y_{2})$. The case of $f(y_{1}) > f(x_{0}) + \rho(x_{0},y_{1})$ and $f(y_{2}) \leq f(x_{0}) + \rho(x_{0},y_{2}) $ is straightforward.
To continue with the proof of $(\mathrm{ii})$, note that $\tilde{f}(y_{0}) < f(y_{0})$. Indeed, if $\tilde{f}(y_{0}) = f(x') - \rho(x',y_{0})$ then $\tilde{f}(y_{0}) - f(x_{0}) = f(x')-f(x_{0})- \rho(x',y_{0}) \leq \rho(x',x_{0}) - \rho(x',y_{0}) \leq \rho(x_{0},y_{0})$. 
By definition, $f(x) - \tilde{f}(y) \leq \rho(x,y)$ for any $x \in \mathcal{X}_{+}$. If $\tilde{f}(y) = f(x')-\rho(x',y)$ for some $x' \in \mathcal{X}_{+}$
then by (i), $\tilde{f}(y) - f(x) = f(x') - f(x) - \rho(x',y) \leq \rho(x,x') - \rho(x',y) \leq \rho(x,y)$. This concludes the proof. \qed

\subsection{The quantum harmonic crystal model revisited.}
\label{appdx2}

In this section, we give an application of the Dobrushin uniqueness theorem (Theorem \ref{Dobrushin}) to the quantum harmonic crystal model. In this lattice model, we associate with each site $j \in \mathbb{Z}^{d}$ a one-particle Hilbert space $L^{2}(\mathbb{R}) = L^{2}(\mathbb{R}, dx_{j})$ where $dx_{j}$ is the Lebesgue measure on $\mathbb{R}$.\\

\indent \textit{Notations.}  Hereafter, we identify $\tau$-periodic functions on $\mathbb{R}$ with functions on the 1-dimensional torus $\mathbb{T}_{\tau}:= \mathbb{R}/(\tau \mathbb{Z})$ which we define by identifying points in $\mathbb{R}$ that differ by $\tau n$ for some $n \in \mathbb{Z}$. The state space is the Banach space $\Omega_{\beta} := \mathcal{C}(\mathbb{T}_{\beta})$ of $\beta$-periodic continuous parametrised paths, endowed with the supremum norm $\Vert \cdot \Vert_{\infty}$ and equipped with the Borel $\sigma$-algebra $\mathcal{B}(\Omega_{\beta})$ of its subsets. We introduce the real Hilbert space $L^{2}(\mathbb{T}_{\beta})$. The standard inner product and norm are denoted by $\langle \cdot\,,\cdot\, \rangle_{\beta}$ and $\Vert \cdot\Vert_{2}$ respectively. By $\mathcal{B}(L^{2}(\mathbb{T}_{\beta}))$ we denote the Borel $\sigma$-algebra of subsets of $L^{2}(\mathbb{T}_{\beta})$. Note that $\Omega_{\beta} \in \mathcal{B}(L^{2}(\mathbb{T}_{\beta}))$ and $\mathcal{B}(\Omega_{\beta}) = \mathcal{B}(L^{2}(\mathbb{T}_{\beta})) \cap \Omega_{\beta}$. We refer to the beginning of Sec. \ref{secnot} for notations related to the configuration spaces.\\
\indent The quantum harmonic crystal is described by the formal translation-invariant Hamiltonian
\begin{equation}
	\label{HquaZ}
	H^{\textrm{qua}} := -  \sum_{j \in \mathbb{Z}^{d}} \frac{1}{2} \frac{d^{2}}{d x_{j}^{2}} + \sum_{j \in \mathbb{Z}^{d}} \frac{1}{2} \alpha x_{j}^{2}  + \sum_{j \in \mathbb{Z}^{d}} \sum_{l \in N_{1}(j)} \frac{1}{2} (x_{j} - x_{l})^{2},
\end{equation}
for some $\alpha>0$. \eqref{HquaZ} may be represented by the family $\{H_{\Lambda}\}_{\Lambda \in \mathcal{S}}$ of local Hamiltonians
\begin{equation}
	\label{HLqt}
	H_{\Lambda} := -  \sum_{j \in \Lambda} \frac{1}{2} \frac{d^{2}}{d x_{j}^{2}} + \sum_{j \in \Lambda} \frac{1}{2} \alpha x_{j}^{2}  + \sum_{j \in \Lambda} \sum_{l \in N_{1}(j) \cap \Lambda} \frac{1}{2} (x_{j} - x_{l})^{2}.
\end{equation}
By standard arguments, \eqref{HLqt} defines a family of bounded below essentially self-adjoint operators acting in  $L^{2}(\mathbb{R}^{\vert \Lambda\vert})$ with discrete spectrum. The definition \eqref{HLqt} corresponds to the free (or zero) boundary conditions. The system described by the family
$\{H_{\Lambda}\}_{\Lambda \in \mathcal{S}}$ of local Schr\"odinger operators can be equivalently described by the family of local path measures $\{\mu_{\Lambda}^{\beta}\}_{\Lambda \in \mathcal{S}}$ defined as follows. The semi-group $\{\exp(-\tau H_{\Lambda}),\,\tau>0\}$ associated with $H_{\Lambda}$ is positivity preserving and of trace class, i.e. $\mathrm{Tr}[\exp(-\tau H_{\Lambda})] < \infty$ for all $\tau >0$. Thus, for every $\beta >0$, the semi-group $\exp(-\beta H_{\Lambda})$ generates a stationary $\beta$-periodic Markov process, see, e.g., \cite[Sec. 3]{AHK}. This stochastic process has a canonical realisation on $(\Omega_{\beta}^{\Lambda},\mathcal{B}(\Omega_{\beta}^{\Lambda}))$ described by the measure $\mu_{\Lambda}^{\beta} \in \mathcal{P}(\Omega_{\beta}^{\Lambda})$, the marginal distributions of which are given by the integral kernels of the operator $\exp(-\tau H_{\Lambda})$, $\tau \in [0,\beta]$. By means of the Feynman-Kac formula, the measure $\mu_{\Lambda}^{\beta}$ on $(\Omega_{\beta}^{\Lambda},\mathcal{B}(\Omega_{\beta}^{\Lambda}))$ is then defined as
\begin{multline}
	\label{FKmea}
	\mu_{\Lambda}^{\beta}(d \underline{\omega}_{\Lambda}) := \frac{1}{Z_{\Lambda}^{\beta}} \exp\left(-\sum_{j \in \Lambda} \frac{1}{2}\alpha \int_{0}^{\beta} \omega_{j}(\tau)^{2} d\tau\right)\\
	\times \exp \left(-\sum_{j \in \Lambda} \sum_{l \in N_{1}(j) \cap \Lambda} \frac{1}{2}
	\int_{0}^{\beta} (\omega_{j}(\tau) - \omega_{l}(\tau))^{2} d\tau \right) \prod_{j \in \Lambda}
	\mu_{0}^{\beta}(d\omega_{j}),
\end{multline}
where $Z_{\Lambda}^{\beta}$ is a normalisation constant, and $\mu_{0}^{\beta}$ denotes the Brownian bridge measure on $(\Omega_{\beta},\mathcal{B}(\Omega_{\beta}))$ defined by means of the conditional Wiener measures with the condition $\omega_{j}(0) = \omega_{j}(\beta)$, see, e.g., \cite[Sec. 6.3.2]{BR2}. Thus defined,
$\{\mu_{\Lambda}^{\beta}\}_{\Lambda \in \mathcal{S}}$ forms the family of local (Euclidean) Gibbs distributions with zero boundary conditions.\\
\indent Next, define the embedding $\iota: \Omega_{\beta} \hookrightarrow L^{2}(\mathbb{T}_{\beta})$,
$\iota(f) = f$. Since $\Vert \iota(f) \Vert_{2} \leq \sqrt{\beta} \Vert f \Vert_{\infty}$, then $\iota$ is a continuous injection. $\mu_{0}^{\beta} \circ \iota^{-1}$ is the image measure of $\mu_{0}^{\beta}$ on $L^{2}(\mathbb{T}_{\beta})$ and $\mu_{0}^{\beta} \circ \iota^{-1}(\Omega_{\beta}) = 1$. The extension of the family of local Gibbs distributions to $((L^{2}(\mathbb{T}_{\beta}))^{\Lambda},\mathcal{B}((L^{2}(\mathbb{T}_{\beta}))^{\Lambda}))$ is then defined similarly to \eqref{FKmea} but with $\mu_{0}^{\beta} \circ \iota^{-1}$. Unless otherwise specified, we will not hereafter distinguish in our notation measures on $(\Omega_{\beta}^{\Lambda},\mathcal{B}(\Omega_{\beta}^{\Lambda}))$ from their extensions to $((L^{2}(\mathbb{T}_{\beta}))^{\Lambda},\mathcal{B}((L^{2}(\mathbb{T}_{\beta}))^{\Lambda}))$.\\

\indent As an application of the Dobrushin uniqueness theorem, we prove the following well-known result, see e.g., \cite{AHK},
\begin{proposition}
	\label{quantc}
	Consider the following formal energy functional (Hamiltonian) with nearest-neighbour interactions defined on $(L^{2}(\mathbb{T}_{\beta}))^{\mathbb{Z}^{d}}$ as
	\begin{equation}
		\label{Hqt}
		h^{\textrm{qua}} (\underline{\omega}) := \sum_{j \in \mathbb{Z}^{d}} \frac{1}{2}\alpha \int_{0}^{\beta} \omega_{j}(\tau)^{2} d\tau + \sum_{j \in \mathbb{Z}^{d}} \sum_{l \in N_{1}(j)} \frac{1}{2} \int_{0}^{\beta} (\omega_{j}(\tau) - \omega_{l}(\tau))^{2} d\tau.
	\end{equation}
	Then, provided that $\alpha>0$, there exists a unique limit (Euclidean) Gibbs distribution at any inverse temperature $\beta>0$ in $\mathcal{P}((L^{2}(\mathbb{T}_{\beta}))^{\mathbb{Z}^{d}})$, associated with $h^{\textrm{qua}}$, with marginal distributions satisfying \eqref{schcond}.
\end{proposition}

\noindent \textbf{Proof of Proposition \ref{quantc}.} Fix $j \in \mathbb{Z}^{d}$ and $\underline{\eta}, \underline{\eta}' \in (L^{2}(\mathbb{T}_{\beta}))^{N_{1}(j)}$ distinct. Set $\eta := \sum_{l \in N_{1}(j)} \eta_{l} \in L^{2}(\mathbb{T}_{\beta})$ and $\eta':=\sum_{l \in N_{1}(j)} \eta_{l}' \in L^{2}(\mathbb{T}_{\beta})$. In view of \eqref{Hqt}, the 1-point Gibbs distribution reads
\begin{equation*}
	\mu_{j}^{\beta}(d\omega \vert \underline{\eta}) :=
	\frac{1}{Z_{j}^{\beta}(\underline{\eta})}
	\exp\left(-\frac{1}{2} \alpha_{d} \Vert \omega \Vert_{2}^{2} + \langle \omega,\eta \rangle_{\beta}\right) \mu_{0}^{\beta}(d\omega),
\end{equation*}
where we set $\alpha_{d}:= \alpha + 2d$, and with
\begin{equation}
\label{Zfunc}
	Z_{j}^{\beta}(\underline{\eta}) := \int_{L^{2}(\mathbb{T}_{\beta})} \exp\left(-\frac{1}{2} \alpha_{d} \Vert \omega \Vert_{2}^{2} + \langle \omega,\eta \rangle_{\beta}\right) \mu_{0}^{\beta}(d\omega).
\end{equation}
We now construct a coupling in $\mathcal{P}(L^{2}(\mathbb{T}_{\beta})\times L^{2}(\mathbb{T}_{\beta}))$ such that the marginals coincide with the 1-point Gibbs distribution above with the different boundary conditions, see \eqref{sigjet}-\eqref{coin-0} below. To do so, introduce the 1-point correlation function defined by
\begin{equation}
	\label{1pointcorr}
	\rho_{j}^{\beta}(f \vert \underline{\eta}) := \int_{L^{2}(\mathbb{T}_{\beta})} \exp\left(i \langle f,
	\omega\rangle_{\beta}\right) \mu_{j}^{\beta}(d\omega \vert \underline{\eta}),\quad f \in  L^{2}(\mathbb{T}_{\beta}).
\end{equation}
Define the Fourier coefficients of $f$ as follows
\begin{equation}
	\label{Fourier2}
	\hat{f}_{\ell} := \frac{1}{\beta} \left\langle f, \exp\left(-\frac{2 i \pi}{\beta} \ell \,\cdot\,\right) \right\rangle_{\beta},\quad \ell \in \mathbb{Z}.
\end{equation}
We claim that, under the above conditions, \eqref{1pointcorr} can be rewritten as
\begin{equation}
	\label{rerhoj}
	\rho_{j}^{\beta}(f \vert \underline{\eta}) =
	\exp\left(\frac{\beta}{2} \sum_{\ell \in \mathbb{Z}}\left(\left(2 \pi \beta^{-1} \ell\right)^{2} + \alpha_{d}\right)^{-1}\left(- \vert \hat{f}_{\ell} \vert^{2} + 2 i \left( \Re(\hat{f}_{\ell}) \Re(\hat{\eta}_{\ell}) + \Im(\hat{f}_{\ell}) \Im(\hat{\eta}_{\ell})\right)\right)\right).
\end{equation}
For reader's convenience, the proof of \eqref{rerhoj} is deferred to the end of this section. Note that \eqref{rerhoj} is the characteristic function of a product of shifted Gaussian measures on $\mathbb{R}^{2}$ when $\ell \neq 0$ and $\mathbb{R}$ when $\ell = 0$, centered at $c_{\ell}(\beta)(\Re(\hat{\eta}_{\ell}),\Im(\hat{\eta}_{\ell}))$ and at $c_{0}(\beta) \hat{\eta}_{0}$ respectively, and with covariance
\begin{equation*}
	c_{\ell}(\beta) := \frac{1}{(2 \pi \beta^{-1} \ell)^{2} + \alpha_{d}}.
\end{equation*}
More precisely, 
\begin{equation*}
	\mu_{j}^{\beta}(\cdot\,\vert \underline{\eta}) = \bigotimes_{\ell \in \mathbb{Z}} \gamma_{\ell}^{\beta},
\end{equation*} 
where, when $\ell =0$ and $\ell \neq 0$ respectively,
\begin{gather*}
	\gamma_{0}^{\beta}(dx) := \exp\left(- \frac{\beta}{2 c_{0}(\beta)} (x - c_{0}(\beta)\hat{\eta}_{0})^{2}\right) \frac{dx}{\sqrt{2\pi c_{0}(\beta)}};\\
	\begin{split}
		\gamma_{\ell}^{\beta}(dx,dy) := \exp\left(-\frac{\beta}{2 c_{\ell}(\beta)}(x - c_{\ell}(\beta)\Re(\hat{\eta}_{\ell}))^{2}\right)
		&\frac{dx}{\sqrt{2\pi c_{\ell}(\beta)}} \\ &\times \exp\left(- \frac{\beta}{2 c_{\ell}(\beta)}(y - c_{\ell}(\beta)
		\Im(\hat{\eta}_{\ell}))^{2}\right) \frac{dy}{\sqrt{2\pi c_{\ell}(\beta)}}.
	\end{split}
\end{gather*}
Note that $x$ and $y$ above denote respectively the real and imaginary parts of $\hat{\omega}_{\ell}$.\\
Subsequently, we define the coupling $\sigma_{j;\underline{\eta},\underline{\eta}'}^{\beta} \in \mathcal{P}(L^{2}(\mathbb{T}_{\beta}) \times L^{2}(\mathbb{T}_{\beta}))$ as
\begin{equation}
	\label{sigjet}
	\sigma_{j;\underline{\eta},\underline{\eta}'}^{\beta}(d\omega, d\omega') := \bigotimes_{\ell \in \mathbb{Z}} \gamma_{\ell}^{\beta}(d \hat{\omega}_{\ell}) \delta\left(\hat{\omega}_{\ell}' - \hat{\omega}_{\ell} - c_{l}(\beta)(\hat{\eta}_{\ell}' - \hat{\eta}_{\ell})\right),
\end{equation}
for any $A,B \in \mathcal{B}(L^{2}(\mathbb{T}_{\beta}))$. Here, $\delta$ denotes the Dirac measure. We can readily check that
\begin{equation}
	\label{coin-0}
	\sigma_{j;\underline{\eta}, \underline{\eta}'}^{\beta}(A \times L^{2}(\mathbb{T}_{\beta})) = \mu_{j}^{\beta}(A \vert \underline{\eta})\quad\textrm{and}\quad \sigma_{j;\underline{\eta}, \underline{\eta}'}^{\beta} (L^{2}(\mathbb{T}_{\beta}) \times B) = \mu_{j}^{\beta}(B \vert \underline{\eta}').
\end{equation}
By Cauchy-Schwarz inequality, we have
\begin{multline*}
	\int_{L^{2}(\mathbb{T}_{\beta})\times L^{2}(\mathbb{T}_{\beta})} \Vert \omega - \omega'\Vert_{2}\, d \sigma_{j;\underline{\eta},\underline{\eta}'}^{\beta}(d \omega, d\omega') \\
	\leq
	\sqrt{\int_{L^{2}(\mathbb{T}_{\beta})\times L^{2}(\mathbb{T}_{\beta})} \Vert \omega - \omega'\Vert_{2}^{2}\, d \sigma_{j;\underline{\eta},\underline{\eta}'}^{\beta}(d \omega, d\omega')}
	= \sqrt{\sum_{\ell \in \mathbb{Z}} c_{\ell}(\beta)^{2} \vert \hat{\eta}_{\ell}- \hat{\eta}_{\ell}' \vert^{2}}.
\end{multline*}
To derive the right-hand side, we used Parseval's identity followed by a direct computation from \eqref{sigjet} and the definition of the $\gamma_{\ell}^{\beta}$'s. Finally, Parseval's identity yields
\begin{equation*}
	\rho_{W}\left(\mu_{j}^{\beta}(\cdot\,\vert \underline{\eta}), \mu_{j}^{\beta}(\cdot\,\vert \underline{\eta}')\right)
	\leq c_{0}(\beta) \Vert \eta - \eta'\Vert_{2} \leq \frac{1}{\alpha_{d}} \sum_{l \in N_{1}(j)} \Vert \eta_{l} - \eta_{l}'\Vert_{2},
\end{equation*}
and the Proposition follows from the condition \eqref{nniupbd} in Remark \ref{nearestn}. \qed

\begin{remark}
	\label{remmm3}
	In lattice models of quantum anharmonic crystals, see, e.g., \cite{Alber1,Alber2,KP} and references therein, the whole system is formally described by
	\begin{equation}
		\label{HquaZ2}
		\tilde{H}^{\textrm{qua}} := -  \sum_{j \in \mathbb{Z}^{d}} \frac{1}{2} \frac{d^{2}}{d x_{j}^{2}} + \sum_{j \in \mathbb{Z}^{d}} \frac{1}{2} \alpha x_{j}^{2} + \sum_{j \in \mathbb{Z}^{d}} g(x_{j})  + \sum_{j \in \mathbb{Z}^{d}} \sum_{l \in N_{1}(j)} \frac{1}{2} (x_{j} - x_{l})^{2}.
	\end{equation}
	\eqref{HquaZ2} may be represented by the corresponding family $\{\tilde{H}_{\Lambda}\}_{\Lambda \in \mathcal{S}}$ of local Hamiltonians
	\begin{equation}
		\label{HLqt2}
		\tilde{H}_{\Lambda} := -  \sum_{j \in \Lambda} \frac{1}{2} \frac{d^{2}}{d x_{j}^{2}} + \sum_{j \in \Lambda} (\frac{1}{2} \alpha x_{j}^{2} + g(x_{j})) + \sum_{j \in \Lambda} \sum_{l \in N_{1}(j) \cap \Lambda} \frac{1}{2} (x_{j} - x_{l})^{2}.
	\end{equation}
	Assuming that $g$ is continuous  and bounded from below, \eqref{HLqt2} defines a family of bounded below essentially self-adjoint operators acting in  $L^{2}(\mathbb{R}^{\vert \Lambda\vert})$ with compact resolvent. By mimicking the arguments above \eqref{FKmea}, $\{\tilde{H}_{\Lambda}\}_{\Lambda \in \mathcal{S}}$ can be equivalently described by the family $\{\tilde{\mu}_{\Lambda}^{\beta}\}_{\Lambda \in \mathcal{S}}$ of measures
	\begin{multline*}
		\tilde{\mu}_{\Lambda}^{\beta}(d \underline{\omega}_{\Lambda}) := \frac{1}{Z_{\Lambda}^{\beta}} \exp[-\sum_{j \in \Lambda} (\frac{1}{2}\alpha \int_{0}^{\beta} \omega_{j}(\tau)^{2} d\tau + \int_{0}^{\beta} g(\omega_{j}(\tau)) d\tau)]\\
		\times \exp [-\sum_{j \in \Lambda} \sum_{l \in N_{1}(j) \cap \Lambda} \frac{1}{2}
		\int_{0}^{\beta} (\omega_{j}(\tau) - \omega_{l}(\tau))^{2} d\tau] \prod_{j \in \Lambda}
		\mu_{0}^{\beta}(d\omega_{j}),
	\end{multline*}
	where $\mu_{0}^{\beta}$ the Brownian bridge measure. Thus defined, $\{\tilde{\mu}_{\Lambda}^{\beta}\}_{\Lambda \in \mathcal{S}}$ on $((L^{2}(\mathbb{T}_{\beta}))^{\Lambda},\mathcal{B}((L^{2}(\mathbb{T}_{\beta}))^{\Lambda}))$ forms the actual family of local (Euclidean) Gibbs distributions with zero boundary conditions.\\
	We refer to \cite{AKRT1,AKRT2} and reference therein for the uniqueness problem in some lattice models of quantum anharmonic crystals with translation-invariant Hamiltonians of type \eqref{HquaZ2}.
\end{remark}

We end this section with the proof of the identity in \eqref{rerhoj}. We first need the following lemma
\begin{lema}
	\label{singh}
	For any $\gamma>0$,
	\begin{equation}
		\label{sinh}
		\lim_{n \rightarrow \infty} \sqrt{\prod_{j=1}^{n-1} \left(2\left(1 - \cos\left(\frac{2\pi}{n} j\right)\right) + \left(\frac{\gamma}{n}\right)^{2}\right)} = 2 \sinh\left(\frac{\gamma}{2}\right).
	\end{equation}
\end{lema}

\noindent \textbf{Proof.} First note the following identity
\begin{equation*}
	\prod_{j=0}^{n-1}\left(2\left(1 - \cos\left(\frac{2\pi}{n} j\right)\right) + \left(\frac{\gamma}{n}\right)^{2}\right) = 2^{n} \prod_{j=0}^{n-1}\left(\cosh(\vartheta_{n}) - \cos\left(\frac{2\pi}{n} j\right)\right),
\end{equation*}
where we set $\vartheta_{n} := \mathrm{arcosh}(1 + \frac{\gamma^{2}}{2 n^{2}})$.
Recall that $\mathrm{arcosh}(x) = \ln(x + \sqrt{x^{2} - 1})$, $x \geq 1$. Next, use
\begin{equation*}
	2^{m-1} \prod_{k=0}^{m-1} \left(\cosh(x) - \cos\left(y + \frac{2\pi}{m} k\right)\right) = \cosh(mx) - \cos(m y),\quad m \in \mathbb{N}.
\end{equation*}
Letting $y=0$ and $m=n$ in the above formula, we then obtain,
\begin{equation*}
	\prod_{j=0}^{n-1}\left(2\left(1 - \cos\left(\frac{2\pi}{n} j\right)\right) + \left(\frac{\gamma}{n}\right)^{2}\right) = 2(\cosh(n \vartheta_{n}) - 1).
\end{equation*}
It remains to use that $\lim_{n \rightarrow \infty} n \vartheta_{n} = \lim_{n \rightarrow \infty} n\,\mathrm{arcosh}(1 + \frac{\gamma^{2}}{2 n^{2}}) = \gamma$, together with the identity $\sinh(\frac{x}{2}) = \mathrm{sign}(x) \sqrt{\frac{\cosh(x) - 1}{2}}$ and the lemma follows. \qed \\

\noindent \textbf{Proof of \eqref{rerhoj}}. Fix $j \in \mathbb{Z}^{d}$ and let $\beta>0$. Let $f \in L^{2}(\mathbb{T}_{\beta})$ and set $\eta := \sum_{l \in N_{1}(j)} \eta_{l} \in L^{2}(\mathbb{T}_{\beta})$. By abuse of notation, we set $\underline{\eta} := (\eta_{k})_{k=1}^{n}$ with $\eta_{k}:= \eta(\frac{\beta}{n} k)$. Also, set $\underline{f} := (f_{k})_{k=1}^{n}$ with $f_{k}:= f(\frac{\beta}{n} k)$. Introduce the sequence of functions $w_{n}^{\beta,\zeta}(\cdot\,\vert \underline{\eta}) = w_{n}^{\beta,\zeta}(\cdot\,,\underline{f}\vert \underline{\eta}): \mathbb{R}^{n} \rightarrow \mathbb{C}$ with $\zeta = 0,1$  defined as
\begin{multline*}
	w_{n}^{\beta,\zeta}(\underline{x}\vert \underline{\eta}) :=
	\left(\frac{2\pi \beta}{n}\right)^{-\frac{n}{2}}
	\exp\left(-\frac{n}{2 \beta}\left( \sum_{k=2}^{n} (x_{k} - x_{k-1})^{2} + (x_{n} - x_{1})^{2} \right)\right) \\
	\times \prod_{k=1}^{n} \exp\left(-\frac{\beta}{2n}\left(\alpha_{d} x_{k}^{2} - 2x_{k}\eta_{k}\right)\right) \exp\left(i \zeta \frac{\beta}{n} \left\langle  \underline{f}, \underline{x}\right\rangle\right).
\end{multline*}
In view of \eqref{Zfunc}-\eqref{1pointcorr}, by definition of Wiener measures, see, e.g.,  \cite[Sec. 6.3.2]{BR2}, we have,
\begin{equation}
	\label{ZLam2}
	\lim_{n \rightarrow \infty} \int_{\mathbb{R}^{n}}
	w_{n}^{\beta,\zeta} (\underline{x}\vert \underline{\eta}) \prod_{k=1}^{n} dx_{k} = \left\{\begin{array}{ll}
		Z_{j}^{\beta}(\underline{\eta}),\quad &\textrm{if $\zeta=0$};\\ Z_{j}^{\beta}(\underline{\eta}) \rho_{j}^{\beta}(\underline{f}\vert \underline{\eta}),\quad &\textrm{if $\zeta=1$}.\\
	\end{array}\right.
\end{equation}
To investigate the limit $n \rightarrow \infty$, we need a suitable rewriting.
Define $A_{n} \in \mathbb{M}_{n}(\mathbb{R})$ as follows
\begin{equation*}
	A_{n}:= \begin{pmatrix}
		\frac{2n}{\beta} & - \frac{n}{\beta} &  0 & \dots & 0 & -\frac{n}{\beta}  \\
		-\frac{n}{\beta} & \frac{2n}{\beta} &  -\frac{n}{\beta} & 0 & \dots & 0 \\
		0 & \ddots & \ddots & \ddots & \ddots & \vdots \\
		\vdots & \ddots & \ddots & \ddots & \ddots & \vdots \\
		0 & \dots &  0 & - \frac{n}{\beta}  & \frac{2n}{\beta} & - \frac{n}{\beta}  \\
		- \frac{n}{\beta}  & 0 & \dots & 0  & - \frac{n}{\beta} & \frac{2n}{\beta}
	\end{pmatrix}.
\end{equation*}
This leads to the following rewriting
\begin{equation}
	\label{smallw}
	w_{n}^{\beta,\zeta}(\underline{x} \vert \underline{\eta}) = \left(\frac{2\pi \beta}{n}\right)^{-\frac{n}{2}}
	\exp\left(-\frac{1}{2}  \left\langle \underline{x}, \left(A_{n} + \frac{\beta}{n} \alpha_{d} \mathbb{I}_{n}\right) \underline{x} \right\rangle\right) \exp\left(\frac{\beta}{n} \left\langle \underline{\eta} + i \zeta \underline{f}, \underline{x} \right\rangle\right).
\end{equation}
Noting that $A_{n}$ is a symmetric circulant matrix, $A_{n}$  can then be diagonalised by the use of discrete Fourier transform. Define
\begin{equation*}
	U_{\ell,k} := \frac{1}{\sqrt{n}} \exp\left(\frac{2i \pi}{n}  \ell k\right),\quad k=1,\ldots, n;\, \ell = 0,\ldots, n-1.
\end{equation*}
Putting $U := (U_{\ell,k=1},\ldots, U_{\ell, k=n})_{0 \leq \ell \leq n-1}$, we have,
\begin{equation}
	\label{unita}
	U A_{n} U^{*} = \mathrm{diag}\left(\lambda_{0}^{(n)},\ldots, \lambda_{n-1}^{(n)}\right),
\end{equation}
where the eigenvalues read
\begin{equation}
	\label{lambdal}
	\lambda_{\ell}^{(n)} := \frac{2n}{\beta}\left(1 - \cos\left(\frac{2\pi \ell}{n}\right)\right),\quad \ell = 0,\ldots, n-1.
\end{equation}
Consider the case when $n$ is odd. Then $\lambda_{\ell}^{(n)} = \lambda_{n-\ell}^{(n)}$, i.e. the eigenvalues
$\lambda_{\ell}^{(n)}$ with $\ell = 1,\ldots, \frac{n-1}{2}$ are two-fold degenerate. Corresponding real-valued eigenvectors lead to the transformation
\begin{equation}
	\label{uvl}
	u_{\ell} = \sqrt{\frac{2}{n}} \sum_{k=1}^{n} x_{k} \cos\left(\frac{2\pi k \ell}{n}\right) \quad \textrm{and}\quad
	v_{\ell} = \sqrt{\frac{2}{n}} \sum_{k=1}^{n} x_{k} \sin\left(\frac{2\pi k \ell}{n}\right),
\end{equation}
and, in the case of $\ell = 0$, we set
\begin{equation*}
	u_{0} := \frac{1}{\sqrt{n}} \sum_{k=1}^{n} x_{k}.
\end{equation*}
Note that, by writing $\omega_{n} := \frac{2\pi}{n}$, the matrix
\begin{equation*}
	O_{n} := \frac{1}{\sqrt{n}} \begin{pmatrix}
		1 & 1 & \dots & 1 & 1 \\
		\sqrt{2} \cos(\omega_{n}) & \sqrt{2} \cos(2 \omega_{n}) & \dots & \sqrt{2} \cos((n-1) \omega_{n}) & 1 \\
		\sqrt{2} \sin(\omega_{n}) & \sqrt{2} \sin(2 \omega_{n}) & \dots & \sqrt{2} \sin((n-1) \omega_{n}) & 0 \\
		\vdots & \vdots & \dots & \vdots & \vdots \\
		\sqrt{2} \cos(\frac{n-1}{2}\omega_{n}) & \sqrt{2} \cos(2\frac{n-1}{2}\omega_{n}) & \dots & \sqrt{2} \cos((n-1) \frac{n-1}{2} \omega_{n}) & 1 \\
		\sqrt{2} \sin(\frac{n-1}{2}\omega_{n}) & \sqrt{2} \sin(2 \frac{n-1}{2}\omega_{n}) & \dots & \sqrt{2} \sin((n-1)\frac{n-1}{2} \omega_{n}) & 0
	\end{pmatrix},
\end{equation*}
is orthogonal, and the inverse transformation is given by
\begin{equation*}
	x_{k} = \frac{1}{\sqrt{n}}\left(u_{0} + \sqrt{2} \sum_{\ell=1}^{\frac{n-1}{2}}\left(u_{\ell} \cos \left(\frac{2 \pi}{n} k \ell\right) +
	v_{\ell} \sin \left(\frac{2 \pi}{n} k \ell\right)\right)\right).
\end{equation*}
For any $\underline{x} \in \mathbb{R}^{n}$, set
\begin{equation*}
	z_{\ell} := \sum_{k=1}^{n} U_{\ell,k} x_{k}, \quad \ell = 0,\ldots, n-1.
\end{equation*}
Note that, in view of \eqref{uvl}, we have the identities
\begin{equation}
	\label{oddin}
	z_{\ell} = \left\{ \begin{array}{ll}
		u_{0}, \quad &\textrm{if $\ell = 0$}; \\
		\frac{1}{\sqrt{2}} (u_{\ell} + i v_{\ell}),\quad &\textrm{if $\ell = 1,\ldots, \frac{n-1}{2}$};\\
		\frac{1}{\sqrt{2}} (u_{n-\ell} - i v_{n-\ell}),\quad &\textrm{if $\ell = \frac{n+1}{2},\ldots, n-1$}.
	\end{array}\right.
\end{equation}
From the above features, we get,
\begin{equation*}
	\langle \underline{x}, A_{n} \underline{x} \rangle = \sum_{k,k'=1}^{n} (A_{n})_{k,k'} \left(\sum_{\ell=0}^{n-1}
	\overline{U_{k,\ell}} z_{\ell}\right) \left(\sum_{\ell'=0}^{n-1} \overline{U_{k',\ell'}} z_{\ell'}\right) = \sum_{\ell, \ell'=0}^{n-1} \overline{z}_{\ell} \left(UA_{n} U^{*}\right)_{\ell, \ell'}z_{\ell'},
\end{equation*}
and by using \eqref{unita} along with \eqref{oddin},
\begin{equation}
	\label{intermr1}
	\langle \underline{x}, A_{n} \underline{x} \rangle = \sum_{\ell = 0}^{n-1} \lambda_{\ell}^{(n)} \vert z_{l} \vert^{2} = \sum_{\ell = 1}^{\frac{n-1}{2}} \lambda_{\ell}^{(n)} \left(u_{\ell}^{2} + v_{\ell}^{2}\right).
\end{equation}
Defining similarly, 
\begin{equation}
	\label{defy}
	\tilde{y}_{\ell} := \sum_{k=1}^{n} U_{\ell,k} y_{k}, \quad \ell=0,\ldots, n-1,
\end{equation}
we also have,
\begin{equation*}
	\langle \underline{y}, \underline{x} \rangle = \sum_{k=1}^{n} \left(\sum_{\ell=0}^{n-1} \overline{U_{k,\ell}} \tilde{y}_{\ell}\right) \left(\sum_{\ell' = 0}^{n-1} \overline{U_{k,\ell'}} z_{\ell'}\right) = \sum_{\ell,\ell'=0}^{n-1} \tilde{y}_{\ell}\overline{z}_{\ell'}  \sum_{k=1}^{n} \frac{1}{\sqrt{n}} U_{\ell'-\ell,k} = \sum_{\ell=0}^{n-1} \tilde{y}_{\ell} \overline{z}_{\ell} ,
\end{equation*}
and by using \eqref{oddin}, it holds
\begin{equation}
	\label{intermr2}
	\langle \underline{y},\underline{x} \rangle = u_{0}\tilde{y}_{0} + \frac{1}{\sqrt{2}} \sum_{\ell = 1}^{\frac{n-1}{2}}
	\left(u_{\ell}(\tilde{y}_{\ell} + \tilde{y}_{n-\ell}) - i v_{\ell}(\tilde{y}_{\ell} - \tilde{y}_{n-\ell})\right).
\end{equation}
In view of \eqref{defy} and using \eqref{intermr2}, note the identity
\begin{equation}
	\label{intermr3}
	\left \langle \underline{\eta} + i \zeta \underline{f}, \underline{x} \right\rangle 
	= u_{0}(\tilde{\eta}_{0} + i \zeta \tilde{f}_{0}) + \sqrt{2} \sum_{\ell =1}^{\frac{n-1}{2}} \left(u_{\ell}\left(\Re(\tilde{\eta}_{\ell}) + i \zeta \Re(\tilde{f}_{\ell})\right) + v_{\ell}\left(\Im(\tilde{\eta}_{\ell}) + i \zeta\Im(\tilde{f}_{\ell})\right)\right).
\end{equation}
From \eqref{smallw}, \eqref{intermr1} along with \eqref{lambdal} and \eqref{intermr3}, we obtain,
\begin{multline*}
	\int_{\mathbb{R}^{n}}  w_{n}^{\beta,\zeta}(\underline{x}\vert \underline{\eta}) \prod_{k=1}^{n} d x_{k} = \\
	\left(\frac{2\pi \beta}{n}\right)^{-\frac{n}{2}} \int_{\mathbb{R}^{n}} du_{0} \left(\prod_{\ell=1}^{\frac{n-1}{2}} d u_{\ell}\right) \left(\prod_{\ell=1}^{\frac{n-1}{2}} d v_{\ell}\right)
	\exp\left(-\frac{1}{2} \frac{\beta}{n}\left(\frac{n}{\beta} \lambda_{0}^{(n)} + \alpha_{d}\right) u_{0}^{2} + \frac{\beta}{n}\left(\tilde{\eta}_{0} + i \zeta \tilde{f}_{0}\right)u_{0}\right) \\
	\times \prod_{\ell=1}^{\frac{n-1}{2}} \exp\left(-\frac{1}{2} \frac{\beta}{n}\left(\frac{n}{\beta} \lambda_{\ell}^{(n)} + \alpha_{d}\right) u_{l}^{2} + \frac{\beta}{n} \sqrt{2} \left(\Re(\tilde{\eta}_{\ell}) + i \zeta \Re(\tilde{f}_{\ell})\right) u_{l}\right) \\
	\times \prod_{\ell=1}^{\frac{n-1}{2}} \exp\left(-\frac{1}{2} \frac{\beta}{n}\left(\frac{n}{\beta} \lambda_{\ell}^{(n)} + \alpha_{d}\right) v_{l}^{2} + \frac{\beta}{n} \sqrt{2} \left(\Im(\tilde{\eta}_{\ell}) + i \zeta \Im(\tilde{f}_{\ell})\right) v_{l}\right).
\end{multline*}
By noticing that $\Re(\tilde{\eta}_{\ell}) = \Re(\tilde{\eta}_{n-\ell})$ and $\Im(\tilde{\eta}_{\ell}) = -\Im(\tilde{\eta}_{n-\ell})$, and similarly, $\Re(\tilde{f}_{\ell}) = \Re(\tilde{f}_{n-\ell})$ and $\Im(\tilde{f}_{\ell}) = -\Im(\tilde{f}_{n-\ell})$, we have, by direct calculation of the Gaussian integrals,
\begin{multline}
	\label{finwr2}
	\int_{\mathbb{R}^{n}}  w_{n}^{\beta,\zeta}(\underline{x}\vert \underline{\eta}) \prod_{k=1}^{n} d x_{k}= \\
	\prod_{\ell=0}^{n-1} \frac{\exp\left(\frac{\beta}{4 n} \left(\frac{n}{\beta} \lambda_{\ell}^{(n)} + \alpha_{d}\right)^{-1} \left(2 \vert \tilde{\eta}_{\ell} \vert^{2} - 2 \zeta^{2} \vert \tilde{f}_{\ell}\vert^{2} + i 4 \zeta \left(\Re(\tilde{\eta}_{\ell}) \Re(\tilde{f}_{\ell}) + \Im(\tilde{\eta}_{\ell})\Im(\tilde{f}_{\ell})\right)\right)\right)}{
		\frac{\beta}{n}\sqrt{\left(\frac{n}{\beta} \lambda_{\ell}^{(n)} + \alpha_{d}\right)}}.
\end{multline}
With a view to taking the limit $n \rightarrow \infty$, see \eqref{ZLam2}, note that 
\begin{equation}
	\label{limm1}
	\lim_{n \rightarrow \infty} \frac{1}{\sqrt{n}} \tilde{f}_{n-\ell} = \lim_{n \rightarrow \infty}
	\frac{1}{\beta} \frac{\beta}{n} \sum_{k=1}^{n} \exp\left(i \frac{2\pi (n- \ell)}{\beta} \frac{\beta k}{n}\right) f\left(\frac{\beta}{n} k\right)  = \hat{f}_{\ell}, 
\end{equation}
and similarly,
\begin{equation}
	\label{limm2}
	\lim_{n \rightarrow \infty} \frac{1}{\sqrt{n}} \tilde{\eta}_{n-\ell} = \hat{\eta}_{\ell}.
\end{equation}
From \eqref{finwr2} with $\zeta=0$ and using the expression \eqref{lambdal} for the eigenvalues, Lemma \ref{singh} yields 
\begin{equation}
	\label{Zbeteta}
	Z_{j}^{\beta}(\underline{\eta}) = \frac{\exp\left(\frac{\beta}{2} \sum_{\ell \in \mathbb{Z}} \left((2\pi \beta^{-1} \ell)^{2} + \alpha_{d}\right)^{-1} \vert \hat{\eta}_{\ell}\vert^{2}\right)}{2 \sinh\left(\frac{1}{2} \beta \sqrt{\alpha_{d}}\right)}.
\end{equation}
Here, we used \eqref{Fourier2}. In view of \eqref{ZLam2} with $\zeta =1$, \eqref{rerhoj}  follows from \eqref{finwr2} with $\zeta=1$ together with \eqref{Zbeteta}, \eqref{limm1}-\eqref{limm2} and \eqref{Fourier2}. \qed


{\small
\begin{thebibliography}{99}
\bibitem{AHK} Albeverio S., H\o egh-Krohn R., \textit{Homogeneous random fields and statistical mechanics}, J. Funct. Anal. \textbf{19}, 242--272 (1975)
\bibitem{AKMR} Albeverio S., Kondratiev Yu.A., Minlos A., Rebenko A.L., \textit{Small mass behaviour of quantum Gibbs states for lattice models with unbounded spins}, J. Stat. Phys. \textbf{92}(5/6), 1153--1172 (1999)
\bibitem{AKRT1} Albeverio S., Kondratiev Yu.G., R\"ockner M., Tsikalenko T.V., \textit{Uniqueness of Gibbs states for quantum lattice systems with non-local interaction}, Proba. Theory. Relat. Fields \textbf{108}, 193--218 (1997)
\bibitem{AKR} Albeverio S., Kondratiev Yu.A., Rebenko A.L., \textit{Peierls argument and long-range order behaviour of  quantum lattice systems with unbounded spins}, J. Stat. Phys \textbf{92}(5/6), 1137--1152 (1998)
\bibitem{AKRT2} Albeverio S., Kondratiev Yu.G., R\"ockner M., Tsikalenko T.V., \textit{Dobrushin's uniqueness for quantum lattice systems with nonlocal interaction}, Commun. Math. Phys. \textbf{189}(2), 621--630 (1997)
\bibitem{Alber1} Albeverio S., Kondratiev Y., Kozitzky Y., R\"ockner M.,  \textit{Euclidean Gibbs states of quantum lattice systems}, Rev. Math. Phys. \textbf{14} 1335 (2002)
\bibitem{Alber2} Albeverio S., Kondratiev Y., Pasurek T., R\"ockner M., \textit{Euclidean Gibbs measures on loop lattices: existence and a priori estimates}, The Annals of Probability \textbf{32}(1A), 153--190 (2004)
\bibitem{Arak1} Araki H., Ion P.D.F., \textit{On the equivalence of KMS and Gibbs conditions for states of quantum lattice systems}, Commun. Math. Phys. \textbf{35}, 1--12 (1974)
\bibitem{BP} Barbu V., Precupanu T., \textit{Convexity and Optimization in Banach Spaces}, Sijthoff \& Noordhoff Int. Publishers, Bucuresti, Romania, 
\bibitem{Bax} Baxter R.J., \textit{Potts model at the critical temperature}, J. Phys. C: Solid State Phys. \textbf{6}(23), L445--L448 (1973)
\bibitem{BK} Bellissard J., H\o egh-Krohn R., \textit{Compactness and the maximal Gibbs state for random Gibbs fields on a lattice}, Commun. Math. Phys. \textbf{84}, 297--327 (1982)
\bibitem{Billi2} Billingsey P., \textit{Probability and Measure}, Third Edition, John Wiley \& Sons, Inc., New York, 1995
\bibitem{Billi} Billingsey P., \textit{Convergence of Probability Measures}, Second Edition, John Wiley \& Sons, Inc., New York, 1999
\bibitem{Bol} Bolley F., \textit{Separability and completeness for the Wasserstein distance}, S\'eminaire de Probabilit\'es XLI, Lectures Notes in Mathematics, Springer, 371--377 (2008)
\bibitem{Bou} Bourbaki N., \textit{Int\'egration Chapitre IX}, \'El\'ement de math\'ematique, Hermann, Paris, 1969
\bibitem{BR2}  Bratelli O., Robinson D.W., \textit{Operator Algebras and Quantum Statistical Mechanics 2}, Second Edition, Springer-Verlag Berlin Heidelberg, 1996
\bibitem{COPP} Cassandro M., Olivieri E., Pellegrinotti A., Presutti E., \textit{Existence and uniqueness of DLR measures for unbounded spin systems}, Z. Wahrsch. verw. Gebiete \textbf{41}, 313--334 (1978)
\bibitem{D0} Dobrushin R.L., \textit{Existence of a phase transition in two- and three-dimensional Ising models}, Teor. Veroyatn. Primen. \textbf{10}, 209--230 (1965) (In Russian)
\bibitem{D1} Dobrushin R.L., \textit{The description of a random field by means of conditional probabilities and
conditions of its regularity}, Teor. Veroyatnost. i Primenen. \textbf{13}(2), 201--229 (1968) (In Russian);
Theory Probab. Appl. \textbf{13}(2), 197--224 (1968)
\bibitem{D2} Dobrushin R.L., \textit{Gibbsian random fields for lattice systems with pairwise interaction},
Funkts. Anal. i Prilozhen. \textbf{2}(4),  31--43 (1968) (In Russian); Funct. Anal. Appl. \textbf{2}(4),  292--301 (1968)
\bibitem{D3} Dobrushin R.L., \textit{The problem of uniqueness of a Gibbsian random field and the problem of phase transition}, Funkts. Anal. i Prilozhen. \textbf{2}(4), 44--57 (1968) (In Russian); Funct. Anal. Appl. \textbf{2}(4), 302--312 (1968)
\bibitem{D4} Dobrushin R.L., \textit{Gibbsian random fields. The general case}, Funkts. Anal. i Prilozhen. \textbf{3}(1), 27--35 (1969) (In Russian); Funct. Anal. Appl. \textbf{3}(1), 22--28 (1969)
\bibitem{D5} Dobrushin R.L., \textit{Prescribing a system of random variables by conditional distributions},
Teor. Veroyatnost. i Primenen.  \textbf{15}(3) 469--497  (1970) (In Russian); Theory Probab. Appl.
\textbf{15}(3), 458--486 (1970)
\bibitem{D6} Dobrushin R.L., Perchersky E.A., \textit{A criterion of the uniqueness of Gibbs fields in the non-compact case}, In Probability Theory  an Mathematical Statistics (Tbilisi, 1982), Lecture Notes in Math. 1021, Springer-Verlag, Berlin, 97--110 (1983)
\bibitem{DS1} Dobrushin R.L., Shlosman S.B., \textit{Constructive criterion for the uniqueness of Gibbs field}, Statistical Physics and Dynamical Systems: Rigorous Results, Progress in Physics Vol.10, pp. 347--370, Birkh\"auser, Boston-Basel-Stuttgart, 1985
\bibitem{DS2} Dobrushin R.L., Shlosman S.B., \textit{Completely analytical Gibbs fields}, Statistical Physics and Dynamical Systems: Rigorous Results, Progress in Physics Vol.10, pp. 371--403, Birkh\"auser, Boston-Basel-Stuttgart, 1985
\bibitem{DS3} Dobrushin R.L., Shlosman S.B., \textit{Completely analytical interactions. Constructive description}, J. Stat. Phys. \textbf{46}, 983--1014 (1987)
\bibitem{Do} Dorlas T.C., \textit{Statistical Mechanics, Fundamentals and Model Solutions}, Second Edition, CRC Press, Taylor \& Francis Group, Boca Raton, 2021
\bibitem{DS} Duneau M., Souillard B., \textit{Cluster properties of lattice and continuous systems}, Comm. Math. Phys. \textbf{47}, 155--166 (1976)
\bibitem{Ed} Edwards D.A., \textit{On the Kantorovich-Rubinstein theorem}, Exposi. Mathematicae \textbf{29}, 387--398 (2011)
\bibitem{FV} Friedli S., Velenik Y., \textit{Statistical mechanics of lattice systems: A concrete mathematical introduction}, Cambridge Univeristy Press, Cambridge, 2017
\bibitem{GMS} Gallavotti G., Miracle-Sole S., \textit{Correlation functions of a lattice system}, Comm. Math. Phys. \textbf{7}, 274--288 (1968)
\bibitem{Gas} Gass S.I., \textit{Linear Programming: Methods and Applications}, First Edition, McGraw-Hill, New York, 1958
\bibitem{Ge} Georgii H.O., \textit{Gibbs Measures and Phase Transitions}, Walter de Gruyter Studies in Mathematics 9, Berlin, New York, 1988
\bibitem{Gro} Gross L., \textit{Decay of correlations in classical lattice models at high temperature}, Comm. Math. Phys. \textbf{68}(1), 9--27 (1979)
\bibitem{Gro1} Gross L., \textit{Absence of second-order phase transitions in the Dobrushin uniqueness region}, J. Stat. Phys. \textbf{25}(1), 57--72 (1981)
\bibitem{GRS} Guerra F., Rosen L., Simon B., \textit{The $P(\varphi)_{2}$ quantum field theory as classical statistical mechanics}, Ann. Math. \textbf{101}, 111--189 (1975)
\bibitem{HHW} Haag R., Hugenholtz N.M., Winnink M., \textit{On the equilibrium states in Quantum Statistical Mechanics}, Commun. Math. Phys. \textbf{5}, 215--236 (1967).
\bibitem{HKW} Hintermann A., Kunz H., Wu F.Y., \textit{Exact results for the Potts model in two dimension}, J. Stat. Phys. \textbf{19}, 623--632 (1978)
\bibitem{HK} H\o egh-Krohn R., \textit{Relativistic quantum statistical mechanics in two-dimensional space-time}, Comm. Math. Phys. \textbf{38}, 195--224 (1974)
\bibitem{Hug} Hugenholtz N.M., \textit{C*- Algebras and Statistical Mechanics}, Proc. Symp. Pure Math., Vol. \textbf{38} Part 2, Ed. R. V. Kadison, (Providence, RI: American Mathematical Society), pp. 407--465, 1982
\bibitem{IS} Iagolnitzer D., Souillard B., \textit{On the analyticity in the potential in classical statistical mechanics}, Comm. Math. Phys. \textbf{60}, 131--152 (1978)
\bibitem{Isr0} Israel R.B., \textit{High temperature analyticity in classical lattice systems}, Comm. Math. Phys. \textbf{50}, 245--257 (1976)
\bibitem{Isr} Israel R.B., \textit{Convexity in the theory of lattice gases}, Princeton Series in Physics, Princeton University Press, Princeton, New Jersey, 1979
\bibitem{Kl} Klein P., \textit{Dobrushin's uniqueness theorem and the decay of correlations in continuous statistical mechanics}, Comm. Math. Phys. \textbf{86}(2), 227--246 (1982)
\bibitem{KP} Kozitsky Y., Pasurek T., \textit{Euclidean Gibbs measures of quantum anharmonic oscillators}, J. Stat. Phys. \textbf{127}, 985--1047 (2007); \textit{Addendum and Corrigendum}, J. Stat. Phys. \textbf{132}, 755--757 (2008)
\bibitem{KW} Kramers H.A., Wannier G.H., \textit{Statistics of the two-dimensional ferromagnet. Part I}, Phys. Rev. \textbf{60}(3), 252--262 (1941)
\bibitem{Ku} K\"unsch H., \textit{Decay of correlations under Dobrushin's uniqueness condition and its applications}, Comm. Math. Phys. \textbf{84}, 207--222 (1982)
\bibitem{LR} Lanford O.E., Ruelle D., \textit{Observables at infinity and states with short-range correlations in statistical mechanics}, Comm. Math. Phys. \textbf{13}, 194--215 (1969)
\bibitem{LP} Lebowitz J., Presutti E., \textit{Statistical mechanics of systems of unbounded spins}, Comm. Math. Phys. \textbf{50}, 195--218 (1976)
\bibitem{Lev} Levin S.L., \textit{Application of Dobrushin's uniqueness theorem to $N$-vector models}, Comm. Math. Phys. \textbf{78}(1), 65--74 (1980).
\bibitem{Liv} Livet F., \textit{The cluster updating Monte Carlo algorithm applied to the 3d Ising problem}, EPL \textbf{16}(2), 139 (1991)
\bibitem{M1} Minlos R.A., \textit{Limiting Gibbs' distribution}, Funkts. Anal. i Prilozhen. \textbf{1}(2), 60--73 (1967) (In Russian); Funct. Anal. Appl. \textbf{1}(2), 140--150  (1967)
\bibitem{M2} Minlos R. A.,  \textit{Regularity of the Gibbs limit distribution},  Funkts. Anal. i Prilozhen. \textbf{1}(3), 40--53 (1967) (In Russian); Funct. Anal. Appl. \textbf{1}(3),  206--217 (1967)
\bibitem{Mu} M\"unster A., \textit{Statistical thermodynamics}, Vol. 1 (Springer/Academic Press), 1969
\bibitem{On} Onsager L., \textit{Crystal Statistics. I. A Two-Dimensional Model with an Order-Disorder Transition}, Phys. Rev. \textbf{65}(3-4), 117--149 (1944)
\bibitem{Pa2} Park Y.M., \textit{The cluster expansion of classical and quantum lattice systems}, J. Stat. Phys \textbf{27}, 553--576 (1982)
\bibitem{Pa} Park Y.M., \textit{Quantum statistical mechanics of unbounded continuous spin systems}, J. Korean Math. Soc. \textbf{22}, 43--74 (1985)
\bibitem{Pa1} Park Y.M., \textit{Quantum statistical mechanics for superstable interactions: Bose-Einstein statistics}, J. Stat. Phys. \textbf{40}, 259--302 (1985)
\bibitem{PaYo1} Park Y.M., Yoo H.J., \textit{A characterization of Gibbs states of lattice boson systems}, J. Stat. Phys. \textbf{75}, 215--239 (1994)
\bibitem{PaYo2} Park Y.M., Yoo H.J., \textit{Uniqueness and clustering properties of Gibbs states for classical and quantum unbounded spin systems}, J. Stat. Phys. \textbf{80}, 223--271 (1995)
\bibitem{Potts} Potts R.B., \textit{Some generalized order-disorder transformation}, Proc. Camb. Philos. Soc. \textbf{48}(1), 106--109 (1952)
\bibitem{Pra} Prakash C., \textit{High-temperature differentiability of lattice Gibbs states by Dobrushin uniqueness techniques}, J. Stat. Phys. \textbf{31}(1), 169--228 (1983)
\bibitem{Rbk} Rebenko A.L., \textit{Euclidean Gibbs states for quantum continuous systems via cluster expansion. II. Bose and Fermi statistics}, Meth. of Func. Anal. and Topology \textbf{5}(2), 86--100 (1999)
\bibitem{Rok} Rockafellar R. T., Convex Analysis, Princeton Landmarks in Mathematics and Physics, Princeton University Press, Princeton, 1970
\bibitem{Roy} Royer G., \textit{Study of Dobrushin's critical coupling in rotator models}, Annales de l'I.H.P., Section A \textbf{44}(1), 29--38 (1986)
\bibitem{Ru2} Ruelle D., \textit{Statistical Mechanics: Rigorous Results}, W.A. Benjamin, New York, Amsterdam, 1969
\bibitem{Ru0} Ruelle D., \textit{Superstable interactions in classical statistical mechanics}, Comm. Math. Phys. \textbf{18}(2), 127--159 (1970)
\bibitem{Ru1} Ruelle D., \textit{Probability estimates for continuous spin systems}, Comm. Math. Phys. \textbf{50}, 189--194 (1976)
\bibitem{Ru3} Ruelle D., \textit{Thermodynamic formalism: The mathematical structures of equilibrium statistical mechanics}, Second Edition, Cambridge University Press, Cambridge, 2004
\bibitem{Sim1} Simon B., \textit{The $P(\varphi)_{2}$ Euclidean field theory}, Princeton University Press, Princeton, 1974
\bibitem{Sim3} Simon B., \textit{A remark on Dobrushin's uniqueness theorem}, Comm. Math. Phys. \textbf{68}, 183--185 (1979)
\bibitem{Sim2} Simon B., \textit{The statistical mechanics of lattice gases, Volume I}, Princeton University Press, Princeton, New Jersey, 1993
\bibitem{Si} Sinai Ya.G., \textit{Theory of phase transitions: Rigorous results}, Pergamon Press, First Edition, 1982
\bibitem{TB} Talapov A.L, Bl\"ote H.W.J., \textit{The magnetization of the 3D Ising model}, J. Phys. A: Math. Gen. \textbf{29}, 5727 (1996)
\bibitem{Val} Vallander S.S., \textit{Calculation of the Wasserstein distance between probability distributions on the line}, Theory Probab. Appl. 18(4) (1973), 784--786
\bibitem{VEVH} Van Enter A.C.D., Van Hemmen J.L., \textit{Absence of phase transitions in certain one-dimensional long-range random systems}, J. Stat. Phys. \textbf{39}, 1--13 (1985)
\bibitem{VEF2S} Van Enter A.C.D., Fernandez R., Schonmann R.H., Shlosman S.B., \textit{Complete analyticity of the 2d Potts model above the critical temperature}, Comm. Math. Phys. \textbf{189}, 373--393 (1997)
\bibitem{Wass} Wasserstein L.N., \textit{Markov processes on a countable product space describing large systems of automata}, Problemy Peredachi Informatsii, \textbf{5}(3), 64--73 (1969) (In Russian)
\bibitem{Wig} Wightman A.S., \textit{Convexity and the notion of equilibrium state in thermodynamics and statistical mechanics}, Introduction to: \textit{The convexity in the theory of lattice gases} by R.B. Israel, Princeton Series in Physics, Princeton University Press, Princeton, New Jersey, 1979, 168pp

\end{thebibliography}}
\end{document}